# L2-Relaxation for Economic Prediction

Zhentao Shi

Department of Economics, The Chinese University of Hong Kong

and

Yishu Wang *

Shenzhen University WeBank Institute of FinTech, Shenzhen University

**Abstract**

We leverage an ensemble of many regressors, the number of which can exceed the sample size, for economic prediction. An underlying latent factor structure implies a dense regression model with highly correlated covariates. We propose the L2-relaxation method for estimating the regression coefficients and extrapolating the out-of-sample (OOS) outcomes. This framework can be applied to policy evaluation using the panel data approach (PDA), where we further establish inference for the average treatment effect. In addition, we extend the traditional single unit setting in PDA to allow for many treated units with a short post-treatment period. Monte Carlo simulations demonstrate that our approach exhibits excellent finite sample performance for both OOS prediction and policy evaluation. We illustrate our method with two empirical examples: (i) predicting China's producer price index growth rate and evaluating the effect of real estate regulations, and (ii) estimating the impact of Brexit on the stock returns of British and European companies.

---

*Zhentao Shi: zhentao.shi@cuhk.edu.hk, 9/F Esther Lee Building, Department of Economics, The Chinese University of Hong Kong, Shatin, New Territories, Hong Kong SAR, China. Yishu Wang: wangy@link.cuhk.edu.hk, Shenzhen University WeBank Institute of FinTech, Huixing Building, Shenzhen University, No.3688 Nanhai Avenue, Nanshan District, Shenzhen, Guangdong, China. We thank Haiqiang Chen and Jianqing Fan for helpful comments. Shi acknowledges partial financial support from the Research Grants Council of Hong Kong (Project No. 14617423) and the National Natural Science Foundation of China (Project No. 72425007).

# 1 Introduction

Various predictive methods have been developed for exploring information and patterns with big data. In a typical high-dimensional linear regression model, the choice of suitable machine learning methods depends on whether the regression coefficients are sparse or dense. The off-the-shelf methods are LASSO (Tibshirani 1996) and ridge (Hoerl & Kennard 1970). Both techniques use shrinkage by adding a penalty term to the loss function to prevent over-fitting. LASSO leads to sparse feature selection, while ridge yields dense estimates.

Another well-known sparse method is the *Dantzig selector* (DS, Candes & Tao 2007). As a LASSO's "cousin" (Meinshausen et al. 2007), DS directly minimizes the L1-norm, subject to a feasible set that controls the tolerance for bias. DS, motivated by applications in compressed sensing (Candes & Tao 2005), inherits from the literature of *basis pursuit*. On the other hand, when the number of regressors is at least of the same order as the sample size, ridge's asymptotics based on Random Matrix Theory works well with linear transformations of independently and identically distributed (i.i.d.) random variables (Hastie et al. 2022), but such assumptions can be restrictive for temporally dependent data. In this paper, we propose the *L2-relaxation for linear regression* as a "cousin" of ridge, which substantially generalizes Shi et al. (2025)'s *L2-relaxation for forecast combination*.[1]

Although sparse methods are popular in applications, Giannone et al. (2021) highlight the importance of dense models and declare that sparsity is an "illusion" in economics and finance. Coefficients in dense models are small, but non-zero. This paper deals with dense models and establishes asymptotic guarantees for L2-relaxation in such an environment. It gets rid of Shi et al. (2025)'s *latent group structure*, which plays an essential role in their theoretical justification. Specifically, we embed the L2-relaxation into the linear regression based on panel data driven by a standard latent factor model without imposing group structures. This setting means that we must seek novel ways to build the theory, as the existing techniques peculiar to the group structure cannot be carried over.

We first establish consistency in Theorem 1 for the L2-relaxation coefficient estimation with high-dimensional panel data when both $N$ and $T$ are large, via the dual problem as a bridge, by virtue of the dual relationship between the sup-norm and the L1-norm. Common techniques for L1-minimization include the *null space property* (NSP, Cohen et al. 2009) and the *compatibility condition* (Bühlmann & van de Geer 2011, Bickel et al. 2009). Under the latent factor structure, however, they may not hold due to multicollinearity.[2] We take a unique route to approach the oracle target associated with the factor loadings via the oracle dual problem, from which we find a low-rank sparse support to avoid the curse of dimensionality. Once we have obtained the consistency of the coefficient estimator, we further derive in Theorem 2 the consistency of the L2-relaxation prediction. We find that the predictive performance is asymptotically irrelevant to the level of heteroskedasticity of the idiosyncratic error in the latent factor model. Monte Carlo simulations show strong finite sample performance of the L2-relaxation in the out-of-sample (OOS) prediction across different factor strength and heteroskedasticity levels.

---

[1] See Appendix A for the geometry of the listed methods illustrated in the $\mathbb{R}^2$ space.
[2] See Appendix B.4.2 and B.6.2 for detailed discussions about these two conditions.

As a use case, we apply L2-relaxation to Hsiao et al. 2012's (HCW, henceforth) panel data approach (PDA) for program evaluation. Similar to the synthetic control method (SCM, Abadie & Gardeazabal 2003, Abadie et al. 2010), PDA estimates the time-varying treatment effects of a single treated unit via the counterfactual prediction from multiple control units. As this paper focuses on linear regression models without constraints, it aligns with PDA, not SCM.[3] The original PDA is designed for low-dimensional data; when the number of control units is large, sparse methods like LASSO have been proposed (Li & Bell 2017, Carvalho et al. 2018). Driven by a latent factor structure, however, the coefficient of PDA is indeed dense, which calls for more suitable estimators for the high-dimensional PDA. When the post-treatment period is short, we further propose using multiple treated units. We illustrate our method with two empirical examples: (i) the effect of real estate policies on the growth rate of China's producer price index (PPI), and (ii) the impact of Brexit on the stock returns of British and European companies. L2-relaxation exhibits its advantages in the predictions and placebo tests in comparison with other methods.

The rest of the paper is organized as follows. Section 2 motivates and formulates the L2-relaxation for regression. Section 3 first introduces the high-dimensional panel data model driven by latent factors, which induces a dense linear regression, and then presents the consistency of the L2-relaxation estimator and prediction. Section 4 applies L2-relaxation in the PDA framework. Section 5 reports the simulation results, and two empirical applications follow in Section 6. All theoretical proofs are relegated to the appendix.

**Notations.** For any positive integer $n$, define $[n] = \{1, 2, \cdots, n\}$. Let $\boldsymbol{I}_n$ to be an $n \times n$ identity matrix, while $\boldsymbol{1}_n$ and $\boldsymbol{0}_n$ respectively denote $n$-dimensional vectors with all elements being 1 and 0. We use a generic $m \times n$ matrix $\boldsymbol{A} = (a_{ij})_{i \in [m], j \in [n]}$ and an $n$-dimensional vector $\boldsymbol{x} = (x_i)_{i \in [n]}$ to illustrate the notations for submatrices and subvectors. For any compatible sets $\mathcal{S} \subset [m]$ and $\mathcal{Q} \subset [n]$, denote the submatrix $\boldsymbol{A}_{\mathcal{S}\mathcal{Q}} = (a_{ij})_{i \in \mathcal{S}, j \in \mathcal{Q}}$, $\boldsymbol{A}_{\mathcal{S}\cdot} = (a_{ij})_{i \in \mathcal{S}, j \in [n]}$, $\boldsymbol{A}_{\cdot\mathcal{Q}} = (a_{ij})_{i \in [m], j \in \mathcal{Q}}$, and subvector $\boldsymbol{x}_{\mathcal{Q}} = (x_i)_{i \in \mathcal{Q}}$. Let $\phi_{\max}(\cdot)$ and $\phi_{\min}(\cdot)$ be the largest and smallest eigenvalues of a Hermitian matrix, respectively. Denote $\|\boldsymbol{A}\|_{\infty} = \max_{i \in [m], j \in [n]} |a_{ij}|$ and $\|\boldsymbol{A}\|_2 = \sqrt{\phi_{\max}(\boldsymbol{A}'\boldsymbol{A})}$ for a matrix $\boldsymbol{A}$, which also apply to a vector. Furthermore, define the L1-norm by $\|\boldsymbol{x}\|_1 = \sum_{i=1}^n |x_i|$. Let $\mathcal{E}_{\mathcal{S}}(\boldsymbol{x}_t) = |\mathcal{S}|^{-1} \sum_{t \in \mathcal{S}} \boldsymbol{x}_t$ be a shorthand notation for the sample average of time series (or random vector) $\{\boldsymbol{x}_t\}$ over any given index set $\mathcal{S}$; let $\Gamma_{\mathcal{S}}(\boldsymbol{x}_t, \boldsymbol{y}_t') = \mathcal{E}_{\mathcal{S}}\left([\boldsymbol{x}_t - \mathcal{E}_{\mathcal{S}}(\boldsymbol{x}_t)][\boldsymbol{y}_t - \mathcal{E}_{\mathcal{S}}(\boldsymbol{y}_t)]'\right)$ be the sample covariance for time series $\{\boldsymbol{x}_t\}$ and $\{\boldsymbol{y}_t\}$ over $\mathcal{S}$. An *absolute constant* is a positive real number invariant with the sample size and bounded away from 0 and $\infty$.

---

[3]Readers interested in L2-relaxation for SCM are referred to Liao et al. (2025).

## 2 L2-relaxation for Linear Regression

Consider a linear regression model

$$y_t = \alpha^0 + \boldsymbol{x}_t'\boldsymbol{\beta}^0 + \epsilon_t, \tag{1}$$

where $\boldsymbol{x}_t = (x_{1t}, \cdots, x_{Nt})'$ contains $N$ regressors, and $\boldsymbol{\beta}^0 = (\beta_1^0, \cdots, \beta_N^0)'$ is the coefficient vector. We learn the model from the training sample indexed by $t \in \mathcal{T}_1$ with the length $|\mathcal{T}_1| = T_1$, and then use the estimated coefficient to conduct OOS prediction in the testing dataset indexed by $t \in \mathcal{T}_2$ with the length $|\mathcal{T}_2| = T_2$.

The most familiar estimator for such a linear model is the ordinary least squares (OLS). Given the training data, denote $\hat{\boldsymbol{\Sigma}} := \Gamma_{\mathcal{T}_1}(\boldsymbol{x}_t, \boldsymbol{x}_t')$ and $\hat{\boldsymbol{\eta}} := \Gamma_{\mathcal{T}_1}(\boldsymbol{x}_t, y_t)$. The optimization problem of OLS implies the Kuhn-Karush-Tucker (KKT) condition $\hat{\boldsymbol{\Sigma}}\boldsymbol{\beta} = \hat{\boldsymbol{\eta}}$. When $\hat{\boldsymbol{\Sigma}}$ is of full rank, which holds in general under the low-dimensional context $N \ll T_1$, we have the OLS estimator $\hat{\boldsymbol{\beta}}^{\mathrm{LS}} = (\hat{\boldsymbol{\Sigma}})^{-1}\hat{\boldsymbol{\eta}}$. When $N > T_1$, however, $\hat{\boldsymbol{\Sigma}}$ must be rank-deficient, and thus multiple solutions satisfy the sample KKT condition. Even if $N < T_1$ but $N$ is close to $T_1$, the solution is numerically unstable as some small eigenvalues of $\hat{\boldsymbol{\Sigma}}$ are close to zero.

Among all solutions to the KKT condition when $N > T_1$, one of them minimizes the L2-norm of the coefficients. Similar to ridge, shrinking the L2-norm of $\boldsymbol{\beta}$ by a reasonable scale can stabilize the numerical solution and avoid in-sample over-fitting. More specifically, consider the following L2-minimization problem:

$$\min_{\boldsymbol{\beta}} \frac{1}{2}\|\boldsymbol{\beta}\|_2^2, \quad \text{s.t. } \hat{\boldsymbol{\Sigma}}\boldsymbol{\beta} = \hat{\boldsymbol{\eta}}. \tag{2}$$

By virtue of "bias-variance trade-off", tolerating some small violation of the equality constraint by expanding the feasible set may lead to better OOS prediction. Following Shi et al. (2025), we relax the sup-norm of the sample moment condition by a tuning parameter $\tau$, to obtain the L2-relaxation problem for linear regression:

$$\min_{\boldsymbol{\beta}} \frac{1}{2}\|\boldsymbol{\beta}\|_2^2, \quad \text{s.t. } \|\hat{\boldsymbol{\eta}} - \hat{\boldsymbol{\Sigma}}\boldsymbol{\beta}\|_\infty \leq \tau, \tag{3}$$

where $\tau \geq 0$ is a tuning parameter. This problem has a unique solution for any given $\tau$, denoted by $\hat{\boldsymbol{\beta}}_\tau$, because its objective function is strictly convex and the feasible set is closed and convex.

The estimator depends on the tuning parameter $\tau$. The following are three special cases. (i) When $\tau = 0$ and $\hat{\boldsymbol{\Sigma}}$ is invertible, the constraint condition has a unique solution, and hence it is reduced to the OLS estimator $\hat{\boldsymbol{\beta}}^{\text{LS}}$. (ii) When $\tau = 0$ but $\hat{\boldsymbol{\Sigma}}$ is rank-deficient, the constraint condition has multiple solutions, among which Hastie et al. (2022)'s ridgeless regression estimator $\hat{\boldsymbol{\beta}} = \hat{\boldsymbol{\Sigma}}^+ \hat{\boldsymbol{\eta}}$, where $(\cdot)^+$ denotes the Moore-Penrose generalized inverse, minimizes the L2-norm. (iii) When $\tau \geq \|\hat{\boldsymbol{\eta}}\|_\infty$, all coefficients are shrunk to zero exactly, which leads to $\hat{\boldsymbol{\beta}}_\tau = \mathbf{0}_N$. In practice, multiple validation methods can be applied to choose a proper tuning parameter for time series data (Bergmeir et al. 2018), for example, $K$-fold cross validation (CV), $K$-block CV, and OOS validation. We will show in Table 1 later that the prediction performance of the validation method is reliable in simulations.

After estimating $\boldsymbol{\beta}^0$ by L2-relaxation, the intercept $\alpha^0$ can be simply recovered by $\hat{\alpha}_\tau = \mathcal{E}_{\mathcal{T}_1}(y_t) - [\mathcal{E}_{\mathcal{T}_1}(\boldsymbol{x}_t)]' \hat{\boldsymbol{\beta}}_\tau$ to yield zero mean residuals. The OOS prediction is

$$\hat{y}_{t,\tau} = \hat{\alpha}_\tau + \boldsymbol{x}_t' \hat{\boldsymbol{\beta}}_\tau = \mathcal{E}_{\mathcal{T}_1}(y_s) + [\boldsymbol{x}_t - \mathcal{E}_{\mathcal{T}_1}(\boldsymbol{x}_s)]' \hat{\boldsymbol{\beta}}_\tau, \quad \text{for } t \in \mathcal{T}_2. \tag{4}$$

In practice, as a step of feature engineering, we suggest standardizing the original data using the in-sample mean and standard deviation before estimation.

## 3 Theoretical Analysis

### 3.1 Latent Factor Structure

In this paper, we study the properties of L2-relaxation under the latent factor structure (Abadie et al. 2010, Hsiao et al. 2012). Let $\{\boldsymbol{x}_t\}$ and $\{y_t\}$ be driven by common factors:

$$\begin{cases} x_{it} = \mu_i + \boldsymbol{\lambda}_i' \boldsymbol{f}_t + u_{it}, & i \in \mathcal{N} := \{1, 2, \cdots, N\}, \\ y_t = \mu_0 + \boldsymbol{\lambda}_0' \boldsymbol{f}_t + u_{0t}, \end{cases} \tag{5}$$

where the intercepts are $\mu_0 = \mathbb{E}(y_t)$ and $\mu_i = \mathbb{E}(x_{it})$, a $q$-dimensional vector $\boldsymbol{f}_t$ of zero mean common factors is multiplied by the loading vector $\boldsymbol{\lambda}_i$, and $u_{it}$ is the idiosyncratic error with zero mean and finite variance $\sigma_i^2$. Without loss of generality, we normalize $\mathbb{E}(\boldsymbol{f}_t \boldsymbol{f}_t') = \boldsymbol{I}_q$. In matrix forms, if define $\boldsymbol{x}_t := (x_{1t}, \cdots, x_{Nt})'$, $\boldsymbol{\mu}_\mathcal{N} := (\mu_1, \cdots, \mu_N)'$, $\boldsymbol{\Lambda} := (\boldsymbol{\lambda}_1, \cdots, \boldsymbol{\lambda}_N)'$ and $\boldsymbol{u}_{\mathcal{N}t} := (u_{1t}, \cdots, u_{Nt})'$, then we rewrite

$$\boldsymbol{x}_t = \boldsymbol{\mu}_\mathcal{N} + \boldsymbol{\Lambda} \boldsymbol{f}_t + \boldsymbol{u}_{\mathcal{N}t}. \tag{6}$$

The unobservable $\boldsymbol{f}_t$ connects the observed target $y_t$ and the controls in $\boldsymbol{x}_t$, which makes the target predictable by the controls. To simplify the analysis, we assume $\mathbb{E}(\boldsymbol{u}_{\mathcal{N}t} | \boldsymbol{f}_t) = \mathbf{0}_N$ and $\mathbb{E}(u_{0t} | \boldsymbol{f}_t, \boldsymbol{u}_{\mathcal{N}t}) = 0$, under which the predictive capacity of $\boldsymbol{x}_t$ on $y_t$ stems solely from the latent factors.

The factor model (5) implies that $\epsilon_t$ in the linear regression (1) satisfies $\mathbb{E}(\epsilon_t) = 0$ and $\mathbb{E}(\boldsymbol{x}_t \epsilon_t) = \mathbf{0}_N$, and the true $\alpha^0$ and $\boldsymbol{\beta}^0$ can be explicitly written as

$$\alpha^0 = \mathbb{E}(y_t) - [\mathbb{E}(\boldsymbol{x}_t)]' \boldsymbol{\beta}^0 = \mu_0 - \boldsymbol{\mu}_\mathcal{N}' \boldsymbol{\beta}^0, \tag{7}$$

$$\boldsymbol{\beta}^0 = \{\mathbb{E}\left[(\boldsymbol{x}_t - \boldsymbol{\mu}_{\mathcal{N}})(\boldsymbol{x}_t - \boldsymbol{\mu}_{\mathcal{N}})'\right]\}^{-1} \mathbb{E}\left[(\boldsymbol{x}_t - \boldsymbol{\mu}_{\mathcal{N}})(y_t - \mu_0)\right] = (\boldsymbol{\Lambda}\boldsymbol{\Lambda}' + \boldsymbol{\Omega})^{-1}\boldsymbol{\Lambda}\boldsymbol{\lambda}_0, \quad (8)$$

if $\boldsymbol{\Lambda}\boldsymbol{\Lambda}' + \boldsymbol{\Omega}$ is invertible, where $\boldsymbol{\Omega} := \mathbb{E}(\boldsymbol{u}_{\mathcal{N}t}\boldsymbol{u}'_{\mathcal{N}t})$ denotes the covariance matrix of $\{\boldsymbol{u}_{\mathcal{N}t}\}$.

If $\boldsymbol{\Omega}$ is invertible, by the Woodbury matrix identity, (8) is equivalent to

$$\boldsymbol{\beta}^0 = \boldsymbol{\Omega}^{-1}\boldsymbol{\Lambda}(\boldsymbol{\Lambda}'\boldsymbol{\Omega}^{-1}\boldsymbol{\Lambda} + \boldsymbol{I}_q)^{-1}\boldsymbol{\lambda}_0, \quad (9)$$

which makes it clear that the coefficient vector $\boldsymbol{\beta}^0$ is dense in general, as highlighted by Liao et al. (2024). Only under some peculiar $\boldsymbol{\Lambda}$ and $\boldsymbol{\Omega}$ should many entries of $\boldsymbol{\beta}^0$ be exactly zero.

We first impose regularity conditions on the factor loadings.

**Assumption 1.** *For all $N > q$, there exist two absolute constants $C$ and $c$ such that*

*(a)* $\|\boldsymbol{\lambda}_0\|_\infty + \|\boldsymbol{\Lambda}\|_\infty \leq C$;

*(b)* $\max_{\mathcal{Q} \subset \mathcal{N}, |\mathcal{Q}|=q} \phi_{\min}\left(\boldsymbol{\Lambda}_{\mathcal{Q}\cdot}\boldsymbol{\Lambda}'_{\mathcal{Q}\cdot}\right) \geq c$.

Assumption 1 (a) bounds the factor loadings to be finite for simplicity. Condition (b) is a natural requirement that $\boldsymbol{\Lambda}$ is of full column rank: there exists a $q$-individual subset that makes $\boldsymbol{\Lambda}_{\mathcal{Q}\cdot}$ non-degenerate. We define $\xi_N := \phi_{\min}(\boldsymbol{\Lambda}'\boldsymbol{\Lambda}/N)$ as the average (over the cross section) factor strength, and we allow weak factors $\xi_N \to 0$.

The idiosyncratic errors are allowed to be heteroskedastic and cross-sectionally dependent. Let $\sigma_{\min}^2 := \phi_{\min}(\boldsymbol{\Omega})$ and $\sigma_{\max}^2 := \phi_{\max}(\boldsymbol{\Omega})$. We assume the following assumption to bound the eigenvalues of $\boldsymbol{\Omega}$.

**Assumption 2.** *There exist absolute constants $\underline{\sigma}^2$ and $\overline{\sigma}^2$ such that $\underline{\sigma}^2 \leq \sigma_{\min}^2 \leq \sigma_{\max}^2 \leq \overline{\sigma}^2$.*

With the above assumptions, we have the following properties about $\boldsymbol{\beta}^0$.

**Proposition 1.** *Under Assumption 1 and 2, we have*

*(a)* $\|\boldsymbol{\beta}^0\|_2 = \mathrm{O}(N^{-\frac{1}{2}}\xi_N^{-\frac{1}{2}})$;

*(b)* $\|\boldsymbol{\lambda}_0 - \boldsymbol{\Lambda}'\boldsymbol{\beta}^0\|_2 = \mathrm{O}(N^{-1}\xi_N^{-1})$.

Proposition 1 has important implications for the mean squared error (MSE) of prediction in the population. Note that for a generic candidate parameter $\boldsymbol{\beta}$, by the latent factor model (5), the MSE can be decomposed as

$$\begin{aligned}\mathrm{MSE}(\boldsymbol{\beta}) :=& \mathbb{E}\left([y_t - \mu_0 - (\boldsymbol{x}_t - \boldsymbol{\mu}_{\mathcal{N}})'\boldsymbol{\beta}]^2\right) \\ =& \sigma_0^2 + \boldsymbol{\beta}'\boldsymbol{\Omega}\boldsymbol{\beta} + \|\boldsymbol{\lambda}_0 - \boldsymbol{\Lambda}'\boldsymbol{\beta}\|_2^2 \leq \sigma_0^2 + \sigma_{\max}^2\|\boldsymbol{\beta}\|_2^2 + \|\boldsymbol{\lambda}_0 - \boldsymbol{\Lambda}'\boldsymbol{\beta}\|_2^2.\end{aligned} \quad (10)$$

In the second row of the above expression, $\sigma_0^2$ is the unpredictable component independent of $\boldsymbol{\beta}$. For the other two terms depending on $\boldsymbol{\beta}$, the quadratic form $\boldsymbol{\beta}'\boldsymbol{\Omega}\boldsymbol{\beta}$ plays the role of variance from the idiosyncratic error, which is bounded above by $\sigma_{\max}^2 \|\boldsymbol{\beta}\|_2^2$, whereas $\|\boldsymbol{\lambda}_0 - \boldsymbol{\Lambda}'\boldsymbol{\beta}\|_2^2$ is the squared bias from the factor structure. For $\boldsymbol{\beta}^0$, Proposition 1 shows that both the variance and bias will vanish as the sample size gets large. Here we derive these properties, which are also needed for SCM and imposed as assumptions in Ferman (2021, Assumption 3.2).

## 3.2 Oracle Problem and Dual

Faced with the high-dimensional problem, L2-relaxation controls the extent of the bias by the feasible set, and minimizes the upper bound of the variance with the objective function. As $q \ll N$, there exist multiple $\boldsymbol{\beta}$ such that $\boldsymbol{\lambda}_0 - \boldsymbol{\Lambda}'\boldsymbol{\beta} = \mathbf{0}_q$; and among these solutions, minimizing $\|\boldsymbol{\beta}\|_2^2$ could derive a qualified upper bound for the total loss. This perspective rationalizes the minimization problem in the population in view of (10):

$$\min_{\boldsymbol{\beta}} \frac{1}{2}\|\boldsymbol{\beta}\|_2^2 \quad \text{s.t.} \quad \boldsymbol{\Lambda}'\boldsymbol{\beta} = \boldsymbol{\lambda}_0, \tag{11}$$

which has a unique closed-form solution

$$\boldsymbol{\beta}^* = \boldsymbol{\Lambda}(\boldsymbol{\Lambda}'\boldsymbol{\Lambda})^{-1}\boldsymbol{\lambda}_0. \tag{12}$$

Being a noiseless approximation for $\boldsymbol{\beta}^0$ by ignoring $\boldsymbol{\Omega}$, this $\boldsymbol{\beta}^*$ will serve as our oracle target in Lemma 1 (a). By the fact that the Moore-Penrose generalized inverse $(\boldsymbol{\Lambda}\boldsymbol{\Lambda}')^+ = \boldsymbol{\Lambda}(\boldsymbol{\Lambda}'\boldsymbol{\Lambda})^{-2}\boldsymbol{\Lambda}'$, we also have $\boldsymbol{\beta}^* = (\boldsymbol{\Lambda}\boldsymbol{\Lambda}')^+\boldsymbol{\Lambda}\boldsymbol{\lambda}_0$, which is exactly the population oracle solution to the ridgeless regression given the factor loading but ignore the idiosyncratic error. In other words, the L2-relaxation shares the same oracle target with the ridgeless counterpart.

To construct an oracle problem for L2-relaxation, given the latent factor model (5), we first decompose the sample gram matrices as

$$\hat{\boldsymbol{\Sigma}} = \hat{\boldsymbol{\Sigma}}^* + \boldsymbol{\Omega} + \hat{\boldsymbol{\Sigma}}^e \quad \text{and} \quad \hat{\boldsymbol{\eta}} = \hat{\boldsymbol{\eta}}^* + \hat{\boldsymbol{\eta}}^e, \tag{13}$$

where

$$\begin{aligned}
\hat{\boldsymbol{\Sigma}}^* &:= \boldsymbol{\Lambda}\hat{\boldsymbol{\Sigma}}_f\boldsymbol{\Lambda}', \quad \hat{\boldsymbol{\Sigma}}^e := (\hat{\boldsymbol{\Omega}} - \boldsymbol{\Omega}) + \boldsymbol{\Lambda}\Gamma_{\mathcal{T}_1}(\boldsymbol{f}_t, \boldsymbol{u}'_{\mathcal{N}t}) + \Gamma_{\mathcal{T}_1}(\boldsymbol{u}_{\mathcal{N}t}, \boldsymbol{f}'_t)\boldsymbol{\Lambda}', \\
\hat{\boldsymbol{\eta}}^* &:= \boldsymbol{\Lambda}\hat{\boldsymbol{\Sigma}}_f\boldsymbol{\lambda}_0, \quad \hat{\boldsymbol{\eta}}^e := \Gamma_{\mathcal{T}_1}(\boldsymbol{u}_{\mathcal{N}t}, u_{0t}) + \boldsymbol{\Lambda}\Gamma_{\mathcal{T}_1}(\boldsymbol{f}_t, u_{0t}) + \Gamma_{\mathcal{T}_1}(\boldsymbol{u}_{\mathcal{N}t}, \boldsymbol{f}'_t)\boldsymbol{\lambda}_0,
\end{aligned} \tag{14}$$

with $\hat{\boldsymbol{\Sigma}}_f := \Gamma_{\mathcal{T}_1}(\boldsymbol{f}_t, \boldsymbol{f}'_t)$ and $\hat{\boldsymbol{\Omega}} := \Gamma_{\mathcal{T}_1}(\boldsymbol{u}_{\mathcal{N}t}, \boldsymbol{u}'_{\mathcal{N}t})$. Observe that in the loss decomposition (10), the predictability comes from the correlation induced by the common factors, whereas the idiosyncratic errors only add variance to the total loss. Hence, we treat $\hat{\boldsymbol{\Sigma}}^*$ and $\hat{\boldsymbol{\eta}}^*$ as the oracle objects. Consider an oracle L2-relaxation problem as

$$\min_{\boldsymbol{\beta}} \frac{1}{2}\|\boldsymbol{\beta}\|_2^2 \quad \text{s.t.} \quad \|\hat{\boldsymbol{\eta}}^* - \hat{\boldsymbol{\Sigma}}^*\boldsymbol{\beta}\|_\infty \leq \tau, \tag{15}$$

for which we denote its unique solution as $\hat{\boldsymbol{\beta}}^*_\tau$ with a given $\tau \geq 0$. The following lemma discusses the solution for $\hat{\boldsymbol{\beta}}^*_\tau$ when $\tau = 0$.

**Lemma 1** (Oracle primal solution). *Under Assumption 1, we have*

(a) $\hat{\boldsymbol{\beta}}_0^* = \boldsymbol{\beta}^*$ *if* $\mathrm{rank}(\hat{\boldsymbol{\Sigma}}_f) = q$;

(b) $\|\boldsymbol{\beta}^*\|_2 = \mathrm{O}(N^{-\frac{1}{2}}\xi_N^{-\frac{1}{2}})$;

(c) $\|\boldsymbol{\beta}^* - \boldsymbol{\beta}^0\|_2 = \mathrm{O}(N^{-\frac{3}{2}}\xi_N^{-\frac{3}{2}}) + \psi_{\max}\mathrm{O}(N^{-\frac{1}{2}}\xi_N^{-1})$, *where* $\psi_{\max} := \sigma_{\max}^2/\sigma_{\min}^2 - 1$.

Lemma 1 (a) shows that the oracle estimator always equals $\boldsymbol{\beta}^*$ when $\tau = 0$ and the sample covariance matrix of the factors $\hat{\boldsymbol{\Sigma}}_f$ is of full rank. Result (b) derives the convergence rate of $\boldsymbol{\beta}^*$, and (c) provides the distance between $\boldsymbol{\beta}^*$ and $\boldsymbol{\beta}^0$. Note that $\psi_{\max} = 0$ if and only if $\boldsymbol{\Omega} = \sigma^2 \boldsymbol{I}_N$, under which the $\mathrm{O}(N^{-\frac{1}{2}}\xi_N^{-1})$ term disappears.

Now that we have locked onto the desirable target $\boldsymbol{\beta}^*$, does the L2-relaxation estimator converge to the target? To establish the closeness between them, we introduce the dual problem as a bridge. The following lemma gives the dual of the primal problem (3).

**Lemma 2** (Dual problem). *For any $\tau \geq 0$, the dual problem of the L2-relaxation primal problem (3) is*

$$\min_{\boldsymbol{\gamma}} \left\{ \frac{1}{2}\boldsymbol{\gamma}'\hat{\boldsymbol{\Sigma}}'\hat{\boldsymbol{\Sigma}}\boldsymbol{\gamma} - \hat{\boldsymbol{\eta}}'\boldsymbol{\gamma} + \tau\|\boldsymbol{\gamma}\|_1 \right\}. \tag{16}$$

*The unique solution of the primal problem (3) can be recovered as* $\hat{\boldsymbol{\beta}}_\tau = \hat{\boldsymbol{\Sigma}}\hat{\boldsymbol{\gamma}}_\tau$, *where* $\hat{\boldsymbol{\gamma}}_\tau$ *is a solution to the above dual problem.*

Based on the oracle primal problem (15), the oracle dual is:

$$\min_{\boldsymbol{\gamma}} \left\{ \frac{1}{2}\boldsymbol{\gamma}'\hat{\boldsymbol{\Sigma}}^{*\prime}\hat{\boldsymbol{\Sigma}}^*\boldsymbol{\gamma} - \hat{\boldsymbol{\eta}}^{*\prime}\boldsymbol{\gamma} + \tau\|\boldsymbol{\gamma}\|_1 \right\}. \tag{17}$$

When $\tau \to 0$, in the limit the above problem reduces to

$$\min_{\boldsymbol{\gamma}} \|\boldsymbol{\gamma}\|_1 \quad \text{s.t.} \quad \hat{\boldsymbol{\Sigma}}^{*\prime}\hat{\boldsymbol{\Sigma}}^*\boldsymbol{\gamma} = \hat{\boldsymbol{\eta}}^*, \tag{18}$$

where the constraint is exactly the first order condition of (17) when $\tau = 0$. Our oracle dual happens to be a typical L1-minimization problem. We are interested in a sparse solution to problem (18), for the purpose of deriving tighter bounds. Define $\mathcal{S}^c := \mathcal{N} \setminus \mathcal{S}$ for any subset $\mathcal{S} \subset \mathcal{N}$. The following lemma solves for a $q$-sparse solution.

**Lemma 3** (Oracle dual solution). *Under Assumption 1, we have*

(a) *if* $\mathrm{rank}(\hat{\boldsymbol{\Sigma}}_f) = q$, *then (18) has a solution* $\hat{\boldsymbol{\gamma}}_0^*$ *such that* $\hat{\boldsymbol{\gamma}}_{0,\mathcal{Q}^*}^* = \left(\hat{\boldsymbol{\Sigma}}_{\cdot\mathcal{Q}^*}^{*\prime}\hat{\boldsymbol{\Sigma}}_{\cdot\mathcal{Q}^*}^*\right)^{-1}\hat{\boldsymbol{\eta}}_{\mathcal{Q}^*}^*$ *and* $\hat{\boldsymbol{\gamma}}_{0,(\mathcal{Q}^*)^c}^* = \boldsymbol{0}_{N-q}$ *with* $\mathcal{Q}^* = \arg\min_{\mathcal{Q} \in \mathscr{Q}_N} \left\|\left(\hat{\boldsymbol{\Sigma}}_{\cdot\mathcal{Q}^*}^{*\prime}\hat{\boldsymbol{\Sigma}}_{\cdot\mathcal{Q}}^*\right)^{-1}\hat{\boldsymbol{\eta}}_{\mathcal{Q}}^*\right\|_1$, *where for all* $N > q$, $\mathscr{Q}_N := \{\mathcal{Q} \subset \mathcal{N} : |\mathcal{Q}| = q \text{ and } \mathrm{rank}(\boldsymbol{\Lambda}_{\mathcal{Q}\cdot}) = q\}$ *must be non-empty*;

(b) *if* $\phi_{\min}(\hat{\boldsymbol{\Sigma}}_f) \geq c_f$ *for some absolute constant* $c_f$, *then any* $\hat{\boldsymbol{\gamma}}_0^*$ *in (a) satisfies that* $\|\hat{\boldsymbol{\gamma}}_0^*\|_1 = \left\|\hat{\boldsymbol{\gamma}}_{0,\mathcal{Q}^*}^*\right\|_1 = \mathrm{O_p}(N^{-1}\xi_N^{-1})$.

*Remark* 1. In compressed sensing, (18) is called a *basis pursuit* problem. A sufficient and necessary condition for exactly recovering every sparse solution from a basis pursuit problem is the well-known NSP (Cohen et al. 2009). However, the NSP may not hold under Assumption 1. See the detailed analysis and a counterexample in Appendix B.4.2. Despite the failure of NSP, a sparse solution still exists, since it is unnecessary for us to recover every sparse solution on a $q$-cardinal subset from the basis pursuit problem (18). Instead, we only need to recover those in Lemma 3 (a). In other words, for our purpose the NSP is no longer a necessary condition, but only a sufficient condition. Lemma 3 (a) guarantees the existence of a $q$-sparse solution.

## 3.3 Consistency

Consistency of the L2-relaxation estimator requests well-behaved noise levels from the sample. Besides all random components in the gram decomposition (14), we further estimate the sample variance of $\{u_{0t}\}$ by $\hat{\sigma}_0^2 := \mathcal{E}_{\mathcal{T}_1}\left([u_{0t} - \mathcal{E}_{\mathcal{T}_1}(u_{0t})]^2\right)$. The following assumption controls the in-sample sampling errors. Define $\varphi_1 := \sqrt{(\log N)/(N \wedge T_1)}$, where $a \wedge b = \min\{a, b\}$ for any $a, b \in \mathbb{R}$. It is required that $T_1$ should diverge at least faster than $\log N$, which is implied by $\varphi_1 \to 0$ as $N, T_1 \to \infty$.

**Assumption 3.** *(a)* $\|\hat{\boldsymbol{\Sigma}}_f - \boldsymbol{I}_q\|_\infty + \|\mathcal{E}_{\mathcal{T}_1}(\boldsymbol{f}_t)\|_\infty = \mathrm{O}_\mathrm{p}(T_1^{-\frac{1}{2}})$;

*(b)* $\|\hat{\boldsymbol{\Omega}} - \boldsymbol{\Omega}\|_\infty + \|\Gamma_{\mathcal{T}_1}(\boldsymbol{f}_t, \boldsymbol{u}'_{\mathcal{N}t})\|_\infty + \|\mathcal{E}_{\mathcal{T}_1}(\boldsymbol{u}_{\mathcal{N}t})\|_\infty = \mathrm{O}_\mathrm{p}(\varphi_1)$;

*(c)* $\|\Gamma_{\mathcal{T}_1}(\boldsymbol{f}_t, u_{0t})\|_\infty = \mathrm{O}_\mathrm{p}(T_1^{-\frac{1}{2}})$, $\|\Gamma_{\mathcal{T}_1}(\boldsymbol{u}_{\mathcal{N}t}, u_{0t})\|_\infty = \mathrm{O}_\mathrm{p}(\varphi_1)$;

*(d)* $|\hat{\sigma}_0^2 - \sigma_0^2| + |\mathcal{E}_{\mathcal{T}_1}(u_{0t})| = \mathrm{O}_\mathrm{p}(T_1^{-\frac{1}{2}})$.

Assumption 3 imposes high-level conditions on the in-sample estimation errors for all components, respectively. For the low-dimensional matrices or scalars, the sampling errors are of the order $T_1^{-1/2}$; for the high-dimensional matrices, the orders are uniformly controlled by $\sqrt{(\log N)/T_1} \leq \varphi_1$. These assumptions can be established from lower-level conditions; especially, the order of $\sqrt{(\log N)/T_1}$ for the sup-norm of sampling errors is commonplace in high-dimensional statistics (Wainwright 2019). Appendix B.5 further shows that the noise levels are controlled for all components in the decomposition (13) and $\hat{\boldsymbol{\eta}} - \hat{\boldsymbol{\Sigma}}'\hat{\boldsymbol{\Sigma}}\hat{\boldsymbol{\gamma}}_0^*$. Notice that these conditions hold under time series dependence that is much more general than linear transformations of i.i.d. random variables. The latter is essential for using Random Matrix Theory to perform asymptotic analysis for the ridge.

Given these assumptions and the preparatory lemmas, we have the consistency of $\hat{\boldsymbol{\beta}}_\tau$ to its oracle target $\boldsymbol{\beta}^*$ under the L2- and L1-norms.

**Theorem 1** (Consistency to oracle)**.** *Under Assumption 1–3, if the tuning parameter $\tau$*

*satisfies* $\tau + \varphi_1/\tau = o(\xi_N)$ *as* $N, T_1 \to \infty$, *then*

(a) $\|\hat{\boldsymbol{\beta}}_\tau - \boldsymbol{\beta}^*\|_2 = O_p(\sqrt{\tau/(N\xi_N)}) = o_p(N^{-\frac{1}{2}})$;

(b) $\|\hat{\boldsymbol{\beta}}_\tau - \boldsymbol{\beta}^*\|_1 = O_p(\sqrt{\tau/\xi_N}) = o_p(1)$.

Recall that when $\tau \geq \|\hat{\boldsymbol{\eta}}\|_\infty$, the high-dimensional constraints will be slack and thus the estimator $\hat{\boldsymbol{\beta}}_\tau = \mathbf{0}_N$ exactly. As a result, the L2-relaxation prediction (4) reduces to the trivial in-sample mean of $\{y_t : t \in \mathcal{T}_1\}$. From Lemma B.1 (b) and B.2 (b), by the triangle inequality $\|\hat{\boldsymbol{\eta}}\|_\infty = O_p(1)$. Therefore, shrinking $\tau$ to zero is necessary for yielding nontrivial predictions. On the other hand, according to Lemma B.3 (b), we request a sufficiently large $\tau$ to bound $\|\hat{\boldsymbol{\eta}} - \hat{\boldsymbol{\Sigma}}'\hat{\boldsymbol{\Sigma}}\hat{\boldsymbol{\gamma}}_0^*\|_\infty = O_p(\varphi_1/\xi_N)$ in the large sample theory. Ultimately, to achieve consistency, the tuning parameter $\tau$ should satisfy $\tau/\xi_N + \varphi_1/(\xi_N\tau) \to 0$.

*Remark* 2. Here the convergence rate is bounded by $\sqrt{\tau}$ rather than $\tau$. This is because the *compatibility condition* (Bühlmann & van de Geer 2011, Bickel et al. 2009), which is critical for the convergence rate of LASSO, may not hold under severe near multicollinearity due to the latent factor structure. We provide detailed analysis in Appendix B.6.2.

Combining Lemma 1 (c) and Theorem 1, the following corollary immediately characterizes the distance between $\hat{\boldsymbol{\beta}}_\tau$ and $\boldsymbol{\beta}^0$.

**Corollary 1.** *Under Assumption 1–3, as* $N, T_1 \to \infty$, *if* $N\xi_N^{\frac{3}{2}} \to \infty$, *then for* $\tau$ *in the range specified in Theorem 1, we have*

(a) $\|\hat{\boldsymbol{\beta}}_\tau - \boldsymbol{\beta}^0\|_2 = o_p(N^{-\frac{1}{2}}) + \psi_{\max}O(N^{-\frac{1}{2}}\xi_N^{-1})$;

(b) $\|\hat{\boldsymbol{\beta}}_\tau - \boldsymbol{\beta}^0\|_1 = o_p(1) + \psi_{\max}O(\xi_N^{-1})$.

As indicated in Proposition 1 (a), any estimator for $\boldsymbol{\beta}^0$, say $\hat{\boldsymbol{\beta}}$, may have nothing better than a trivial estimator $\mathbf{0}_N$ when $N$ is large, if the L2-distance $\|\hat{\boldsymbol{\beta}} - \boldsymbol{\beta}^0\|_2$ cannot reach an order that converges faster than $1/\sqrt{N\xi_N}$. From this corollary, if we further have $\psi_{\max}/\xi_N \to 0$, then we can establish both L2 and L1 consistency of the L2-relaxation estimator to the true $\boldsymbol{\beta}^0$, which is possible if the cross-sectional dependence and heteroskedasticity of the idiosyncratic errors are not too strong.

Consistency is a desirable property though, we are ultimately interested in the performance of the OOS prediction and the statistical inference based on the L2-relaxation estimator. In the next section, we will find that the conditions in Theorem 1 suffice for the convergence of the prediction outcomes, whereas $\psi_{\max}$ is asymptotically irrelevant for prediction performance.

### 3.4 Prediction Performance

We estimate the model with the training sample in $\mathcal{T}_1$ and make OOS prediction with a new dataset in $\mathcal{T}_2$ according to (4). The prediction error $e_{t,\tau}$ can be decomposed by

$$e_{t,\tau} = y_t - \hat{y}_{t,\tau} = \epsilon_t^* - \mathcal{E}_{\mathcal{T}_1}(\epsilon_s^*) - [\boldsymbol{x}_t - \mathcal{E}_{\mathcal{T}_1}(\boldsymbol{x}_s)]'(\hat{\boldsymbol{\beta}}_\tau - \boldsymbol{\beta}^*). \tag{19}$$

Here $\epsilon_t^*$ is the oracle error term with known $\boldsymbol{\beta}^*$ and $\alpha^* = \mu_0 - \boldsymbol{\mu}_{\mathcal{N}}'\boldsymbol{\beta}^*$, defined as:

$$\epsilon_t^* := y_t - \mu_0 - (\boldsymbol{x}_t - \boldsymbol{\mu}_{\mathcal{N}})'\boldsymbol{\beta}^* = (\boldsymbol{\lambda}_0 - \boldsymbol{\Lambda}'\boldsymbol{\beta}^*)'\boldsymbol{f}_t + u_{0t} - \boldsymbol{u}_{\mathcal{N}t}'\boldsymbol{\beta}^* = u_{0t} - \boldsymbol{u}_{\mathcal{N}t}'\boldsymbol{\beta}^*, \tag{20}$$

where the last equality follows by the fact $\boldsymbol{\Lambda}'\boldsymbol{\beta}^* = \boldsymbol{\lambda}_0$, as $\boldsymbol{\beta}^*$ satisfies the constraint condition of problem (11).

Theorem 1 immediately bounds the in-sample mean predicted squared error (MPSE) $\mathcal{E}_{\mathcal{T}_1}(e_{t,\tau}^2)$. On the other hand, additional conditions concerning the OOS noise levels are needed for the OOS MPSE $\mathcal{E}_{\mathcal{T}_2}(e_{t,\tau}^2)$. Define $\varphi_2 := \sqrt{\log N/T_2}$.

**Assumption 4.** (a) $\|\mathcal{E}_{\mathcal{T}_2}(\boldsymbol{f}_t\boldsymbol{f}_t') - \boldsymbol{I}_q\|_\infty + \|\mathcal{E}_{\mathcal{T}_2}(\boldsymbol{f}_t)\|_\infty = \mathrm{O}_{\mathrm{p}}(T_2^{-\frac{1}{2}})$;

(b) $\|\mathcal{E}_{\mathcal{T}_2}(\boldsymbol{u}_{\mathcal{N}t}\boldsymbol{u}_{\mathcal{N}t}') - \boldsymbol{\Omega}\|_\infty + \|\mathcal{E}_{\mathcal{T}_2}(\boldsymbol{f}_t\boldsymbol{u}_{\mathcal{N}t}')\|_\infty + \|\mathcal{E}_{\mathcal{T}_2}(\boldsymbol{u}_{\mathcal{N}t})\|_\infty = \mathrm{O}_{\mathrm{p}}(\varphi_2)$;

(c) $\|\mathcal{E}_{\mathcal{T}_2}(\boldsymbol{f}_t u_{0t})\|_\infty = \mathrm{O}_{\mathrm{p}}(T_2^{-\frac{1}{2}})$, and $\|\mathcal{E}_{\mathcal{T}_2}(\boldsymbol{u}_{\mathcal{N}t}u_{0t})\|_\infty = \mathrm{O}_{\mathrm{p}}(\varphi_2)$;

(d) $|\mathcal{E}_{\mathcal{T}_2}(u_{0t}^2) - \sigma_0^2| + |\mathcal{E}_{\mathcal{T}_2}(u_{0t})| = \mathrm{O}_{\mathrm{p}}(T_2^{-\frac{1}{2}})$.

Assumption 4 imposes high-level conditions on the data in $\mathcal{T}_2$ parallel to their counterparts in Assumption 3. The only difference here is that we do not further assume $\varphi_2 \to 0$ as $N, T_2 \to \infty$. We allow $T_2$ to diverge to infinity as slowly as $\log N$. This moderate requirement accommodates empirical scenarios where the OOS period is relatively short.

The following theorem derives the oracle inequalities that guarantee the in-sample and OOS prediction performance.

**Theorem 2** (Oracle inequalities). *Suppose Assumptions 1–3 hold and $\tau + \varphi_1/\tau = \mathrm{o}(\xi_N)$. As $N, T_1, T_2 \to \infty$, we have*

(a) $\mathcal{E}_{\mathcal{T}_1}(e_{t,\tau}^2) \leq \mathcal{E}_{\mathcal{T}_1}(\epsilon_t^{*2}) + \mathrm{o}_{\mathrm{p}}(1)$, *where* $\mathcal{E}_{\mathcal{T}_1}(\epsilon_t^{*2}) = \sigma_0^2 + \mathrm{o}_{\mathrm{p}}(1)$;

(b) *in addition, if Assumption 4 holds and $N\varphi_2 \to \infty$, then $\mathcal{E}_{\mathcal{T}_2}(e_{t,\tau}^2) \leq \mathcal{E}_{\mathcal{T}_2}(\epsilon_t^{*2}) + \mathrm{o}_{\mathrm{p}}\left(1 + \varphi_2/\sqrt{\xi_N}\right)$, where $\mathcal{E}_{\mathcal{T}_2}(\epsilon_t^{*2}) = \sigma_0^2 + \mathrm{O}_{\mathrm{p}}(\varphi_2/\xi_N)$.*

Theorem 2 (a) establishes the consistency of the in-sample prediction for its oracle target $\mathcal{E}_{\mathcal{T}_1}(\epsilon_t^{*2})$. In (b), the extra term $\mathrm{o}_{\mathrm{p}}(\varphi_2/\sqrt{\xi_N})$ can be regarded as the expense of OOS prediction; this term vanishes as long as $\varphi_2 = \mathrm{O}(\sqrt{\xi_N})$, which does not have to shrink to zero asymptotically if $\xi_N$ is bounded away from zero.

As pointed out in the comments following Corollary 1, the asymptotic bounds in Theorem 2 are irrelevant to $\psi_{\max}$. In other words, heteroskedasticity does not undermine the large sample performance of the prediction. To understand the asymptotic irrelevance of $\psi_{\max}$, note that the oracle MSE equals $\mathbb{E}(\epsilon_t^{*2}) = \sigma_0^2 + \boldsymbol{\beta}^{*\prime}\boldsymbol{\Omega}\boldsymbol{\beta}^*$, and recall that the population MSE is minimized by the true $\boldsymbol{\beta}^0$ as $\mathbb{E}(\epsilon_t^2) = \sigma_0^2 + \boldsymbol{\beta}^{0\prime}\boldsymbol{\Omega}\boldsymbol{\beta}^0 + \|\boldsymbol{\lambda}_0 - \boldsymbol{\Lambda}'\boldsymbol{\beta}^0\|_2^2$. Comparing (12) with (9), as $\boldsymbol{\Omega}$ does not play a role in $\boldsymbol{\beta}^*$, the oracle error $\mathbb{E}(\epsilon_t^{*2})$ is robust to heteroskedasticity.

On the contrary, elementary calculation shows the gap

$$\mathbb{E}(\epsilon_t^{*2}) - \mathbb{E}(\epsilon_t^2) = \boldsymbol{\beta}^{*\prime}\boldsymbol{\Omega}\boldsymbol{\beta}^* - \left(\boldsymbol{\beta}^{0\prime}\boldsymbol{\Omega}\boldsymbol{\beta}^0 + \|\boldsymbol{\lambda}_0 - \boldsymbol{\Lambda}'\boldsymbol{\beta}^0\|_2^2\right)$$
$$\leq \sigma_{\min}^2 \left[\psi_{\max}\|\boldsymbol{\beta}^*\|_2^2 + \left(\|\boldsymbol{\beta}^*\|_2 + \|\boldsymbol{\beta}^0\|_2\right)\|\boldsymbol{\beta}^* - \boldsymbol{\beta}^0\|_2\right]$$
$$= \psi_{\max} O(N^{-\frac{1}{2}}\xi_N^{-\frac{1}{2}} + N^{-1}\xi_N^{-\frac{3}{2}}) + O(N^{-2}\xi_N^{-2}) \quad (21)$$

depends on $\psi_{\max}$ in the finite sample. We will show by simulations in Appendix D.1 that the finite sample OOS MPSE of L2-relaxation prediction distributes around $\mathbb{E}(\epsilon_t^{*2})$ across different heteroskedasticity levels, whereas $\mathbb{E}(\epsilon_t^2)$ deviates from $\mathbb{E}(\epsilon_t^{*2})$ according to $\psi_{\max}$.

In summary, in this section we have established the asymptotic theory for L2-relaxation in the high-dimensional dense regression model driven by latent factors. We first define the primal and dual oracle targets as stated in Lemma 1 and 3, respectively. The dual problem is a device for us to carry out the proofs by utilizing inequalities concerning the L1-penalized problems. Given a $q$-sparse solution to the oracle dual, which is the key component in our construction of the asymptotics, we derive the consistency of the L2-relaxation estimator to its oracle target in Theorem 1. Finally, the in-sample and OOS prediction errors are well bounded in Theorem 2.

## 4 Application: Panel Data Approach

### 4.1 Single Treated Unit

Our method can be directly applied to the PDA for program evaluation. Let $y_t$ be the treated unit, and $\boldsymbol{x}_t$ be the control units. In the time dimension, let $\mathcal{T}_1 = \{1, 2, \cdots, T_1\}$ be the pre-treatment period, and $\mathcal{T}_2 = \{T_1 + 1, \cdots, T\}$ be the post-treatment period. For any control unit $i \in \mathcal{N}$, we observe untreated $x_{it} = x_{it}^0$ for the entire time span $\mathcal{T} := \mathcal{T}_1 \cup \mathcal{T}_2$. The treated unit appears as

$$y_t = \begin{cases} y_t^0, & t \in \mathcal{T}_1 \\ y_t^1 = y_t^0 + \Delta_t, & t \in \mathcal{T}_2 \end{cases}, \quad (22)$$

where the post-treatment counterfactual $\{y_t^0 : t \in \mathcal{T}_2\}$ cannot be observed. The treatment effects $\{\Delta_t : t \in \mathcal{T}_2\}$ can be time-varying, and the average treatment effect (ATE) is defined as $\Delta_{T_2} := T_2^{-1} \sum_{t \in \mathcal{T}_2} \mathbb{E}(\Delta_t)$. Here we say "counterfactual" as if the treatment did not happen and all untreated outcomes $\{(y_t^0, \boldsymbol{x}_t) : t \in \mathcal{T}\}$ followed the same model (5).

We are interested in predicting the post-treatment counterfactual outcomes $\{y_t^0 : t \in \mathcal{T}_2\}$, estimating the treatment effects $\{\Delta_t : t \in \mathcal{T}_2\}$, and furthermore conducting statistical inference concerning the ATE. It can be implemented with the pre-treatment L2-relaxation estimator $\hat{\boldsymbol{\beta}}_\tau$, which leads to the estimated treatment effect

$$\hat{\Delta}_t = y_t^1 - \hat{y}_{t,\tau} = \Delta_t + e_{t,\tau}, \quad t \in \mathcal{T}_2, \tag{23}$$

and thus the estimated ATE $\bar{\Delta} = \mathcal{E}_{\mathcal{T}_2}(\hat{\Delta}_t)$. Here we suppress $\tau$ in the notation when there is no confusion.

Based on (19), we denote $d_t^* := \Delta_t - \mathbb{E}(\Delta_t) + \epsilon_t^*$ as the sum of the demeaned treatment effect and the oracle prediction error. Recall in (20) that $\epsilon_t^*$ depends on $\mathcal{N}$. Then suppose that $\{\epsilon_t^*\}$ and $\{d_t^*\}$ have finite and positive long-run variances (LRV) $\rho_{\epsilon^*}^2$ and $\rho_{d^*}^2$ for any sufficiently large $N$, respectively, and they can be consistently estimated by the heteroskedasticity and autocorrelation consistent (HAC) estimators. Here for simplicity, we estimate the LRVs by $\hat{\rho}_{\epsilon^*}^2 := \sum_{l=-h_1}^{h_1} \mathcal{E}_{\mathcal{T}_1}\left([\epsilon_t^* - \mathcal{E}_{\mathcal{T}_1}(\epsilon_s^*)][\epsilon_{t+l}^* - \mathcal{E}_{\mathcal{T}_1}(\epsilon_s^*)]\right)$ and $\hat{\rho}_{d^*}^2 := \sum_{l=-h_2}^{h_2} \mathcal{E}_{\mathcal{T}_2}\left([d_t^* - \mathcal{E}_{\mathcal{T}_2}(d_s^*)][d_{t+l}^* - \mathcal{E}_{\mathcal{T}_2}(d_s^*)]\right)$, where $h_1$ and $h_2$ are the respective truncation lag orders. Formally, we make the following assumptions.

**Assumption 5.** *When $N, T_1, T_2 \to \infty$, the following conditions are satisfied:*

(a) $\sum_{t \in \mathcal{T}_2} [\mathbb{E}(\Delta_t) - \Delta_{T_2}]^2 = \mathrm{O}(\sqrt{T_2})$ and $\mathcal{E}_{\mathcal{T}_2}\left([d_t^* - \mathcal{E}_{\mathcal{T}_2}(d_s^*)]^2\right) = \mathrm{O}_\mathrm{p}(1)$;

(b) $\sup_N |\hat{\rho}_{\epsilon^*}^2 - \rho_{\epsilon^*}^2| + \sup_N |\hat{\rho}_{d^*}^2 - \rho_{d^*}^2| \xrightarrow{\mathrm{P}} 0$, for $h_1, h_2 \to \infty$ and $h_1/T_1^{1/4} + h_2/T_2^{1/4} \to 0$;

(c) $\sqrt{T_1}\mathcal{E}_{\mathcal{T}_1}(\epsilon_t^*) \xrightarrow{\mathrm{d}} N\left(0, \rho_{(1)}^2\right)$, where $\rho_{(1)}^2 := \lim_{N \to \infty} \rho_{\epsilon^*}^2$ is finite and positive;

(d) $\sqrt{T_2}\mathcal{E}_{\mathcal{T}_2}(d_t^*) \xrightarrow{\mathrm{d}} N\left(0, \rho_{(2)}^2\right)$, where $\rho_{(2)}^2 := \lim_{N \to \infty} \rho_{d^*}^2$ is finite and positive;

(e) $\sum_{s \in \mathcal{T}_1} \sum_{t \in \mathcal{T}_2} \mathbb{E}(d_t^* \epsilon_s^*) = \mathrm{o}(T_1 \wedge T_2)$.

Assumption 5 makes sure that $\epsilon_t^*$ and $d_t^*$ are well-behaved. Condition (a) controls the deviation of $\{\mathbb{E}(\Delta_t) : t \in \mathcal{T}_2\}$ from their average $\Delta_{T_2}$ and the sample variance of $d_t^*$, where the latter naturally holds when $\varphi_2 = \mathrm{O}(\xi_N)$ in view of Theorem 2 (b). Condition (b) adopts those in Newey & West (1987) for the consistency of LRV estimators. Conditions (c) and (d) are two high-level assumptions by the sequential central limit theorem (CLT) for $\{\epsilon_t^*\}$ and $\{d_t^*\}$ as $(T_1, N \to \infty)_{\mathrm{seq}}$ introduced in Phillips & Moon (1999); the joint convergence is implied if the condition in Phillips & Moon (1999, Lemma 5(b)) is satisfied. Assumption 5 (e) allows the dependence between the corresponding variables in $\mathcal{T}_1$ and $\mathcal{T}_2$, while restricts their overall magnitude.

*Remark* 3. Assumption 5 (e) is satisfied by many stationary time series. Denote $a_n \asymp b_n$ if $a_n$ has the same asymptotic order as $b_n$ asymptotically. Regardless of the treatment effects as they are assumed to be exogenous, if $T_2 < T_1$, then as $T_1, T_2 \to \infty$, for $\mathbb{E}(\epsilon_s^* \epsilon_t^*) \asymp \kappa^{|t-s|}$ with $\kappa \in (0, 1)$, we have

$$\sum_{s \in \mathcal{T}_1} \sum_{t \in \mathcal{T}_2} \mathbb{E}(\epsilon_s^* \epsilon_t^*) \asymp \sum_{l=1}^{T_2} l\kappa^l + \sum_{l=T_2+1}^{T_1} T_2 \kappa^l + \sum_{l=T_1+1}^{T-1} (T-l)\kappa^l = \frac{\kappa}{(1-\kappa)^2} + \mathrm{o}(1);$$

and for $\mathbb{E}(\epsilon_s^* \epsilon_t^*) \asymp |t-s|^{-\kappa}$ with $\kappa > 1$,

$$\sum_{s \in \mathcal{T}_1} \sum_{t \in \mathcal{T}_2} \mathbb{E}(\epsilon_s^* \epsilon_t^*) \asymp \sum_{l=1}^{T_2} l^{1-\kappa} + \sum_{l=T_2+1}^{T_1} T_2 l^{-\kappa} + \sum_{l=T_1+1}^{T-1} (T-l) l^{-\kappa} \leq \sum_{l=1}^{T-1} l^{1-\kappa},$$

which converges when $\kappa > 2$, while it diverges more slowly than $T$ when $\kappa \in (1, 2]$.

The following theorem derives the limiting distribution of the $t$-statistic for ATE.

**Theorem 3.** *Suppose Assumptions 1–5 hold, and asymptotically $N, T_1, T_2 \to \infty$ with $(\log N + T_1^{1/p})\varphi_1 = \mathrm{o}(\xi_N^2)$ for some $p > 2$ and $T_2 = \mathrm{O}(N \wedge T_1)$. If $(\log N + T_1^{1/p})\tau + \varphi_1/\tau = \mathrm{o}(\xi_N)$, and $h_1, h_2 \to \infty$ with $h_1 = \mathrm{O}(T_1^{1/(2p)})$ and $h_2 = \mathrm{O}(T_2^{1/(2p)})$, then*

$$\hat{Z} := \frac{\bar{\Delta} - \Delta_{T_2}}{\sqrt{\hat{\rho}_{(1)}^2/T_1 + \hat{\rho}_{(2)}^2/T_2}} \xrightarrow{d} N(0,1), \qquad (24)$$

*where $\hat{\rho}_{(1)}^2 := \sum_{l=-h_1}^{h_1} \mathcal{E}_{\mathcal{T}_1}(e_{t,\tau} e_{t+l,\tau})$ and $\hat{\rho}_{(2)}^2 := \sum_{l=-h_2}^{h_2} \mathcal{E}_{\mathcal{T}_2}\left((\hat{\Delta}_t - \bar{\Delta})(\hat{\Delta}_{t+l} - \bar{\Delta})\right)$.*

Two comments are in order. Firstly, the denominator of the $t$-statistic in (24) uses the HAC estimator with the uniform kernel for simplicity. The result is compatible with the Bartlett kernel (Newey & West 1987) and can be extended to those in Andrews (1991). The choice of the truncation lag order in the LRV estimator has been well studied in the literature (Newey & West 1994). Secondly, in this section, we assume $\{\Delta_t\}$ to be a random time series for more general settings. Deterministic treatment effects can be accommodated by our framework, where $\epsilon_t^*$ is the only random component in $d_t^*$.

## 4.2 Many Treated Units

In some policy evaluation applications, especially those that have been recently implemented, the lengths of post-treatment periods are relatively short. Inference remains possible if there are many treated units, despite a finite $T_2$. Still, let all control units $\{x_{it} : i \in \mathcal{N}, t \in \mathcal{T}\}$ follow the factor model (6) with $\mathbb{E}(\boldsymbol{u}_{\mathcal{N}t}|\boldsymbol{f}_t) = \boldsymbol{0}_N$. For a $t \in \mathcal{T}_2$, we consider a cross section of treated individuals $\{y_{it}\}$ indexed by $i \in \mathcal{M}$, with $|\mathcal{M}| = M$. For any treated unit, again we observe

$$y_{it} = \begin{cases} y_{it}^0, & i \in \mathcal{M}, t \in \mathcal{T}_1 \\ y_{it}^1 = y_{it}^0 + \Delta_{it}, & i \in \mathcal{M}, t \in \mathcal{T}_2 \end{cases}, \qquad (25)$$

while the counterfactual outcomes $\{y_{it}^0 : i \in \mathcal{M}, t \in \mathcal{T}_2\}$ are unobservable. The untreated counterfactual of each treated unit is driven by latent factors as

$$y_{it}^0 = \mu_i + \boldsymbol{\lambda}_i' \boldsymbol{f}_t + u_{it}, \quad i \in \mathcal{M}, \qquad (26)$$

where the idiosyncratic error $\boldsymbol{u}_{\mathcal{M}t} := (u_{it})_{i \in \mathcal{M}}$ satisfies $\mathbb{E}(\boldsymbol{u}_{\mathcal{M}t}|\boldsymbol{f}_t, \boldsymbol{u}_{\mathcal{N}t}) = \boldsymbol{0}_M$. In this section, following the *difference-in-difference* and the *conformal inference* (Chernozhukov et al. 2021), we regard the treatment effects as deterministic but allowed to vary across time and individuals without assuming any functional form. We are interested in the ATE $\Delta_{M,t} := \frac{1}{M} \sum_{i \in \mathcal{M}} \Delta_{it}$ for $t \in \mathcal{T}_2$.

Similarly, we predict the post-treatment counterfactual outcome for each treated unit $i \in \mathcal{M}$ by $\hat{y}_{it,\tau} = \mathcal{E}_{\mathcal{T}_1}(y_{is}) + [\boldsymbol{x}_t - \mathcal{E}_{\mathcal{T}_1}(\boldsymbol{x}_s)]' \hat{\boldsymbol{\beta}}_{i,\tau}$, where $\hat{\boldsymbol{\beta}}_{i,\tau}$ is the L2-relaxation estimator obtained from the pre-treatment sample. The prediction error is

$$e_{it,\tau} = y_{it}^0 - \hat{y}_{it,\tau} = \epsilon_{it}^* - \mathcal{E}_{\mathcal{T}_1}(\epsilon_{is}^*) - [\boldsymbol{x}_t - \mathcal{E}_{\mathcal{T}_1}(\boldsymbol{x}_s)]' (\hat{\boldsymbol{\beta}}_{i,\tau} - \boldsymbol{\beta}_i^*), \tag{27}$$

where $\boldsymbol{\beta}_i^* = \boldsymbol{\Lambda}(\boldsymbol{\Lambda}'\boldsymbol{\Lambda})^{-1}\boldsymbol{\lambda}_i$ and $\epsilon_{it}^* = u_{it} - \boldsymbol{u}_{\mathcal{N}t}'\boldsymbol{\beta}_i^*$. Then the individual treatment effect can be estimated by

$$\hat{\Delta}_{it} = y_{it}^1 - \hat{y}_{it,\tau} = \Delta_{it} + e_{it,\tau}, \quad i \in \mathcal{M}, \tag{28}$$

and the cross-sectional ATE estimator is $\bar{\hat{\Delta}}_t = M^{-1} \sum_{i \in \mathcal{M}} \hat{\Delta}_{it}$.

When $\mathcal{T}_2$ is finite, for example $T_2 = 1$, the standard error (s.e.) of ATE can only be estimated from the pre-treatment data, so we can borrow such s.e. to construct the inference under the assumption that the data generating processes (DGPs) are the same under $\mathcal{T}_1$ and $\mathcal{T}_2$. Specifically, if $\{(\boldsymbol{u}_{\mathcal{N}t}, \boldsymbol{u}_{\mathcal{M}t})\}$ is stationary over time, then $\mathbb{E}(\epsilon_{it}^*\epsilon_{jt}^*)$ does not depend on $t$. For any sufficiently large $M$, define $V_{\epsilon^*}^2 := \text{Var}\left(\frac{1}{\sqrt{M}} \sum_{i \in \mathcal{M}} \epsilon_{it}^*\right) = \frac{1}{M} \sum_{i \in \mathcal{M}} \sum_{j \in \mathcal{M}} \mathbb{E}(\epsilon_{it}^*\epsilon_{jt}^*)$ as the cross-sectional LRV. Denote $\boldsymbol{\Omega}_{\mathcal{M}} =: \mathbb{E}(\boldsymbol{u}_{\mathcal{M}t}\boldsymbol{u}_{\mathcal{M}t}')$. We impose the following assumptions under the case of multiple treated units.

**Assumption 6.** *(a) $\max_{i \in \mathcal{M}} \|\boldsymbol{\lambda}_i\|_\infty \leq C$ for an absolute constant $C$;*

*(b) $\|\Gamma_{\mathcal{T}_1}(\boldsymbol{f}_t, \boldsymbol{u}_{\mathcal{M}t}')\|_\infty + \|\Gamma_{\mathcal{T}_1}(\boldsymbol{u}_{\mathcal{N}t}, \boldsymbol{u}_{\mathcal{M}t}')\|_\infty = \text{O}_\text{p}(\varphi_1)$;*

*(c) $\|\Gamma_{\mathcal{T}_1}(\boldsymbol{u}_{\mathcal{M}t}, \boldsymbol{u}_{\mathcal{M}t}') - \boldsymbol{\Omega}_{\mathcal{M}}\|_\infty + \|\mathcal{E}_{\mathcal{T}_1}(\boldsymbol{u}_{\mathcal{M}t})\|_\infty = \text{O}_\text{p}(\varphi_1)$;*

*(d) $\|\boldsymbol{x}_t - \boldsymbol{\mu}_{\mathcal{N}}\|_\infty = \text{O}_\text{p}(\sqrt{\log N})$ for any $t \in \mathcal{T}_2$;*

*(e) $M^{-1/2} \sum_{i \in \mathcal{M}} \epsilon_{it}^* \xrightarrow{\text{d}} N(0, V^2)$ for any $t \in \mathcal{T}_2$ as $M, N \to \infty$, where $V^2 := \lim_{M,N \to \infty} V_{\epsilon^*}^2$ is finite and positive.*

Assumption 6's (a)–(c) are the counterparts of Assumptions 1 (a) and 3 (c)–(d) for the case of a single treated unit. For $M = \text{O}(N)$, the orders in (b)–(d) naturally hold for high-dimensional matrices and vectors. Assumption 6 (e) is a high-level condition of asymptotic normality.

*Remark* 4. We provide some lower-level discussions about Assumption 6 (e). Notice that the decomposition

$$\frac{1}{\sqrt{M}} \sum_{i \in \mathcal{M}} \epsilon_{it}^* = \frac{1}{\sqrt{M}} \sum_{i \in \mathcal{M}} u_{it} - \frac{1}{\sqrt{N}} \sum_{j \in \mathcal{N}} \ell_{(M,N),j} u_{jt},$$

where $\ell_{(M,N),j} := \sqrt{\frac{M}{N}} \left(\frac{1}{M}\sum_{i\in\mathcal{M}}\boldsymbol{\lambda}_i\right)'(\boldsymbol{\Lambda}'\boldsymbol{\Lambda}/N)^{-1}\boldsymbol{\lambda}_j$. For the first term, for any $t \in \mathcal{T}_2$, if $\boldsymbol{u}_{\mathcal{M}t} := \boldsymbol{\Omega}_{\mathcal{M}}^{1/2}\boldsymbol{\nu}_{\mathcal{M}t}$ with the eigenvalues of $\boldsymbol{\Omega}_{\mathcal{M}}$ bounded away from zero and infinity, and $\boldsymbol{\nu}_{\mathcal{M}t}$ is cross-sectionally independent with zero mean and $\mathbb{E}(\boldsymbol{\nu}_{\mathcal{M}t}\boldsymbol{\nu}'_{\mathcal{M}t}) = \boldsymbol{I}_M$, then we can invoke the Lindeberg-Feller CLT for $\frac{1}{\sqrt{M}}\sum_{i\in\mathcal{M}} u_{it} \xrightarrow{d} N(0, V_0^2)$ where $V_0^2 := \lim_{M\to\infty} \boldsymbol{1}'_M \boldsymbol{\Omega}_{\mathcal{M}} \boldsymbol{1}_M/M \in (0, \infty)$. For $M = \mathrm{O}(N\xi_N^2)$, as $\lim_{N\to\infty} \frac{1}{N}\sum_{j\in\mathcal{N}} \ell^2_{(M,N),j} < \infty$ under Assumption 1 and 6 (a), CLT also holds for the second term if $\boldsymbol{u}_{\mathcal{N}t} := \boldsymbol{\Omega}^{1/2}\boldsymbol{\nu}_{\mathcal{N}t}$ with $\boldsymbol{\nu}_{\mathcal{N}t}$ defined similarly to $\boldsymbol{\nu}_{\mathcal{M}t}$.

The following theorem provides the limiting distribution of the $t$-statistic for the cross-sectional ATE.

**Theorem 4.** *Under Assumption 1–3 and 6, as $M, N, T_1 \to \infty$ with $(M \log N)\varphi_1 = \mathrm{o}(\xi_N^2)$, if $(M \log N)\tau + \varphi_1/\tau = \mathrm{o}(\xi_N)$, then*

$$\hat{Z}_t := \frac{\sqrt{M}(\bar{\Delta}_t - \Delta_{M,t})}{\hat{V}} \xrightarrow{d} N(0, 1), \quad t \in \mathcal{T}_2, \tag{29}$$

where $\hat{V}^2 := \frac{1}{M}\sum_{i\in\mathcal{M}}\sum_{j\in\mathcal{M}} \mathcal{E}_{\mathcal{T}_1}(e_{it,\tau}e_{jt,\tau})$.

The condition in Theorem 4 implies that $M = \mathrm{o}\left(\xi_N^2\sqrt{N/(\log N)^3}\right)$ is dominated by $\sqrt{N}$. This is reasonable since only when $N$ is much larger than $M$ can we accommodate the estimation errors in the synthetic counterfactual for those many treated units.

## 5 Simulations

In this section, we carry out Monte Carlo simulations to check the finite sample behaviors of the estimation and inferential procedures. Section 5.1 runs DGPs driven by latent factors with strong or weak factor loadings with homoskedastic or heteroskedastic idiosyncratic errors. Section 5.2 shows the testing performance of the single-treated-unit and many-treated-unit PDA, respectively.

### 5.1 Estimation and Prediction

Consider the factor model (5) with 4 factors for the DGPs, where we set $\mu_0 = 0$ and $\boldsymbol{\mu}_{\mathcal{N}} = \boldsymbol{0}_N$ for simplicity. For the factor $\boldsymbol{f}_t$, we specify $q = 4$ as follows: (i) $f_{1t}$ i.i.d. $\sim N(0, 1)$; (ii) AR(1): $f_{2t} = 0.9 f_{2,t-1} + v_{2t}$ where $v_{2t}$ i.i.d. $\sim N(0, 0.19)$; (iii) MA(2): $f_{3t} = v_{3t} + 0.8 v_{3,t-1} + 0.4 v_{3,t-2}$ where $v_{3t}$ i.i.d. $\sim N(0, 5/9)$; and (iv) ARMA(1,1): $f_{4t} = 0.5 f_{4,t-1} + v_{4t} + 0.5 v_{4,t-1}$ where $v_{4t}$ i.i.d. $\sim N(0, 3/7)$.[4] For $y_t$, all entries of $\boldsymbol{\lambda}_0$ are independently drawn from Uniform $([-0.5, -0.3] \cup [0.3, 0.5])$, and the idiosyncratic error $u_{0t}$ i.i.d. $\sim N(0, 0.5)$.

We experiment with 6 ($= 2 \times 3$) types of DGPs for $\boldsymbol{x}_t$. Firstly, we generate two sets of factor loadings: (i) the *strong factors*: $\boldsymbol{\lambda}_i$s for all $i \in \mathcal{N}$ have the same DGP with $\boldsymbol{\lambda}_0$; and (ii) the

---

[4] The variances for $\{v_{jt} : j = 2, 3, 4\}$ are specified that the unconditional variances are 1 in all models.

*weak factors*: $\boldsymbol{\lambda}_i$s for $i \in \{1, \cdots, 4\}$ share the previous DGP, whereas for $i \in \{5, \cdots, N\}$ from Uniform$(-0.15, 0.15)$. Secondly, we consider three DGPs for $\boldsymbol{u}_{\mathcal{N}t}$: (i) *homoskedasticity*: $u_{it}$ i.i.d. $\sim N(0, 0.5)$ across $i$ and $t$; (ii) *mild heteroskedasticity*: $u_{it}$ i.i.d. $\sim N(0, \sigma_i^2)$ for all $i \in \mathcal{N}$ where $\sigma_i^2$ is independently drawn from Uniform$(0.3, 0.7)$; (iii) *severe heteroskedasticity*: similar to the previous case, except that $\sigma_i^2$ is drawn from Uniform$(0.1, 0.9)$.

For each simulated sample, we first conduct the L2-relaxation estimation and OOS prediction with a sequence of tuning parameters. For each DGP and sample size, we choose the $\tau$ that minimizes the average OOS MPSE across all simulated samples as the infeasible best tuning parameter for the finite sample analysis. Observing such selected $\tau$ in Table 1, as the sample size goes large, the infeasible best $\tau$ shrinks toward zero. In practice, multiple validation methods can be applied to choose the tuning parameter. To illustrate, we further show the simulated performance of the OOS validation with 20% validation sample sequentially split from the end of the training sample. From Table 1, we observe that the performance of the feasible L2-relaxation (under validated $\tau$) closely tracks that under the infeasible best $\tau$. This implies that the validation method properly picks the desirable tuning parameter.

Table 1: Average simulated MPSE (net of the unpredictable $\sigma_0^2$, $N = 100$)

| DGP | $T_1, T_2$ | L2-relaxation | | | Ridge | | LASSO | | PCA | |
|---|---|---|---|---|---|---|---|---|---|---|
| | | Best $\tau$ | Infeas. | Valid. | Infeas. | Valid. | Infeas. | Valid. | $q = 4$ | PC$_{p1}$ |
| | | | | | DGP: *strong factors* | | | | | |
| Homo. | 50 | 0.11 | 0.120 | 0.222 | 0.144 | 0.230 | 0.207 | 0.322 | 0.106 | 0.106 |
| | 100 | 0.09 | 0.062 | 0.108 | 0.082 | 0.117 | 0.126 | 0.184 | 0.055 | 0.055 |
| | 200 | 0.07 | 0.041 | 0.059 | 0.055 | 0.067 | 0.084 | 0.103 | 0.037 | 0.037 |
| Mild hetero. | 50 | 0.11 | 0.122 | 0.227 | 0.145 | 0.232 | 0.208 | 0.333 | 0.107 | 0.107 |
| | 100 | 0.09 | 0.063 | 0.111 | 0.083 | 0.118 | 0.125 | 0.181 | 0.055 | 0.055 |
| | 200 | 0.07 | 0.041 | 0.058 | 0.055 | 0.067 | 0.082 | 0.102 | 0.037 | 0.037 |
| Sev. hetero. | 50 | 0.11 | 0.126 | 0.230 | 0.147 | 0.234 | 0.208 | 0.341 | 0.108 | 0.108 |
| | 100 | 0.09 | 0.065 | 0.116 | 0.083 | 0.120 | 0.118 | 0.173 | 0.056 | 0.056 |
| | 200 | 0.07 | 0.042 | 0.060 | 0.055 | 0.067 | 0.072 | 0.091 | 0.037 | 0.037 |
| | | | | | DGP: *weak factors* | | | | | |
| Homo. | 50 | 0.07 | 0.554 | 0.674 | 0.576 | 0.691 | 0.661 | 0.808 | 0.618 | 0.767 |
| | 100 | 0.06 | 0.431 | 0.518 | 0.459 | 0.524 | 0.514 | 0.621 | 0.455 | 0.696 |
| | 200 | 0.04 | 0.347 | 0.378 | 0.374 | 0.398 | 0.403 | 0.447 | 0.338 | 0.644 |
| Mild hetero. | 50 | 0.07 | 0.561 | 0.684 | 0.581 | 0.693 | 0.668 | 0.820 | 0.634 | 0.770 |
| | 100 | 0.06 | 0.436 | 0.525 | 0.462 | 0.529 | 0.519 | 0.624 | 0.474 | 0.700 |
| | 200 | 0.04 | 0.349 | 0.382 | 0.375 | 0.400 | 0.405 | 0.450 | 0.358 | 0.649 |
| Sev. hetero. | 50 | 0.07 | 0.573 | 0.696 | 0.586 | 0.702 | 0.674 | 0.834 | 0.667 | 0.768 |
| | 100 | 0.06 | 0.441 | 0.533 | 0.459 | 0.528 | 0.506 | 0.610 | 0.514 | 0.703 |
| | 200 | 0.04 | 0.343 | 0.382 | 0.364 | 0.391 | 0.377 | 0.423 | 0.401 | 0.644 |

Table 1 also provides evidence that the prediction performance of L2-relaxation is not affected by heteroskedasticity, as both infeasible and validated MPSEs do not vary significantly across different levels of heteroskedasticity. This result echoes what we discussed following Theorem 2. In Appendix D.1, we show additional simulation results on the L2-relaxation coefficient estimation and OOS MPSE across different levels of heteroskedasticity to support our theoretical analysis.

We compare the OOS prediction performance of L2-relaxation to ridge, LASSO and principal component analysis (PCA, respectively with true $q = 4$ and the number of factors determined by the $PC_{p1}$ criterion in Bai & Ng 2002). In Table 1, PCA serves as the benchmark method in our simulation, especially when the factors are strong; however, it is unsatisfactory under weak factors. For the DGPs with strong factors, L2-relaxation performs better than ridge and LASSO, and approaches PCA; with weak factors, L2-relaxation is the best in most DGPs. In the meantime, we notice that LASSO is not a suitable method, even for the weak factor cases in which only 4 coefficients are large while the others are small.

## 5.2 Panel Data Approach

For the single-treated-unit PDA, we consider 9 ($= 3 \times 3$) DGPs for the treatment effect as follows. D1: $\Delta_t = 0$; D2: $\Delta_t$ i.i.d. $\sim N(0,1)$; D3: $\Delta_t = 0.3\Delta_{t-1} + \nu_t$ where $\nu_t$ i.i.d. $\sim N(0, 0.91)$; D4–D6 are generated by adding 0.3 respectively on D1–D3 ($\Delta = 0.3$); and D7–D9 by adding 0.5 on D1–D3 ($\Delta = 0.5$). The null hypothesis $H_0 : \Delta = 0$ is true under D1–D3, but false under D4–D9. Note that for any time-variant treatment effects, the unconditional variance equals 1; however, D3, D6, and D9 share a long-run variance larger than 1 due to the time dependence. The untreated data are generated by the DGPs with strong factors in Section 5.1.

Next, for the PDA with many treated units, we still use the DGPs with strong factors in Section 5.1 except two differences for the treated units: (1) for any $i \in \mathcal{M}$, all entries of $\boldsymbol{\lambda}_i$ are independently drawn from Uniform $([-0.5, -0.3] \cup [0.3, 0.5])$; and (2) for any $i \in \mathcal{M}$, $u_{it}$ i.i.d. $\sim N(0, 0.5)$. For the treatment effects, consider 9 ($= 3 \times 3$) sets with a single OOS time period $\mathcal{T}_2 = \{T_1 + 1\}$ so that $T_2 = 1$. For the cases where the null hypothesis $H_0 : \Delta = 0$ is true, D1: $\Delta_{i,T_1+1} = 0$ for all $i \in \mathcal{M}$; D2: $\{\Delta_{i,T_1+1} : i \in \mathcal{M}\}$ are independently drawn from $N(0, M^{-3/2})$; and D3: add a cross-sectional network with correlation of 0.3 on the adjacent units based on D2. When the null hypothesis is false, D4–D6 are generated by adding 0.3 respectively on D1–D3 ($\Delta = 0.3$); and D7–D9 by adding 0.5 on D1–D3 ($\Delta = 0.5$).

Table 2 shows the simulated testing size and power with a 5% nominal size for the L2-relaxation PDA with single and multiple treated units. The empirical size of the test approaches the nominal size well, and the power gets large when the sample size increases. We also plot the distributions of the simulated test statistics under the null hypothesis in

Table 2: Simulated size and power of L2-relaxation PDA

(a) Single treated unit ($N = 100$)

| DGP | $T_1, T_2$ | Size | | | Power | | | | | |
|---|---|---|---|---|---|---|---|---|---|---|
| | | D1 | D2 | D3 | D4 | D5 | D6 | D7 | D8 | D9 |
| Homo. | 50 | 0.142 | 0.122 | 0.152 | 0.570 | 0.392 | 0.360 | 0.872 | 0.714 | 0.598 |
| | 100 | 0.100 | 0.078 | 0.094 | 0.812 | 0.570 | 0.474 | 0.986 | 0.908 | 0.840 |
| | 200 | 0.072 | 0.048 | 0.060 | 0.980 | 0.836 | 0.730 | 1.000 | 1.000 | 0.996 |
| Mild hetero. | 50 | 0.140 | 0.124 | 0.146 | 0.562 | 0.388 | 0.352 | 0.874 | 0.718 | 0.602 |
| | 100 | 0.104 | 0.082 | 0.096 | 0.812 | 0.562 | 0.480 | 0.986 | 0.912 | 0.836 |
| | 200 | 0.074 | 0.050 | 0.060 | 0.980 | 0.836 | 0.740 | 1.000 | 1.000 | 0.994 |
| Sev. hetero. | 50 | 0.144 | 0.124 | 0.146 | 0.556 | 0.402 | 0.352 | 0.876 | 0.706 | 0.614 |
| | 100 | 0.108 | 0.086 | 0.100 | 0.816 | 0.564 | 0.476 | 0.986 | 0.912 | 0.844 |
| | 200 | 0.076 | 0.050 | 0.064 | 0.982 | 0.838 | 0.738 | 1.000 | 1.000 | 0.994 |

(b) Multiple treated units ($T_2 = 1$)

| DGP | $N, T_1$ | $M$ | Size | | | Power | | | | | |
|---|---|---|---|---|---|---|---|---|---|---|---|
| | | | D1 | D2 | D3 | D4 | D5 | D6 | D7 | D8 | D9 |
| Homo. | 50 | 30 | 0.120 | 0.116 | 0.112 | 0.674 | 0.684 | 0.690 | 0.960 | 0.960 | 0.964 |
| | 100 | 40 | 0.084 | 0.092 | 0.082 | 0.814 | 0.844 | 0.792 | 0.994 | 0.994 | 0.990 |
| | 200 | 50 | 0.052 | 0.052 | 0.054 | 0.832 | 0.852 | 0.836 | 1.000 | 1.000 | 1.000 |
| Mild hetero. | 50 | 30 | 0.120 | 0.118 | 0.120 | 0.676 | 0.682 | 0.692 | 0.964 | 0.966 | 0.968 |
| | 100 | 40 | 0.086 | 0.088 | 0.084 | 0.814 | 0.844 | 0.790 | 0.992 | 0.996 | 0.990 |
| | 200 | 50 | 0.050 | 0.052 | 0.050 | 0.832 | 0.858 | 0.834 | 1.000 | 1.000 | 1.000 |
| Sev. hetero. | 50 | 30 | 0.124 | 0.128 | 0.124 | 0.678 | 0.686 | 0.692 | 0.962 | 0.962 | 0.968 |
| | 100 | 40 | 0.092 | 0.098 | 0.084 | 0.812 | 0.842 | 0.792 | 0.992 | 0.994 | 0.992 |
| | 200 | 50 | 0.052 | 0.052 | 0.050 | 0.834 | 0.862 | 0.834 | 1.000 | 1.000 | 1.000 |

Note: the nominal size is 5%.

## 6 Empirical Examples

We demonstrate the application of L2-relaxation prediction with two empirical examples. Section 6.1 evaluates the effect of real estate regulation policies on PPI growth rates in China as an example of the single-treated-unit PDA; Section 6.2 assesses the effects of Brexit on the stock returns of British and European companies as the multiple treated units.

## 6.1 PPI and Real Estate Regulation in China

China's PPI has fallen to negative growth since October 2022. As one of the major indicators of price levels, PPI is susceptible to multiple global and domestic factors. Global events, such as fluctuations in commodity prices and disruptions in supply chains, affect many countries, and thus can be pinned down by common factors. If we can control these global effects, we would single out the influence of China's domestic policies that have suppressed the PPI.

In the first two decades of the 21st century, the real estate industry was a pillar of China's growth model, and the demand spilled over into other upperstream and downstream industries. However, 2021 was a turning point for this industry, marked by over 650 policy announcements within one year that caused an abrupt market cool-down and pushed multiple property developers into credit crises. Announced in August 2020, the "Three Red Lines" policy was implemented in January 2021, which tightly restricted the liability ratio of property developers, and consequently triggered the default of highly leveraged companies. From February to June, over 16 major cities tightened their housing purchase and mortgage restrictions, which significantly dampened the demand.

We collect the PPI data from the CEIC database for all available countries between 2011 and 2024. The pre-treatment data are those reported before August 2020 ($T_1 = 115$); and the post-treatment period starts from June 2021 ($T_2 = 43$), when the majority of real estate regulation policies have been fully implemented. We compute the monthly PPI growth rates (year-over-year, YoY), and use those of all other countries ($N = 64$) to fit China's PPI growth rate. The gap between the pre-treatment and post-treatment periods is the interval between the initiation and full implementation of the regulatory policies. The PDA framework can accommodate such practical scenarios when multiple intermediate periods exist, as long as $\mathcal{T}_1$ is untreated and $\mathcal{T}_2$ is treated.

To further analyze the mechanism of the policy effect, we additionally regress on the sectoral PPI growth rate of China's building material industry. The building material industry is, among all available sectoral PPIs, the most related industry to the real estate market. The sudden deterioration of the real estate industry exposed excess capacity in its upperstream industries such as steel, since construction (including real estate and infrastructure) has long been the dominant driver of steel consumption in China. Overcapacity in the building material industry remains a primary contributor to the decline in domestic PPI.

Firstly, we use the pre-treatment subsample to compare the OOS prediction performance and placebo test among L2-relaxation, ridge, LASSO (5-block CV for these three methods), forward selection (FS, stopping rule proposed by Shi & Huang 2023), SCM, PCA (number of factors determined by PC$_{p1}$ criterion in Bai & Ng 2002) and OLS. We further split this pre-treatment subsample into a training sample (before June 2018, 90 months) and a testing sample (from July 2018 to July 2020, 25 months). Table 3 summarizes the pre-treatment prediction and placebo testing results. Among all methods, L2-relaxation performs the best in the OOS prediction for the PPI growth rates of both overall industries and the building material industry according to rooted MPSE (RMPSE) and OOS $R^2$. We additionally use the pre-treatment OOS prediction to conduct the placebo test. It can be observed that L2-relaxation, ridge, and PCA do not reject the null hypothesis of zero ATE in either the overall PPI or the building material sectoral PPI before announcing the "Three Red Lines" policy. It should be noticed that, although performing the third best in the OOS prediction, LASSO always rejects the null hypothesis, indicating severe bias in its prediction.

Table 3: OOS prediction performance and placebo test

| Rank | Method | RMPSE | OOS $R^2$ | ATE | $t$-stat. | $p$-value |
|---|---|---|---|---|---|---|
| | | | *Overall* | | | |
| 1 | L2-relaxation | 0.7895 | 0.8686 | -0.0244 | -0.0929 | 0.9260 |
| 2 | Ridge | 0.9195 | 0.8218 | -0.4187 | -1.6182 | 0.1056 |
| 3 | LASSO | 1.3598 | 0.6103 | 0.9359 | 2.8157 | 0.0049 *** |
| 4 | PCA ($PC_{p1}$) | 1.5009 | 0.5252 | 0.3391 | 0.6402 | 0.5220 |
| 5 | FS | 1.6191 | 0.4474 | 0.1421 | 0.2988 | 0.7651 |
| 6 | SCM | 1.6860 | 0.4008 | 1.1073 | 2.4558 | 0.0141 ** |
| 7 | OLS | 1.7745 | 0.3362 | 0.1516 | 0.2719 | 0.7857 |
| | | | *Building material industry* | | | |
| 1 | L2-relaxation | 1.2754 | 0.8591 | -0.4105 | -1.1162 | 0.2643 |
| 2 | PCA ($PC_{p1}$) | 1.5728 | 0.7857 | 0.6656 | 1.5313 | 0.1257 |
| 3 | LASSO | 1.7872 | 0.7233 | 1.2075 | 2.7465 | 0.0060 *** |
| 4 | Ridge | 2.0195 | 0.6467 | 0.6539 | 0.9170 | 0.3591 |
| 5 | SCM | 2.2898 | 0.5457 | 1.7973 | 3.8214 | 0.0001 *** |
| 6 | OLS | 2.4171 | 0.4938 | 2.0711 | 5.4600 | 0.0000 *** |
| 7 | FS | 2.5411 | 0.4405 | 2.1830 | 5.4955 | 0.0000 *** |

Note: * $p$-value < 0.1, ** $p$-value < 0.05, *** $p$-value < 0.01.

Figure 1 (a)–(b) show the OOS L2-relaxation prediction estimated from the training sample for the placebo test; panels (c)–(d) exhibit the L2-relaxation counterfactual prediction trained from the entire pre-treatment dataset. Comparing each column, regardless of the overall PPI or the building material industry's PPI, the predictions remain consistent and robust to the choice of training samples. Focusing on panels (c)–(d), the shaded months are the intermediate periods between the announcement and the full implementation of the policies. Starting from June 2021, the treatment effects have been significantly negative. For the overall PPI, the ATE over the post-treatment period is -6.40(%). The derived $t$-statistic is -3.6087, which rejects the zero mean ATE null hypothesis with $p$-value as small as 0.0003. While for the building material industry, the ATE rises to -16.51(%), as it is the main channel of the effect. Correspondingly, the $t$-statistic is -5.8243, which also rejects the null hypothesis with $p$-value virtually 0. While the realized overall PPI growth is under the negative regime after October 2022, the counterfactual outcomes suggest lifting out of the negative regime in May 2024. The counterfactuals in the building material industry also paint a much brighter picture.

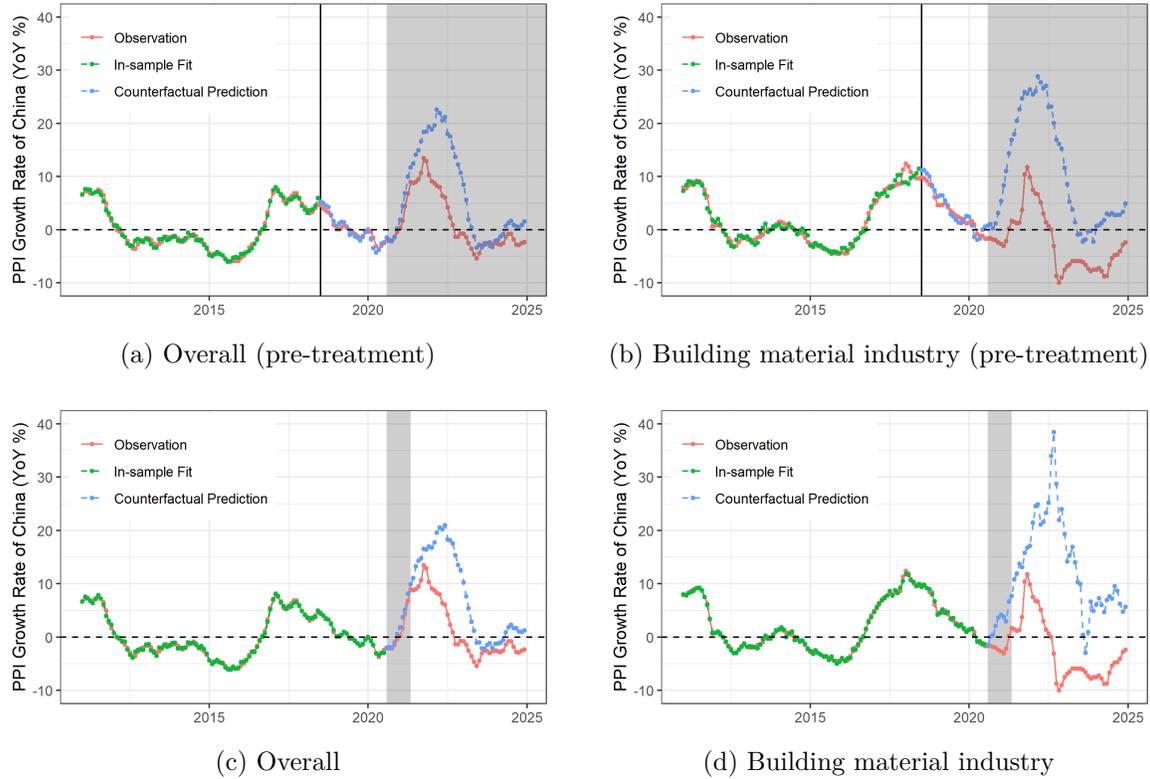

Figure 1: Counterfactual prediction for PPI growth rate of China

## 6.2 Stock Returns of U.K. and E.U. Companies After Brexit

The portmanteau "Brexit" refers to the withdrawal of the United Kingdom (U.K.) from the European Union (E.U.). The referendum was held on June 24, 2016, and the outcome shocked the financial markets. To estimate this shock, we select the equities listed in the New York Stock Exchange, the American Stock Exchange, and NASDAQ. Although the event may have larger effects on the U.K. and E.U. stock markets, here we choose the American markets for consistency in the market reactions to U.K./E.U. corporations from those outside of the two regions. We acquire the daily stock return data from *Center for Research in Security Prices* (CRSP) and company fundamentals from *Compustat North America*. The pre-treatment sample ranges from June 24, 2015, to June 23, 2016 ($T_1 = 253$ trading days), while the post-treatment period is June 24 to July 25, 2016 ($T_2 = 21$ trading days). We treat the equities with headquarters located in the U.K. and E.U. as two treated groups, respectively. After removing securities with missing values, we have 52 U.K. stocks and 99 E.U. stocks. We construct a donor pool of 300 equities whose headquarters are located in the rest of the world to maintain $N > T_1$.

We estimate the ATE by separately predicting the counterfactual outcome of each treated unit with L2-relaxation (5-block CV). The upper panel in Figure 2 plots the daily ATEs on the stock returns of U.K. companies after the referendum, while the lower panel demonstrates those of the E.U. equities. The dashed blue lines enclose 95% confidence intervals (CI) for the null hypothesis of zero ATE. This exercise reveals the substantial impact of the referendum that caught the market by surprise. Our method predicts a shock of 0.04 average drop among U.K. stock returns on the first trading day, followed by another 0.02 decline on the second trading day. As the market digested the information, the daily returns mostly got back to the 95% CI in the proceeding days. The pattern and magnitude of the stocks of E.U. companies were similar, and the returns were less volatile after the initial drop.

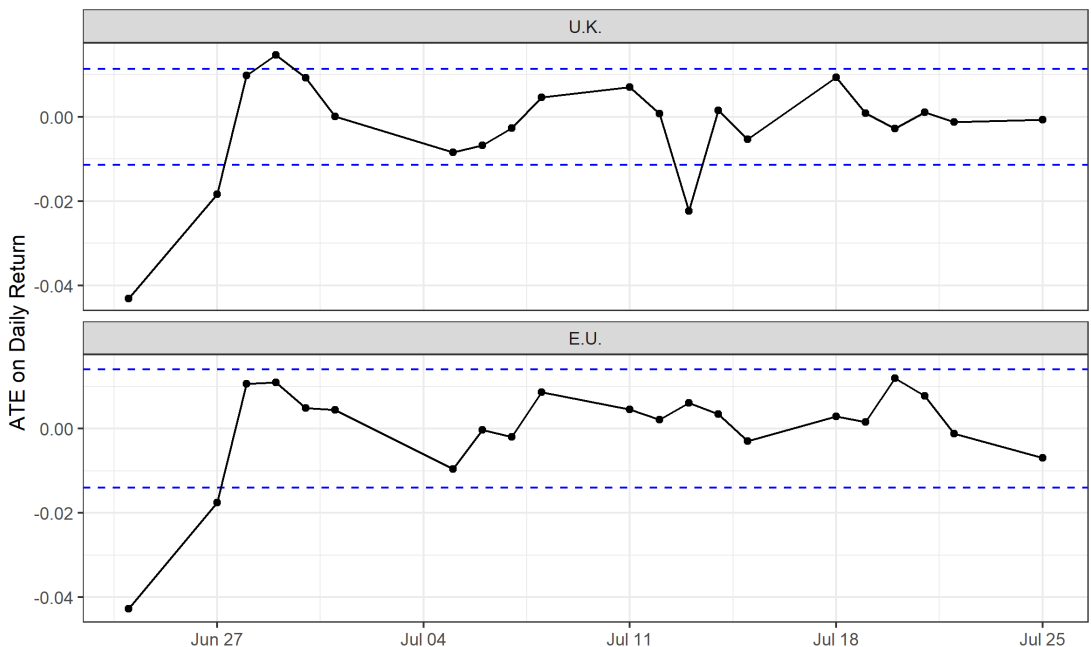

Figure 2: ATE on daily stock returns after Brexit

The two real-data applications show that our procedure not only helps predict the counterfactuals, but also quantifies the statistical uncertainty. Compared with the *event studies* that have been widely used in empirical research, our method provides finer controls via the large donor pool to isolate many co-moving latent factors. It can be readily used in a variety of case studies. In Appendix E, we further revisit two examples: the handover in Hong Kong (Hsiao et al. 2012) and the luxury watch imports (Shi & Huang 2023).

# 7 Conclusion

In this paper, we propose using the L2-relaxation to estimate many regression coefficients and to conduct economic predictions. By employing the latent factor structure, where all coefficients are generally non-zero and the variables are highly correlated, we can consistently estimate the generic coefficients and predict the outcomes. We further apply our method to policy evaluation with PDA when many control units are available. Based on the L2-relaxation counterfactual prediction as if the policy had not taken effect, we establish the inference for ATE. Additionally, we extend PDA to the setting of many treated units with a short post-treatment period. Replication data and code have been made available at: https://github.com/ishwang1/L2relax-PDA.

# References

Abadie, A., Diamond, A. & Hainmueller, J. (2010), 'Synthetic control methods for comparative case studies: Estimating the effect of california's tobacco control program', *Journal of the American Statistical Association* **105**(490), 493–505.

Abadie, A. & Gardeazabal, J. (2003), 'The economic costs of conflict: A case study of the basque country', *American Economic Review* **93**(1), 113–132.

Andrews, D. W. K. (1991), 'Heteroskedasticity and autocorrelation consistent covariance matrix estimation', *Econometrica* **59**(3), 817–858.

Bai, J. & Ng, S. (2002), 'Determining the number of factors in approximate factor models', *Econometrica* **70**(1), 191–221.

Bergmeir, C., Hyndman, R. J. & Koo, B. (2018), 'A note on the validity of cross-validation for evaluating autoregressive time series prediction', *Computational Statistics & Data Analysis* **120**, 70–83.

Bickel, P. J., Ritov, Y. & Tsybakov, A. B. (2009), 'Simultaneous analysis of lasso and dantzig selector', *The Annals of Statistics* **37**(4), 1705–1732.

Bühlmann, P. & van de Geer, S. (2011), *Statistics for high-dimensional data: Methods, theory and applications*, Springer Science & Business Media.

Candes, E. J. & Tao, T. (2005), 'Decoding by linear programming', *IEEE Transactions on Information Theory* **51**(12), 4203–4215.

Candes, E. J. & Tao, T. (2007), 'The dantzig selector: Statistical estimation when $p$ is much larger than $n$', *The Annals of Statistics* **35**(6), 2313–2351.

Carvalho, C., Masini, R. & Medeiros, M. C. (2018), 'ArCo: An artificial counterfactual

approach for high-dimensional panel time-series data', *Journal of Econometrics* **207**(2), 352–380.

Chernozhukov, V., Wüthrich, K. & Zhu, Y. (2021), 'An exact and robust conformal inference method for counterfactual and synthetic controls', *Journal of the American Statistical Association* **116**(536), 1849–1864.

Cohen, A., Dahmen, W. & DeVore, R. (2009), 'Compressed sensing and best *k*-term approximation', *American Mathematical Society* **22**, 211–231.

Ferman, B. (2021), 'On the properties of the synthetic control estimator with many periods and many controls', *Journal of the American Statistical Association* **116**(536), 1764–1772.

Giannone, D., Lenza, M. & Primiceri, G. E. (2021), 'Economic predictions with big data: The illusion of sparsity', *Econometrica* **89**(5), 2409–2437.

Hastie, T., Montanari, A., Rosset, S. & Tibshirani, R. J. (2022), 'Surprises in high-dimensional ridgeless least squares interpolation', *The Annals of Statistics* **50**(2), 949–986.

Hoerl, A. E. & Kennard, R. W. (1970), 'Ridge regression: Biased estimation for nonorthogonal problems', *Technometrics* **12**(1), 55–67.

Hsiao, C., Ching, S. H. & Wan, K. S. (2012), 'A panel data approach for program evaluation: Measuring the benefits of political and economic integration of hong kong with mainland china', *Journal of Applied Econometrics* **27**(5), 705–740.

Li, K. T. & Bell, D. R. (2017), 'Estimation of average treatment effects with panel data: Asymptotic theory and implementation', *Journal of Econometrics* **197**(1), 65–75.

Liao, C., Shi, Z. & Zheng, Y. (2025), 'A relaxation approach to synthetic control', *arXiv preprint arXiv:2508.01793* .

Liao, Y., Ma, X., Neuhierl, A. & Shi, Z. (2024), 'Does noise hurt economic forecasts?', *Working Paper* .

Meinshausen, N., Rocha, G. & Yu, B. (2007), 'Discussion: A tale of three cousins: Lasso, l2boosting and dantzig', *The Annals of Statistics* **35**(6), 2373–2384.

Newey, W. & West, K. (1987), 'A simple, positive semi-definite, heteroskedasticity and autocorrelation consistent covariance matrix', *Econometrica* **55**(3), 703–08.

Newey, W. & West, K. (1994), 'Automatic lag selection in covariance matrix estimation', *The Review of Economic Studies* **61**(4), 631–653.

Phillips, P. C. B. & Moon, H. R. (1999), 'Linear regression limit theory for nonstationary panel data', *Econometrica* **67**(5), 1057–1111.

Shi, Z. & Huang, J. (2023), 'Forward-selected panel data approach for program evaluation', *Journal of Econometrics* **234**(2), 512–535.

Shi, Z., Su, L. & Xie, T. (2025), '$\ell_2$-relaxation: With applications to forecast combination and portfolio analysis', *The Review of Economics and Statistics* **107**(2), 523–538.

Tibshirani, R. (1996), 'Regression shrinkage and selection via the lasso', *Journal of the Royal Statistical Society Series B: Statistical Methodology* **58**(1), 267–288.

Wainwright, M. J. (2019), *Random matrices and covariance estimation*, Cambridge Series in Statistical and Probabilistic Mathematics, Cambridge University Press, pp. 159—193.

# APPENDIX

**Additional Notations.** Besides the notations defined in Section 1, here we introduce some additional notations for convenience in proofs. Denote $\mathbb{1}(\cdot)$ for the indicator function, and define $\mathrm{sgn}(x) = \mathbb{1}(x > 0) - \mathbb{1}(x < 0)$ for all $x \in \mathbb{R}$ as the sign function. For $n$-dimensional vector $\boldsymbol{x} = (x_i)_{i \in [n]}$, define $\mathrm{supp}(\boldsymbol{x}) = \{i \in [n] : x_i \neq 0\}$ and $\|\boldsymbol{x}\|_0 = \sum_{i=1}^n \mathbb{1}(x_i \neq 0)$. For $m \times n$ matrix $\boldsymbol{A} = (a_{ij})_{i \in [m], j \in [n]}$, define the columnwise L2-norm $\|\boldsymbol{A}\|_{c2} = \max_{j \in [n]} \|\boldsymbol{A}_{\cdot j}\|_2$.

# A    Regularization Schemes for Linear Regression

Consider a demeaned linear regression model

$$\boldsymbol{y} = \boldsymbol{X}\boldsymbol{\beta} + \boldsymbol{\epsilon},$$

where $\boldsymbol{y}$ is a $n$-dimensional vector of responses, $\boldsymbol{X}$ is a $n \times p$ design matrix, $\boldsymbol{\beta}$ is a $p$-dimensional vector of unknown coefficients, and $\boldsymbol{\epsilon}$ is a $n$-dimensional vector of errors. Here

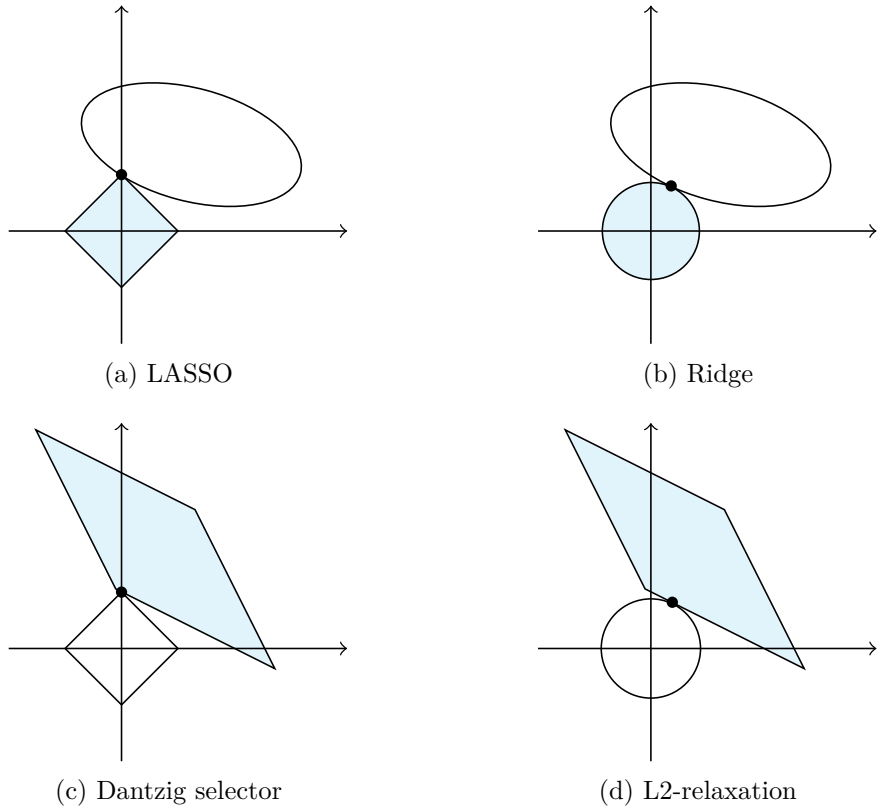

(a) LASSO    (b) Ridge

(c) Dantzig selector    (d) L2-relaxation

Note: in each graph, the coordinate system presents a 2-dimensional parameter space. The white closed region is a contour of the criterion function, and the light blue closed region is the constraint. The constrained optimization is explicit for DS and L2-relaxation. The penalized optimization problem of LASSO or ridge can be equivalently written as a constrained optimization.

Figure A.1: Diagram of the geometry in $\mathbb{R}^2$ space

we formulate the optimization problems of four machine learning methods:

$$\text{LASSO:} \quad \min_{\boldsymbol{\beta}} \frac{1}{2n} \|\boldsymbol{y} - \boldsymbol{X}\boldsymbol{\beta}\|_2^2 + \lambda \|\boldsymbol{\beta}\|_1, \tag{A.1}$$

$$\text{DS:} \quad \min_{\boldsymbol{\beta}} \|\boldsymbol{\beta}\|_1, \quad \text{s.t.} \ \|\boldsymbol{X}'(\boldsymbol{y} - \boldsymbol{X}\boldsymbol{\beta})/n\|_\infty \leq \tau, \tag{A.2}$$

$$\text{Ridge:} \quad \min_{\boldsymbol{\beta}} \frac{1}{2n} \|\boldsymbol{y} - \boldsymbol{X}\boldsymbol{\beta}\|_2^2 + \lambda \|\boldsymbol{\beta}\|_2^2, \tag{A.3}$$

$$\text{L2-relaxation:} \quad \min_{\boldsymbol{\beta}} \frac{1}{2} \|\boldsymbol{\beta}\|_2^2, \quad \text{s.t.} \ \|\boldsymbol{X}'(\boldsymbol{y} - \boldsymbol{X}\boldsymbol{\beta})/n\|_\infty \leq \tau. \tag{A.4}$$

Figure A.1 compares the geometry of the above methods illustrated in the $\mathbb{R}^2$ space. The figures demonstrate that LASSO and DS yield sparse solutions, while ridge and L2-relaxation solve for the dense ones.

# B  Proofs and Discussions for Section 3

We first show some simple facts under Assumption 1–2 without proof:

(i) $\text{rank}(\boldsymbol{\Lambda}) = q$; $\xi_N = O(1)$ and $\xi_N > 0$.

(ii) $\phi_{\min}(\boldsymbol{\Lambda}'\boldsymbol{\Omega}^{-1}\boldsymbol{\Lambda}/N + N^{-1}\boldsymbol{I}_q) > \phi_{\min}(\boldsymbol{\Lambda}'\boldsymbol{\Omega}^{-1}\boldsymbol{\Lambda}/N) \geq \phi_{\min}(\boldsymbol{\Lambda}'\boldsymbol{\Lambda}/N)/\sigma_{\max}^2$, which guarantees the invertibility of $\boldsymbol{\Lambda}'\boldsymbol{\Omega}^{-1}\boldsymbol{\Lambda}/N + N^{-1}\boldsymbol{I}_q$ in (9) and also $\boldsymbol{\Lambda}'\boldsymbol{\Omega}^{-1}\boldsymbol{\Lambda}/N$;

(iii) by the fact between the columnwise L2-norm and the spectral norm, $\|\boldsymbol{\Omega}\|_{c2} \leq \sigma_{\max}^2$.

## B.1  Proof of Proposition 1

**Part (a).** The explicit expression of $\boldsymbol{\beta}^0$ in (9) yields

$$\begin{aligned}
\|\boldsymbol{\beta}^0\|_2^2 &\leq N^{-1} \sigma_{\min}^{-2} \left\| \frac{\boldsymbol{\Omega}^{-\frac{1}{2}}\boldsymbol{\Lambda}}{\sqrt{N}} \left( \frac{\boldsymbol{\Lambda}'\boldsymbol{\Omega}^{-1}\boldsymbol{\Lambda}}{N} + N^{-1}\boldsymbol{I}_q \right)^{-1} \boldsymbol{\lambda}_0 \right\|_2^2 \\
&\leq N^{-1} \sigma_{\min}^{-2} \boldsymbol{\lambda}_0' \left( \frac{\boldsymbol{\Lambda}'\boldsymbol{\Omega}^{-1}\boldsymbol{\Lambda}}{N} + N^{-1}\boldsymbol{I}_q \right)^{-1} \boldsymbol{\lambda}_0 \\
&\leq N^{-1} \xi_N^{-1} (\sigma_{\max}^2/\sigma_{\min}^2) \|\boldsymbol{\lambda}_0\|_2^2 = O(N^{-1}\xi_N^{-1}),
\end{aligned}$$

where the last inequality follows by fact (ii).

**Part (b).** Furthermore, since

$$\begin{aligned}
\boldsymbol{\lambda}_0 - \boldsymbol{\Lambda}'\boldsymbol{\beta}^0 &= \left[ \boldsymbol{I}_q - \frac{\boldsymbol{\Lambda}'\boldsymbol{\Omega}^{-1}\boldsymbol{\Lambda}}{N} \left( \frac{\boldsymbol{\Lambda}'\boldsymbol{\Omega}^{-1}\boldsymbol{\Lambda}}{N} + N^{-1}\boldsymbol{I}_q \right)^{-1} \right] \boldsymbol{\lambda}_0 \\
&= N^{-1} \left( \frac{\boldsymbol{\Lambda}'\boldsymbol{\Omega}^{-1}\boldsymbol{\Lambda}}{N} + N^{-1}\boldsymbol{I}_q \right)^{-1} \boldsymbol{\lambda}_0,
\end{aligned}$$

we have

$$\|\boldsymbol{\lambda}_0 - \boldsymbol{\Lambda}'\boldsymbol{\beta}^0\|_2 \leq N^{-1} \left\|\left(\frac{\boldsymbol{\Lambda}'\boldsymbol{\Omega}^{-1}\boldsymbol{\Lambda}}{N} + N^{-1}\boldsymbol{I}_q\right)^{-1}\right\|_2 \|\boldsymbol{\lambda}_0\|_2 = \mathrm{O}(N^{-1}\xi_N^{-1}).$$

## B.2 Proof of Lemma 1

**Part (a).** Consider the constraint condition of problem (15), when $\tau = 0$, by (14) we have $\boldsymbol{\Lambda}\hat{\boldsymbol{\Sigma}}_f\boldsymbol{\Lambda}'\boldsymbol{\beta} = \boldsymbol{\Lambda}\hat{\boldsymbol{\Sigma}}_f\boldsymbol{\lambda}_0$. As $\boldsymbol{\Lambda}'\boldsymbol{\Lambda}/N$ and $\hat{\boldsymbol{\Sigma}}_f$ are nonsingular, the above equation is equivalent to $\boldsymbol{\Lambda}'\boldsymbol{\beta} = \boldsymbol{\lambda}_0$ after we pre-multiply both sides by $\hat{\boldsymbol{\Sigma}}_f^{-1}(\boldsymbol{\Lambda}'\boldsymbol{\Lambda})^{-1}\boldsymbol{\Lambda}'$. Hence, under Assumption 1 (b), problem (15) is equivalent to (11), which has a unique solution $\boldsymbol{\beta}^*$ as in (12).

**Part (b).** By (12), we have

$$\|\boldsymbol{\beta}^*\|_2^2 = N^{-1}\boldsymbol{\lambda}_0'\left(\frac{\boldsymbol{\Lambda}'\boldsymbol{\Lambda}}{N}\right)^{-1}\boldsymbol{\lambda}_0 \leq N^{-1}\left[\phi_{\min}\left(\frac{\boldsymbol{\Lambda}'\boldsymbol{\Lambda}}{N}\right)\right]^{-1}\|\boldsymbol{\lambda}_0\|_2^2 = \mathrm{O}(N^{-1}\xi_N^{-1}).$$

**Part (c).** By the fact that $\phi_{\min}(\boldsymbol{\Lambda}'\boldsymbol{\Omega}^{-1}\boldsymbol{\Lambda}/N) \geq \xi_N/\sigma_{\max}^2$, as $N$ goes large, the eigenvalues of $\boldsymbol{\Lambda}'\boldsymbol{\Omega}^{-1}\boldsymbol{\Lambda}/N$ will dominate those of $N^{-1}\boldsymbol{I}_q$. By this virtue, we construct a limiting approximation

$$\boldsymbol{\beta}^N := N^{-1}\boldsymbol{\Omega}^{-1}\boldsymbol{\Lambda}\left(\frac{\boldsymbol{\Lambda}'\boldsymbol{\Omega}^{-1}\boldsymbol{\Lambda}}{N}\right)^{-1}\boldsymbol{\lambda}_0, \tag{B.1}$$

as a bridge between $\boldsymbol{\beta}^*$ and $\boldsymbol{\beta}^0$.

For the *homoskedasticity* case, observing that $\boldsymbol{\beta}^N\big|_{\boldsymbol{\Omega}=\sigma^2\boldsymbol{I}_N} = \boldsymbol{\beta}^*$ for any positive and finite $\sigma^2$, then we simply have $\|\boldsymbol{\beta}^* - \boldsymbol{\beta}^0\|_2 = \|\boldsymbol{\beta}^N - \boldsymbol{\beta}^0\|_2$.

While for the *heteroskedasticity* setting, by the triangle inequality, we have

$$\|\boldsymbol{\beta}^* - \boldsymbol{\beta}^0\|_2 \leq \|\boldsymbol{\beta}^N - \boldsymbol{\beta}^0\|_2 + \|\boldsymbol{\beta}^N - \boldsymbol{\beta}^*\|_2.$$

By the Woodbury matrix identity, for the first term,

$$\boldsymbol{\beta}^N - \boldsymbol{\beta}^0 = N^{-2}\boldsymbol{\Omega}^{-1}\boldsymbol{\Lambda}\left(\frac{\boldsymbol{\Lambda}'\boldsymbol{\Omega}^{-1}\boldsymbol{\Lambda}}{N}\right)^{-1}\left(\frac{\boldsymbol{\Lambda}'\boldsymbol{\Omega}^{-1}\boldsymbol{\Lambda}}{N} + N^{-1}\boldsymbol{I}_q\right)^{-1}\boldsymbol{\lambda}_0,$$

and it follows that

$$\|\boldsymbol{\beta}^N - \boldsymbol{\beta}^0\|_2 \leq N^{-\frac{3}{2}}\sigma_{\min}^{-1}\left\|\frac{\boldsymbol{\Omega}^{-\frac{1}{2}}\boldsymbol{\Lambda}}{\sqrt{N}}\left(\frac{\boldsymbol{\Lambda}'\boldsymbol{\Omega}^{-1}\boldsymbol{\Lambda}}{N}\right)^{-1}\right\|_2 \left\|\left(\frac{\boldsymbol{\Lambda}'\boldsymbol{\Omega}^{-1}\boldsymbol{\Lambda}}{N} + N^{-1}\boldsymbol{I}_q\right)^{-1}\right\|_2 \|\boldsymbol{\lambda}_0\|_2$$
$$= \mathrm{O}(N^{-\frac{3}{2}}\xi_N^{-\frac{3}{2}}).$$

Regarding the second term, $\|\boldsymbol{\beta}^N - \boldsymbol{\beta}^*\|_2$, we first define $\boldsymbol{\Psi} := \sigma_{\max}^2\boldsymbol{\Omega}^{-1} - \boldsymbol{I}_N$. Obviously $\phi_{\max}(\boldsymbol{\Psi}) \leq \sigma_{\max}^2[\phi_{\min}(\boldsymbol{\Omega})]^{-1} - 1 = \psi_{\max}$ and $\boldsymbol{\Omega}^{-1} = (\sigma_{\max}^2)^{-1}(\boldsymbol{I}_N + \boldsymbol{\Psi})$. Similarly, by the Woodbury matrix identity,

$$\boldsymbol{\beta}^N - \boldsymbol{\beta}^* = \boldsymbol{\Psi}\boldsymbol{\beta}^* - N^{-1}\boldsymbol{\Omega}^{-1}\boldsymbol{\Lambda}\left(\frac{\boldsymbol{\Lambda}'\boldsymbol{\Omega}^{-1}\boldsymbol{\Lambda}}{N}\right)^{-1}\frac{\boldsymbol{\Lambda}'\boldsymbol{\Psi}\boldsymbol{\Lambda}}{N}\left(\frac{\boldsymbol{\Lambda}'\boldsymbol{\Lambda}}{N}\right)^{-1}\boldsymbol{\lambda}_0,$$

with $\|\boldsymbol{\Psi}\boldsymbol{\beta}^*\|_2 \leq \psi_{\max}\|\boldsymbol{\beta}^*\|_2 = \psi_{\max}O(N^{-\frac{1}{2}}\xi_N^{-\frac{1}{2}})$, and the latter term being bounded by

$$\frac{1}{\sqrt{N}\sigma_{\min}}\left\|\frac{\boldsymbol{\Omega}^{-\frac{1}{2}}\boldsymbol{\Lambda}}{\sqrt{N}}\left(\frac{\boldsymbol{\Lambda}'\boldsymbol{\Omega}^{-1}\boldsymbol{\Lambda}}{N}\right)^{-1}\right\|_2\left\|\frac{\boldsymbol{\Lambda}'\boldsymbol{\Psi}\boldsymbol{\Lambda}}{N}\left(\frac{\boldsymbol{\Lambda}'\boldsymbol{\Lambda}}{N}\right)^{-1}\right\|_2\|\boldsymbol{\lambda}_0\|_2 = \psi_{\max}O(N^{-\frac{1}{2}}\xi_N^{-1}),$$

which dominates the former one as $N$ goes large, by the fact that $\xi_N = O(1)$. This result is compatible for the *homoskedasticity* case by letting $\psi_{\max} = 0$.

## B.3  Proof of Lemma 2

Rewrite problem (3) in terms of linear constraints as

$$\min_{\boldsymbol{\beta}} \frac{1}{2}\|\boldsymbol{\beta}\|_2^2, \quad \text{s.t. } \hat{\eta}_i - \hat{\boldsymbol{\Sigma}}_{i\cdot}\boldsymbol{\beta} \leq \tau \text{ and } -(\hat{\eta} - \hat{\boldsymbol{\Sigma}}_{i\cdot}\boldsymbol{\beta}) \leq \tau \text{ for all } i \in \mathcal{N}. \tag{B.2}$$

The Lagrangian function is

$$\mathcal{L}(\boldsymbol{\beta};\boldsymbol{\gamma}_1,\boldsymbol{\gamma}_2) = \frac{1}{2}\|\boldsymbol{\beta}\|_2^2 + \boldsymbol{\gamma}_1'(\hat{\boldsymbol{\eta}} - \hat{\boldsymbol{\Sigma}}\boldsymbol{\beta} - \tau\mathbf{1}_N) - \boldsymbol{\gamma}_2'(\hat{\boldsymbol{\eta}} - \hat{\boldsymbol{\Sigma}}\boldsymbol{\beta} + \tau\mathbf{1}_N)$$
$$= (\boldsymbol{\gamma}_1 - \boldsymbol{\gamma}_2)'\hat{\boldsymbol{\eta}} - \tau(\boldsymbol{\gamma}_1 + \boldsymbol{\gamma}_2)'\mathbf{1}_N - (\boldsymbol{\gamma}_1 - \boldsymbol{\gamma}_2)'\hat{\boldsymbol{\Sigma}}\boldsymbol{\beta} + \frac{1}{2}\|\boldsymbol{\beta}\|_2^2,$$

with all elements in $\boldsymbol{\gamma}_1$ and $\boldsymbol{\gamma}_2$ being non-negative; and the conjugate function of the objective function is

$$\varphi^*(\boldsymbol{b}) = \sup_{\boldsymbol{\beta}}\left\{\boldsymbol{b}'\boldsymbol{\beta} - \frac{1}{2}\|\boldsymbol{\beta}\|_2^2\right\} = \frac{1}{2}\|\boldsymbol{b}\|_2^2.$$

Then the Lagrangian dual function is

$$g(\boldsymbol{\gamma}_1,\boldsymbol{\gamma}_2) = \inf_{\boldsymbol{\beta}} \mathcal{L}(\boldsymbol{\beta};\boldsymbol{\gamma}_1,\boldsymbol{\gamma}_2)$$
$$= (\boldsymbol{\gamma}_1 - \boldsymbol{\gamma}_2)'\hat{\boldsymbol{\eta}} - \tau(\boldsymbol{\gamma}_1 + \boldsymbol{\gamma}_2)'\mathbf{1}_N - \sup_{\boldsymbol{\beta}}\left\{(\boldsymbol{\gamma}_1 - \boldsymbol{\gamma}_2)'\hat{\boldsymbol{\Sigma}}\boldsymbol{\beta} - \frac{1}{2}\|\boldsymbol{\beta}\|_2^2\right\}$$
$$= (\boldsymbol{\gamma}_1 - \boldsymbol{\gamma}_2)'\hat{\boldsymbol{\eta}} - \tau(\boldsymbol{\gamma}_1 + \boldsymbol{\gamma}_2)'\mathbf{1}_N - \varphi^*\left(\hat{\boldsymbol{\Sigma}}(\boldsymbol{\gamma}_1 - \boldsymbol{\gamma}_2)\right)$$
$$= (\boldsymbol{\gamma}_1 - \boldsymbol{\gamma}_2)'\hat{\boldsymbol{\eta}} - \tau(\boldsymbol{\gamma}_1 + \boldsymbol{\gamma}_2)'\mathbf{1}_N - \frac{1}{2}\|\hat{\boldsymbol{\Sigma}}(\boldsymbol{\gamma}_1 - \boldsymbol{\gamma}_2)\|_2^2. \tag{B.3}$$

Let $\boldsymbol{\gamma} := \boldsymbol{\gamma}_1 - \boldsymbol{\gamma}_2$. Note that when $\tau > 0$, for any $i \in \mathcal{N}$, two inequality constraints in problem (B.2) cannot be binding simultaneously, which indicates that $\gamma_{1i}\gamma_{2i} = 0$ must hold. Hence, we have

$$\|\boldsymbol{\gamma}\|_1 = \sum_{i=1}^N |\gamma_{1i} - \gamma_{2i}| = \sum_{i=1}^N (\gamma_{1i} + \gamma_{2i}) = (\boldsymbol{\gamma}_1 + \boldsymbol{\gamma}_2)'\mathbf{1}_N,$$

which implies that the dual problem has the form of

$$\max_{\boldsymbol{\gamma}}\left\{-\frac{1}{2}\|\hat{\boldsymbol{\Sigma}}\boldsymbol{\gamma}\|_2^2 + \hat{\boldsymbol{\eta}}'\boldsymbol{\gamma} - \tau\|\boldsymbol{\gamma}\|_1\right\}.$$

For $\tau = 0$, the Lagrangian dual function becomes $g(\boldsymbol{\gamma}) = \hat{\boldsymbol{\eta}}'\boldsymbol{\gamma} - \frac{1}{2}\|\hat{\boldsymbol{\Sigma}}\boldsymbol{\gamma}\|_2^2$, and it follows that the corresponding dual problem turns out to be $\max_{\boldsymbol{\gamma}} \left\{-\frac{1}{2}\|\hat{\boldsymbol{\Sigma}}\boldsymbol{\gamma}\|_2^2 + \hat{\boldsymbol{\eta}}'\boldsymbol{\gamma}\right\}$, which is compatible to problem (16) as a special case by letting $\tau = 0$.

The relationship between solutions of the primal problem and the dual problem can be derived from the first order condition of $\mathcal{L}(\boldsymbol{\beta}; \boldsymbol{\gamma}_1, \boldsymbol{\gamma}_2)$ with respect to $\boldsymbol{\beta}$, which is $\hat{\boldsymbol{\beta}} - \hat{\boldsymbol{\Sigma}}\hat{\boldsymbol{\gamma}} = \boldsymbol{0}_N$.

## B.4 Proof and Discussion of Lemma 3

### B.4.1 Proof of Lemma 3

**Part (a).** If $\hat{\boldsymbol{\Sigma}}_f$ is nonsingular, then the constraint condition of problem (18) is equivalent to
$$\boldsymbol{\Lambda}'\boldsymbol{\gamma} = \hat{\boldsymbol{\Sigma}}_f^{-1}(\boldsymbol{\Lambda}'\boldsymbol{\Lambda})^{-1}\boldsymbol{\lambda}_0. \tag{B.4}$$
The following proof is based on $\text{rank}(\hat{\boldsymbol{\Sigma}}_f) = q$.

For any $s$-sparse solution to problem (18), $\hat{\boldsymbol{\gamma}}_0^{(s)} = (\hat{\gamma}_{0,1}^{(s)}, \cdots, \hat{\gamma}_{0,N}^{(s)})'$ with $\|\hat{\boldsymbol{\gamma}}_0^{(s)}\|_0 = s$, define $\mathcal{S} := \text{supp}(\hat{\boldsymbol{\gamma}}_0^{(s)})$. If $\text{rank}(\boldsymbol{\Lambda}_{\mathcal{S}\cdot}) < s$, then there must exist a $\boldsymbol{v} = (v_1, \cdots, v_N)'$ such that $\boldsymbol{v}_{\mathcal{S}} \in \text{Null}\left(\boldsymbol{\Lambda}_{\mathcal{S}\cdot}'\right) \setminus \{\boldsymbol{0}_s\}$ and $\boldsymbol{v}_{\mathcal{S}^c} = \boldsymbol{0}_{N-s}$. Note that for any $\theta \neq 0$ with $|\theta| \leq \min_{i \in \mathcal{S}, v_i \neq 0} |\hat{\gamma}_{0,i}^{(s)}/v_i|$, the linear combination $\hat{\boldsymbol{\gamma}}_0^{(s)} + \theta\boldsymbol{v}$ also satisfies the constraint condition (B.4), since
$$\boldsymbol{\Lambda}'(\hat{\boldsymbol{\gamma}}_0^{(s)} + \theta\boldsymbol{v}) = \boldsymbol{\Lambda}'\hat{\boldsymbol{\gamma}}_0^{(s)} + \theta\left(\boldsymbol{\Lambda}_{\mathcal{S}\cdot}'\boldsymbol{v}_{\mathcal{S}} + \boldsymbol{\Lambda}_{\mathcal{S}^c\cdot}'\boldsymbol{v}_{\mathcal{S}^c}\right) = \boldsymbol{\Lambda}'\hat{\boldsymbol{\gamma}}_0^{(s)};$$
and for all $i \in \mathcal{S}$, $\hat{\gamma}_{0,i}^{(s)} + \theta v_i$ must either be zero or share the same sign with $\hat{\gamma}_{0,i}^{(s)}$, therefore,
$$\left\|\hat{\boldsymbol{\gamma}}_{0,\mathcal{S}}^{(s)} + \theta\boldsymbol{v}_{\mathcal{S}}\right\|_1 = \sum_{i \in \mathcal{S}} \left(\hat{\gamma}_{0,i}^{(s)} + \theta v_i\right) \text{sgn}\left(\hat{\gamma}_{0,i}^{(s)} + \theta v_i\right) = \sum_{i \in \mathcal{S}} \left(\hat{\gamma}_{0,i}^{(s)} + \theta v_i\right) \text{sgn}\left(\hat{\gamma}_{0,i}^{(s)}\right)$$
$$= \left\|\hat{\boldsymbol{\gamma}}_{0,\mathcal{S}}^{(s)}\right\|_1 + \theta \sum_{i \in \mathcal{S}} v_i \text{sgn}\left(\hat{\gamma}_{0,i}^{(s)}\right). \tag{B.5}$$

By the definition of $\hat{\boldsymbol{\gamma}}_0^{(s)}$, the inequality
$$\left\|\hat{\boldsymbol{\gamma}}_0^{(s)}\right\|_1 \leq \left\|\hat{\boldsymbol{\gamma}}_0^{(s)} + \theta\boldsymbol{v}\right\|_1 = \left\|\hat{\boldsymbol{\gamma}}_0^{(s)}\right\|_1 + \theta \sum_{i \in \mathcal{S}} v_i \text{sgn}\left(\hat{\gamma}_{0,i}^{(s)}\right)$$
must hold for all positive and negative $\theta \in \left[-\min_{i \in \mathcal{S}, v_i \neq 0} |\hat{\gamma}_{0,i}^{(s)}/v_i|, 0\right) \cup \left(0, \min_{i \in \mathcal{S}, v_i \neq 0} |\hat{\gamma}_{0,i}^{(s)}/v_i|\right]$. Hence, we must have
$$\sum_{i \in \mathcal{S}} v_i \text{sgn}\left(\hat{\gamma}_{0,i}^{(s)}\right) = 0. \tag{B.6}$$

The above result has a geometric interpretation indicating that $\text{Null}(\boldsymbol{\Lambda}_{\mathcal{S}\cdot}')$ is parallel to a hyper-surface of $\{\boldsymbol{x} \in \mathbb{R}^s : \|\boldsymbol{x}\|_1 = a\}$ for any $a > 0$ in a $\mathbb{R}^s$-space, and thus we have multiple $s$-sparse solutions to problem (18). Together with this fact, if we choose $\theta_{\boldsymbol{v}}^{(s)} :=$ $-\arg\min_{\theta\in\Theta_{\boldsymbol{v}}^{(s)}}|\theta|$ where $\Theta_{\boldsymbol{v}}^{(s)}:=\left\{\hat{\gamma}_{0,i}^{(s)}/v_i:i\in\mathcal{S},v_i\neq 0\right\}$, then among all $i\in\mathcal{S}$ there must exist one $\hat{\gamma}_{0,i}^{(s)}+\theta_{\boldsymbol{v}}^{(s)}v_i=0$. Therefore, by (B.5) and (B.6), we have

$$\left\|\hat{\boldsymbol{\gamma}}_0^{(s)}+\theta_{\boldsymbol{v}}^{(s)}\boldsymbol{v}\right\|_1 = \left\|\hat{\boldsymbol{\gamma}}_0^{(s)}\right\|_1 + \theta_{\boldsymbol{v}}^{(s)}\sum_{i\in\mathcal{S}}v_i\text{sgn}\left(\hat{\gamma}_{0,i}^{(s)}\right) = \left\|\hat{\boldsymbol{\gamma}}_0^{(s)}\right\|_1,$$

which implies that $\hat{\boldsymbol{\gamma}}_0^{(s)}+\theta_{\boldsymbol{v}}^{(s)}\boldsymbol{v}$ is also a solution to problem (18), but $\|\hat{\boldsymbol{\gamma}}_0^{(s)}+\theta_{\boldsymbol{v}}^{(s)}\boldsymbol{v}\|_0 < s$. In other words, we can always find a sparser solution if $\text{rank}(\boldsymbol{\Lambda}_{\mathcal{S}\cdot}) < s$.

The above argument makes clear that there must exist a $s$-sparse solution $\hat{\boldsymbol{\gamma}}_0^{(s)}$ such that $\text{rank}(\boldsymbol{\Lambda}_{\mathcal{S}\cdot}) = s$. From now on, we focus on such solutions. Note that $\text{rank}(\boldsymbol{\Lambda}_{\mathcal{S}\cdot}) \leq q$ for all $\mathcal{S}\subset\mathcal{N}$. Therefore, there must exist a solution to problem (18), denoted as $\hat{\boldsymbol{\gamma}}_0^*$, such that

$$\|\hat{\boldsymbol{\gamma}}_0^*\|_0 = \text{rank}\left(\boldsymbol{\Lambda}_{\text{supp}(\hat{\gamma}_0^*),\cdot}\right) \leq q.$$

If $\|\hat{\boldsymbol{\gamma}}_0^*\|_0 = \text{rank}\left(\boldsymbol{\Lambda}_{\text{supp}(\hat{\gamma}_0^*),\cdot}\right) < q$, then there must be a subset $\mathcal{Q}^+\subset[\text{supp}(\hat{\boldsymbol{\gamma}}_0^*)]^c$ with $|\mathcal{Q}^+| = q-\|\hat{\boldsymbol{\gamma}}_0^*\|_0$ such that $\text{rank}(\boldsymbol{\Lambda}_{\mathcal{Q}^*\cdot}) = q$ for $\mathcal{Q}^* = \text{supp}(\hat{\boldsymbol{\gamma}}_0^*)\cup\mathcal{Q}^+$; otherwise $\text{rank}(\boldsymbol{\Lambda}) = q$ cannot hold. As a result, we must have $\hat{\boldsymbol{\gamma}}_{0,(\mathcal{Q}^*)^c}^* = \mathbf{0}_{N-q}$, as $(\mathcal{Q}^*)^c\subset[\text{supp}(\hat{\boldsymbol{\gamma}}_0^*)]^c$. While for $\|\hat{\boldsymbol{\gamma}}_0^*\|_0 = \text{rank}(\boldsymbol{\Lambda}_{\mathcal{Q}^*\cdot}) = q$ with $\mathcal{Q}^* = \text{supp}(\hat{\boldsymbol{\gamma}}_0^*)$, the result is trivial. To sum up, there is a solution $\hat{\boldsymbol{\gamma}}_0^*$ to problem (18) such that $\hat{\boldsymbol{\gamma}}_{0,(\mathcal{Q}^*)^c}^* = \mathbf{0}_{N-q}$ with respect to some $\mathcal{Q}^*\in\mathscr{Q}_N$, where $\mathscr{Q}_N\neq\emptyset$ is guaranteed by $\text{rank}(\boldsymbol{\Lambda}) = q$ for all $N > q$.

For any $\boldsymbol{\gamma}\in\mathbb{R}^N$ satisfying $\boldsymbol{\gamma}_{\mathcal{Q}^c} = \mathbf{0}_{N-q}$ with respect to $\mathcal{Q}\in\mathscr{Q}_N$, if it solves the constraint condition (B.4), then

$$\boldsymbol{\Lambda}_{\mathcal{Q}\cdot}'\boldsymbol{\gamma}_{\mathcal{Q}} = \boldsymbol{\Lambda}'\boldsymbol{\gamma} = \hat{\boldsymbol{\Sigma}}_f^{-1}(\boldsymbol{\Lambda}'\boldsymbol{\Lambda})^{-1}\boldsymbol{\lambda}_0, \tag{B.7}$$

which is equivalent to

$$\hat{\boldsymbol{\Sigma}}_{\cdot\mathcal{Q}}^{*\prime}\hat{\boldsymbol{\Sigma}}_{\cdot\mathcal{Q}}^*\boldsymbol{\gamma}_{\mathcal{Q}} = \boldsymbol{\Lambda}_{\mathcal{Q}\cdot}\hat{\boldsymbol{\Sigma}}_f\boldsymbol{\Lambda}'\boldsymbol{\Lambda}\hat{\boldsymbol{\Sigma}}_f\boldsymbol{\Lambda}_{\mathcal{Q}\cdot}'\boldsymbol{\gamma}_{\mathcal{Q}} = \boldsymbol{\Lambda}_{\mathcal{Q}\cdot}\hat{\boldsymbol{\Sigma}}_f\boldsymbol{\lambda}_0 = \hat{\boldsymbol{\eta}}_{\mathcal{Q}}^*,$$

by the fact that $\boldsymbol{\Lambda}_{\mathcal{Q}\cdot}$ is of full rank.

Therefore, for all $\boldsymbol{\gamma}$ satisfying the constraint condition of problem (18) and $\boldsymbol{\gamma}_{\mathcal{Q}^c} = \mathbf{0}_{N-q}$ with respect to some $\mathcal{Q}\in\mathscr{Q}_N$, we can explicitly solve

$$\boldsymbol{\gamma}_{\mathcal{Q}} = \left(\hat{\boldsymbol{\Sigma}}_{\cdot\mathcal{Q}}^{*\prime}\hat{\boldsymbol{\Sigma}}_{\cdot\mathcal{Q}}^*\right)^{-1}\hat{\boldsymbol{\eta}}_{\mathcal{Q}}^*,$$

as $\hat{\boldsymbol{\Sigma}}_{\cdot\mathcal{Q}}^{*\prime}\hat{\boldsymbol{\Sigma}}_{\cdot\mathcal{Q}}^*$ is nonsingular for all $\mathcal{Q}\in\mathscr{Q}_N$. Recall that $\hat{\boldsymbol{\gamma}}_0^*$ is a solution to problem (18) such that $\hat{\boldsymbol{\gamma}}_{0,(\mathcal{Q}^*)^c}^* = \mathbf{0}_{N-q}$ with respect to $\mathcal{Q}^*\in\mathscr{Q}_N$; hence $\mathcal{Q}^*$ minimizes $\left\{\|\boldsymbol{\gamma}_{\mathcal{Q}}\|_1:\mathcal{Q}\in\mathscr{Q}_N\right\}$.

**Part (b).** For all $N > q$, for any $\mathcal{Q}\subset\mathcal{N}$ with $|\mathcal{Q}| = q$, by the definition of $\mathscr{Q}_N$ we have $\phi_{\min}\left(\boldsymbol{\Lambda}_{\mathcal{Q}\cdot}\boldsymbol{\Lambda}_{\mathcal{Q}\cdot}'\right) > 0$ if and only if $\mathcal{Q}\in\mathscr{Q}_N$. Therefore, as $\mathscr{Q}_N\neq\emptyset$, we must have

$$\xi_q := \max_{\mathcal{Q}\subset\mathcal{N},|\mathcal{Q}|=q}\phi_{\min}\left(\boldsymbol{\Lambda}_{\mathcal{Q}\cdot}\boldsymbol{\Lambda}_{\mathcal{Q}\cdot}'\right) = \max_{\mathcal{Q}\in\mathscr{Q}_N}\phi_{\min}\left(\boldsymbol{\Lambda}_{\mathcal{Q}\cdot}\boldsymbol{\Lambda}_{\mathcal{Q}\cdot}'\right) > 0.$$

From (B.7), we solve the constraint condition of problem (18) with

$$\boldsymbol{\gamma}_{\mathcal{Q}} = N^{-1} \left( \boldsymbol{\Lambda}_{\mathcal{Q}\cdot} \boldsymbol{\Lambda}'_{\mathcal{Q}\cdot} \right)^{-1} \boldsymbol{\Lambda}_{\mathcal{Q}\cdot} \hat{\boldsymbol{\Sigma}}_f^{-1} \left( \frac{\boldsymbol{\Lambda}' \boldsymbol{\Lambda}}{N} \right)^{-1} \boldsymbol{\lambda}_0.$$

Then for any $\hat{\boldsymbol{\gamma}}_0^*$ in part (a), we bound

$$\left\| \hat{\boldsymbol{\gamma}}_{0,\mathcal{Q}^*}^* \right\|_1 \leq \|\boldsymbol{\gamma}_{\mathcal{Q}}\|_1 \leq \sqrt{q} \|\boldsymbol{\gamma}_{\mathcal{Q}}\|_2 \leq N^{-1} c_f^{-1} \xi_N^{-1} \left\| \left( \boldsymbol{\Lambda}_{\mathcal{Q}\cdot} \boldsymbol{\Lambda}'_{\mathcal{Q}\cdot} \right)^{-1} \boldsymbol{\Lambda}_{\mathcal{Q}\cdot} \right\|_2 \|\boldsymbol{\lambda}_0\|_2.$$

Since the above inequality holds for all $\mathcal{Q} \in \mathcal{Q}_N$, it follows that

$$\left\| \hat{\boldsymbol{\gamma}}_{0,\mathcal{Q}^*}^* \right\|_1 \leq N^{-1} c_f^{-1} \xi_N^{-1} \left\{ \min_{\mathcal{Q} \in \mathcal{Q}_N} \left[ \phi_{\min} \left( \boldsymbol{\Lambda}_{\mathcal{Q}\cdot} \boldsymbol{\Lambda}'_{\mathcal{Q}\cdot} \right) \right]^{-1} \|\boldsymbol{\Lambda}_{\mathcal{Q}\cdot}\|_2 \right\} \|\boldsymbol{\lambda}_0\|_2$$

$$\leq q N^{-1} c_f^{-1} \xi_N^{-1} \left[ \max_{\mathcal{Q} \in \mathcal{Q}_N} \phi_{\min} \left( \boldsymbol{\Lambda}_{\mathcal{Q}\cdot} \boldsymbol{\Lambda}'_{\mathcal{Q}\cdot} \right) \right]^{-1} \|\boldsymbol{\Lambda}\|_{\infty} \|\boldsymbol{\lambda}_0\|_2$$

$$= \xi_q^{-1} \mathrm{O}_{\mathrm{p}}(N^{-1} \xi_N^{-1}),$$

where the second inequality holds by the Gershgorin circle theorem. The result is then simply implied by $\xi_q^{-1} = \mathrm{O}(1)$ under Assumption 1 (b).

### B.4.2 Discussion about Lemma 3

In our proof for Lemma 3, we do not use the full implication of Assumption 1 (b); instead, we only need $\xi_N > 0$. Hence, our theory is more general in that we even allow the worst scenario such that $\phi_{\min}\left(\boldsymbol{\Lambda}_{\mathcal{Q}\cdot}\boldsymbol{\Lambda}'_{\mathcal{Q}\cdot}\right)$ shrinks to zero for all $\mathcal{Q} \subset \mathcal{N}$ at a proper speed. Since Assumption 1 (b) is a moderate one, we keep it to simplify the expression. If we allow $\xi_q$ to slowly go to zero, the general result $\|\hat{\boldsymbol{\gamma}}_0^*\|_1 = \mathrm{O}_{\mathrm{p}}(N^{-1}\xi_N^{-1}\xi_q^{-1})$ is desirable among all $q$-sparse solutions to the constraint condition of problem (18).

As mentioned after Lemma 3, a sufficient and necessary condition for exactly recovering every sparse solution from a basis pursuit problem is the NSP. If $\mathrm{rank}(\hat{\boldsymbol{\Sigma}}_f) = q$, then from the constraint condition (B.4) in the proof of Lemma 3 (a), the measurement matrix of problem (18) reduces to simply $\boldsymbol{\Lambda}'$.

We say $\boldsymbol{\Lambda}'$ satisfies *weak NSP* if there exists a $\mathcal{Q} \in \mathcal{Q}_N$ such that for all $\boldsymbol{v} \in \mathrm{Null}(\boldsymbol{\Lambda}') \setminus \{\boldsymbol{0}_N\}$, we have $\|\boldsymbol{v}_{\mathcal{Q}}\|_1 \leq \|\boldsymbol{v}_{(\mathcal{Q})^c}\|_1$. Here we add the qualifier "weak" as the solution does not have to be unique, in contrast to the standard NSP. If this weak NSP holds for some $\mathcal{Q}^*$, then we immediately have a $q$-sparse solution to problem (18) supported on $\mathcal{Q}^*$. However, the weak NSP may not hold only with Assumption 1. See a toy counterexample as following.

**Example 1.** Let $q = 2$ and $N = 4$, and we verify for all $\mathcal{Q} \in \mathcal{Q}_4$: there exists a $\boldsymbol{v} \in \mathrm{Null}(\boldsymbol{\Lambda}') \setminus \{\boldsymbol{0}_N\}$, such that $\|\boldsymbol{v}_{\mathcal{Q}}\|_1 > \|\boldsymbol{v}_{\mathcal{Q}^c}\|_1$. For

$$\boldsymbol{\Lambda}' = \begin{pmatrix} 1 & \sqrt{2}/2 & 0 & -\sqrt{2}/2 \\ 0 & \sqrt{2}/2 & 1 & \sqrt{2}/2 \end{pmatrix},$$

it can be easily checked: (i) $\boldsymbol{v} = (\sqrt{2}, -1, 0, 1)'$ for $\mathcal{Q} = \{1, 2\}, \{1, 4\}, \{2, 4\}$; (ii) $\boldsymbol{v} = (\sqrt{2}/2, -1, \sqrt{2}/2, 0)'$ for $\mathcal{Q} = \{1, 2\}, \{1, 3\}, \{2, 3\}$; and (iii) $\boldsymbol{v} = (0, 1, -\sqrt{2}, 1)'$ for $\mathcal{Q} = \{2, 3\}, \{2, 4\}, \{3, 4\}$.

## B.5 Noise Level

Note that Lemma 1 (a) and 3 (a) require the condition rank$(\hat{\boldsymbol{\Sigma}}_f) = q$, and Lemma 3 (b) further assume that the smallest eigenvalue of $\hat{\boldsymbol{\Sigma}}_f$ is bounded from zero. To satisfy these conditions, together with Assumption 3 (a), we show that the smallest eigenvalue of $\hat{\boldsymbol{\Sigma}}_f$ is indeed bounded away from zero with probability approaching 1 (w.p.a.1) as $T_1 \to \infty$ in the following lemma.

**Lemma B.1.** *Under Assumption 1 and 3,*

*(a) there exists an absolute constant $c_f$, such that as $T_1 \to \infty$, $\mathbb{P}\left\{\phi_{\min}(\hat{\boldsymbol{\Sigma}}_f) \geq c_f\right\} \to 1$;*

*(b) $\|\hat{\boldsymbol{\Sigma}}^*\|_{c2} = O_p(\sqrt{N})$, $\|\hat{\boldsymbol{\eta}}^*\|_\infty = O_p(1)$.*

*Proof.* **Part (a).** Under Assumption 3 (a), the Gershgorin circle theorem yields

$$\|\hat{\boldsymbol{\Sigma}}_f - \boldsymbol{I}_q\|_2 \leq q\|\hat{\boldsymbol{\Sigma}}_f - \boldsymbol{I}_q\|_\infty = O_p(T_1^{-\frac{1}{2}}). \tag{B.8}$$

Then by the fact that $\phi_{\min}(\hat{\boldsymbol{\Sigma}}_f) = 1 + \phi_{\min}\left(\hat{\boldsymbol{\Sigma}}_f - \boldsymbol{I}_q\right) \geq 1 - \|\hat{\boldsymbol{\Sigma}}_f - \boldsymbol{I}_q\|_2$, the conclusion follows for any constant $c_f \in (0, 1)$ as $T_1 \to \infty$.

**Part (b).** With the intermediate result (B.8) in the proof of part (a), by the triangle inequality, we know that as $T_1 \to \infty$,

$$\|\hat{\boldsymbol{\Sigma}}_f\|_2 \leq 1 + \|\hat{\boldsymbol{\Sigma}}_f - \boldsymbol{I}_q\|_2 = O_p(1). \tag{B.9}$$

The Cauchy-Schwarz inequality gives

$$\|\hat{\boldsymbol{\Sigma}}^*\|_{c2} = \max_{j \in \mathcal{N}} \sqrt{\sum_{i=1}^N \left(\boldsymbol{\lambda}_i'\hat{\boldsymbol{\Sigma}}_f\boldsymbol{\lambda}_j\right)^2} \leq q\sqrt{N}\|\hat{\boldsymbol{\Sigma}}_f\|_2\|\boldsymbol{\Lambda}\|_\infty^2 = O_p(\sqrt{N}),$$

$$\|\hat{\boldsymbol{\eta}}^*\|_\infty = \max_{i \in \mathcal{N}} \left|\boldsymbol{\lambda}_i'\hat{\boldsymbol{\Sigma}}_f\boldsymbol{\lambda}_0\right| \leq q\|\hat{\boldsymbol{\Sigma}}_f\|_2\|\boldsymbol{\Lambda}\|_\infty\|\boldsymbol{\lambda}_0\|_\infty = O_p(1). \qquad \square$$

Besides eliminating sampling errors' effects in the oracle components, we further control the noise level of $\hat{\boldsymbol{\Sigma}}^e$ and $\hat{\boldsymbol{\eta}}^e$ in the following lemma.

**Lemma B.2.** *If Assumptions 1–3 hold, then*

*(a) $\|\boldsymbol{\Omega} + \hat{\boldsymbol{\Sigma}}^e\|_{c2} = O_p(\sqrt{N}\varphi_1)$;*

*(b) $\|\hat{\boldsymbol{\eta}}^e\|_\infty = O_p(\varphi_1)$.*

*Proof.* **Part (a).** By the triangular inequality, we have

$$\|\hat{\boldsymbol{\Sigma}}^e\|_\infty \leq \|\hat{\boldsymbol{\Omega}} - \boldsymbol{\Omega}\|_\infty + 2\|\boldsymbol{\Lambda}\Gamma_{\mathcal{T}_1}(\boldsymbol{f}_t, \boldsymbol{u}'_{\mathcal{N}t})\|_\infty.$$

By Assumption 3 (b), the first term on the right-hand-side of the above inequality has the order $O_p(\varphi_1)$. And the second term is bounded by

$$\|\boldsymbol{\Lambda}\Gamma_{\mathcal{T}_1}(\boldsymbol{f}_t, \boldsymbol{u}'_{\mathcal{N}t})\|_\infty = \max_{i,j\in\mathcal{N}} |\boldsymbol{\lambda}'_i \Gamma_{\mathcal{T}_1}(\boldsymbol{f}_t, u_{jt})| \leq q\|\boldsymbol{\Lambda}\|_\infty \|\Gamma_{\mathcal{T}_1}(\boldsymbol{f}_t, \boldsymbol{u}'_{\mathcal{N}t})\|_\infty = O_p(\varphi_1).$$

Therefore, $\|\hat{\boldsymbol{\Sigma}}^e\|_\infty = O_p(\varphi_1)$ implies $\|\hat{\boldsymbol{\Sigma}}^e\|_{c2} = O_p(\sqrt{N}\varphi_1)$. Furthermore, by the fact that $\|\boldsymbol{\Omega}\|_{c2} \leq \sigma^2_{\max}$, we have

$$\|\boldsymbol{\Omega} + \hat{\boldsymbol{\Sigma}}^e\|_{c2} \leq \sigma^2_{\max} + \|\hat{\boldsymbol{\Sigma}}^e\|_{c2},$$

where the second term dominates the first one as $N$ goes large, since $\sqrt{N}\varphi_1 \geq \sqrt{\log N}$. Therefore, $\|\boldsymbol{\Omega} + \hat{\boldsymbol{\Sigma}}^e\|_{c2} = O_p(\sqrt{N}\varphi_1)$.

**Part (b).** The triangular inequality yields

$$\|\hat{\boldsymbol{\eta}}^e\|_\infty \leq \|\Gamma_{\mathcal{T}_1}(\boldsymbol{u}_{\mathcal{N}t}, u_{0t})\|_\infty + \|\boldsymbol{\Lambda}\Gamma_{\mathcal{T}_1}(\boldsymbol{f}_t, u_{0t})\|_\infty + \|\Gamma_{\mathcal{T}_1}(\boldsymbol{u}_{\mathcal{N}t}, \boldsymbol{f}'_t)\boldsymbol{\lambda}_0\|_\infty.$$

By Assumption 3 (c), the first term on the right-hand-side of the above inequality has the order $O_p(\varphi_1)$. The other two terms are bounded by

$$\|\boldsymbol{\Lambda}\Gamma_{\mathcal{T}_1}(\boldsymbol{f}_t, u_{0t})\|_\infty = \max_{i\in\mathcal{N}} |\boldsymbol{\lambda}'_i \Gamma_{\mathcal{T}_1}(\boldsymbol{f}_t, u_{0t})| \leq q\|\boldsymbol{\Lambda}\|_\infty \|\Gamma_{\mathcal{T}_1}(\boldsymbol{f}_t, u_{0t})\|_\infty = O_p(T_1^{-\frac{1}{2}}),$$

$$\|\Gamma_{\mathcal{T}_1}(\boldsymbol{u}_{\mathcal{N}t}, \boldsymbol{f}'_t)\boldsymbol{\lambda}_0\|_\infty = \max_{i\in\mathcal{N}} \left|[\Gamma_{\mathcal{T}_1}(\boldsymbol{f}_t, u_{it})]' \boldsymbol{\lambda}_0\right| \leq q\|\Gamma_{\mathcal{T}_1}(\boldsymbol{f}_t, \boldsymbol{u}'_{\mathcal{N}t})\|_\infty \|\boldsymbol{\lambda}_0\|_\infty = O_p(\varphi_1).$$

Therefore, $\|\hat{\boldsymbol{\eta}}^e\|_\infty = O_p(\varphi_1)$ given $\varphi_1 \gg 1/\sqrt{T_1}$. □

The basic inequality and the empirical process of the L2-relaxation dual problem (16) are shown in the following lemma. Let $\hat{\boldsymbol{\delta}}_\tau := \hat{\boldsymbol{\gamma}}_\tau - \hat{\boldsymbol{\gamma}}^*_0$ for any $\hat{\boldsymbol{\gamma}}^*_0$ defined in Lemma 3 (a), where $\hat{\boldsymbol{\gamma}}_\tau$ solves the L2-relaxation dual problem (16) for tuning parameter $\tau$.

**Lemma B.3.** *For any $\hat{\boldsymbol{\gamma}}^*_0$ in Lemma 3 (a),*

*(a) the basic inequality holds for any $\tau \geq 0$ as*

$$\frac{1}{2}\|\hat{\boldsymbol{\Sigma}}\hat{\boldsymbol{\delta}}_\tau\|^2_2 + \tau\|\hat{\boldsymbol{\gamma}}_\tau\|_1 \leq (\hat{\boldsymbol{\eta}} - \hat{\boldsymbol{\Sigma}}'\hat{\boldsymbol{\Sigma}}\hat{\boldsymbol{\gamma}}^*_0)'\hat{\boldsymbol{\delta}}_\tau + \tau\|\hat{\boldsymbol{\gamma}}^*_0\|_1; \quad (B.10)$$

*(b) under Assumption 1–3, we have $\|\hat{\boldsymbol{\eta}} - \hat{\boldsymbol{\Sigma}}'\hat{\boldsymbol{\Sigma}}\hat{\boldsymbol{\gamma}}^*_0\|_\infty = O_p(\varphi_1/\xi_N)$ as $N \wedge T_1 \to \infty$.*

*Proof.* **Part (a).** The following inequality holds by the definition of $\hat{\boldsymbol{\gamma}}_\tau$:

$$\frac{1}{2}\hat{\boldsymbol{\gamma}}'_\tau \hat{\boldsymbol{\Sigma}}'\hat{\boldsymbol{\Sigma}}\hat{\boldsymbol{\gamma}}_\tau - \hat{\boldsymbol{\eta}}'\hat{\boldsymbol{\gamma}}_\tau + \tau\|\hat{\boldsymbol{\gamma}}_\tau\|_1 \leq \frac{1}{2}(\hat{\boldsymbol{\gamma}}^*_0)'\hat{\boldsymbol{\Sigma}}'\hat{\boldsymbol{\Sigma}}\hat{\boldsymbol{\gamma}}^*_0 - \hat{\boldsymbol{\eta}}'\hat{\boldsymbol{\gamma}}^*_0 + \tau\|\hat{\boldsymbol{\gamma}}^*_0\|_1.$$

Plug in $\hat{\boldsymbol{\gamma}}_\tau = \hat{\boldsymbol{\gamma}}_0^* + \hat{\boldsymbol{\delta}}_\tau$ and the left-hand-side of the above inequality becomes

$$\frac{1}{2}(\hat{\boldsymbol{\gamma}}_0^* + \hat{\boldsymbol{\delta}}_\tau)'\hat{\boldsymbol{\Sigma}}'\hat{\boldsymbol{\Sigma}}(\hat{\boldsymbol{\gamma}}_0^* + \hat{\boldsymbol{\delta}}_\tau) - \hat{\boldsymbol{\eta}}'(\hat{\boldsymbol{\gamma}}_0^* + \hat{\boldsymbol{\delta}}_\tau) + \tau\|\hat{\boldsymbol{\gamma}}_\tau\|_1$$
$$=\frac{1}{2}(\hat{\boldsymbol{\gamma}}_0^*)'\hat{\boldsymbol{\Sigma}}'\hat{\boldsymbol{\Sigma}}\hat{\boldsymbol{\gamma}}_0^* - \hat{\boldsymbol{\eta}}'\hat{\boldsymbol{\gamma}}_0^* + \frac{1}{2}\hat{\boldsymbol{\delta}}_\tau'\hat{\boldsymbol{\Sigma}}'\hat{\boldsymbol{\Sigma}}\hat{\boldsymbol{\delta}}_\tau - (\hat{\boldsymbol{\eta}} - \hat{\boldsymbol{\Sigma}}'\hat{\boldsymbol{\Sigma}}\hat{\boldsymbol{\gamma}}_0^*)'\hat{\boldsymbol{\delta}}_\tau + \tau\|\hat{\boldsymbol{\gamma}}_\tau\|_1.$$

The basic inequality (B.10) follows after rearrangement.

**Part (b).** The gram decomposition (13) implies

$$\hat{\boldsymbol{\eta}} - \hat{\boldsymbol{\Sigma}}'\hat{\boldsymbol{\Sigma}}\hat{\boldsymbol{\gamma}}_0^* = \hat{\boldsymbol{\eta}}^* + \hat{\boldsymbol{\eta}}^e - (\hat{\boldsymbol{\Sigma}}^* + \boldsymbol{\Omega} + \hat{\boldsymbol{\Sigma}}^e)'(\hat{\boldsymbol{\Sigma}}^* + \boldsymbol{\Omega} + \hat{\boldsymbol{\Sigma}}^e)\hat{\boldsymbol{\gamma}}_0^*$$
$$=\hat{\boldsymbol{\eta}}^e - \left[(\boldsymbol{\Omega} + \hat{\boldsymbol{\Sigma}}^e)'(\boldsymbol{\Omega} + \hat{\boldsymbol{\Sigma}}^e) + \hat{\boldsymbol{\Sigma}}^{*'}(\boldsymbol{\Omega} + \hat{\boldsymbol{\Sigma}}^e) + (\boldsymbol{\Omega} + \hat{\boldsymbol{\Sigma}}^e)'\hat{\boldsymbol{\Sigma}}^*\right]\hat{\boldsymbol{\gamma}}_0^*,$$

where $\hat{\boldsymbol{\Sigma}}^{*'}\hat{\boldsymbol{\Sigma}}^*\hat{\boldsymbol{\gamma}}_0^* = \hat{\boldsymbol{\eta}}^*$ holds by the fact that $\hat{\boldsymbol{\gamma}}_0^*$ satisfies the constraint condition of problem (18). The triangle inequality continues with

$$\|\hat{\boldsymbol{\eta}} - \hat{\boldsymbol{\Sigma}}'\hat{\boldsymbol{\Sigma}}\hat{\boldsymbol{\gamma}}_0^*\|_\infty \leq \|\hat{\boldsymbol{\eta}}^e\|_\infty + \left\|(\boldsymbol{\Omega} + \hat{\boldsymbol{\Sigma}}^e)'(\boldsymbol{\Omega}_{.\mathcal{Q}^*} + \hat{\boldsymbol{\Sigma}}^e_{.\mathcal{Q}^*})\right\|_\infty \left\|\hat{\boldsymbol{\gamma}}^*_{0,\mathcal{Q}^*}\right\|_1$$
$$+ \left\|\hat{\boldsymbol{\Sigma}}^{*'}(\boldsymbol{\Omega}_{.\mathcal{Q}^*} + \hat{\boldsymbol{\Sigma}}^e_{.\mathcal{Q}^*})\right\|_\infty \left\|\hat{\boldsymbol{\gamma}}^*_{0,\mathcal{Q}^*}\right\|_1 + \left\|(\boldsymbol{\Omega} + \hat{\boldsymbol{\Sigma}}^e)'\hat{\boldsymbol{\Sigma}}^*_{.\mathcal{Q}^*}\right\|_\infty \left\|\hat{\boldsymbol{\gamma}}^*_{0,\mathcal{Q}^*}\right\|_1$$
$$\leq \|\hat{\boldsymbol{\eta}}^e\|_\infty + \|\boldsymbol{\Omega} + \hat{\boldsymbol{\Sigma}}^e\|_{c2}\left(\|\boldsymbol{\Omega} + \hat{\boldsymbol{\Sigma}}^e\|_{c2} + 2\|\hat{\boldsymbol{\Sigma}}^*\|_{c2}\right)\left\|\hat{\boldsymbol{\gamma}}^*_{0,\mathcal{Q}^*}\right\|_1,$$

where the second inequality holds by the Cauchy-Schwarz inequality. As $N, T_1 \to \infty$, by Lemma B.1 (a), the condition of Lemma 3 (b) holds w.p.a.1; therefore, by Lemma 3 (b), B.1 (b) and B.2, we have the result. $\square$

## B.6 Proofs and Discussions of Theorem 1 and Corollary 1

### B.6.1 Proof of Theorem 1

**Part (a).** Given Lemma 1 (a), we obtain $\boldsymbol{\beta}^* = \hat{\boldsymbol{\Sigma}}^*\hat{\boldsymbol{\gamma}}_0^*$ by applying Lemma 2 to the oracle problems when $\tau \to 0$. We decompose the estimation error by

$$\hat{\boldsymbol{\beta}}_\tau - \boldsymbol{\beta}^* = \hat{\boldsymbol{\Sigma}}\hat{\boldsymbol{\gamma}}_\tau - \left[\hat{\boldsymbol{\Sigma}} - (\boldsymbol{\Omega} + \hat{\boldsymbol{\Sigma}}^e)\right]\hat{\boldsymbol{\gamma}}_0^* = \hat{\boldsymbol{\Sigma}}\hat{\boldsymbol{\delta}}_\tau + \left(\boldsymbol{\Omega}_{.\mathcal{Q}^*} + \hat{\boldsymbol{\Sigma}}^e_{.\mathcal{Q}^*}\right)\hat{\boldsymbol{\gamma}}^*_{0,\mathcal{Q}^*}.$$

For the first term, given the admissible range of the tuning parameter $\tau$, we have $\varphi_1/(\xi_N\tau) \to 0$ and thus from Lemma B.3 (b),

$$\mathbb{P}\left\{\|\hat{\boldsymbol{\eta}} - \hat{\boldsymbol{\Sigma}}'\hat{\boldsymbol{\Sigma}}\hat{\boldsymbol{\gamma}}_0^*\|_\infty \leq \tau\right\} \to 1.$$

Under the event $\|\hat{\boldsymbol{\eta}} - \hat{\boldsymbol{\Sigma}}'\hat{\boldsymbol{\Sigma}}\hat{\boldsymbol{\gamma}}_0^*\|_\infty \leq \tau$, the basic inequality (B.10) gives

$$\frac{1}{2}\|\hat{\boldsymbol{\Sigma}}\hat{\boldsymbol{\delta}}_\tau\|_2^2 \leq \tau\|\hat{\boldsymbol{\delta}}_\tau\|_1 - \tau\|\hat{\boldsymbol{\gamma}}_\tau\|_1 + \tau\|\hat{\boldsymbol{\gamma}}_0^*\|_1 \leq \tau\|\hat{\boldsymbol{\gamma}}_\tau - \hat{\boldsymbol{\delta}}_\tau\|_1 + \tau\|\hat{\boldsymbol{\gamma}}_0^*\|_1 = 2\tau\|\hat{\boldsymbol{\gamma}}_0^*\|_1,$$

where the second inequality holds by the triangle inequality. Hence, Lemma 3 (b) yields

$$\|\hat{\boldsymbol{\Sigma}}\hat{\boldsymbol{\delta}}_\tau\|_2^2 \leq 4\tau\|\hat{\boldsymbol{\gamma}}_0^*\|_1 = \mathrm{O}_\mathrm{p}\left(\tau/(N\xi_N)\right).$$

The second term, in view of Lemma 3 (b) and Lemma B.2 (a), can be bounded by the Cauchy-Schwarz inequality as

$$\left\|\left(\boldsymbol{\Omega}_{\cdot\mathcal{Q}^*} + \hat{\boldsymbol{\Sigma}}_{\cdot\mathcal{Q}^*}^e\right)\hat{\boldsymbol{\gamma}}_{0,\mathcal{Q}^*}^*\right\|_2 \leq \sqrt{\sum_{i=1}^N \left\|\boldsymbol{\Omega}_{i,\mathcal{Q}^*} + \hat{\boldsymbol{\Sigma}}_{i,\mathcal{Q}^*}^e\right\|_2^2}\left\|\hat{\boldsymbol{\gamma}}_{0,\mathcal{Q}^*}^*\right\|_2 = \sqrt{\sum_{j\in\mathcal{Q}^*} \left\|\boldsymbol{\Omega}_{\cdot j} + \hat{\boldsymbol{\Sigma}}_{\cdot j}^e\right\|_2^2}\left\|\hat{\boldsymbol{\gamma}}_{0,\mathcal{Q}^*}^*\right\|_2$$

$$\leq \sqrt{q}\|\boldsymbol{\Omega} + \hat{\boldsymbol{\Sigma}}^e\|_{c2}\left\|\hat{\boldsymbol{\gamma}}_{0,\mathcal{Q}^*}^*\right\|_1 = \mathrm{O}_\mathrm{p}\left(\frac{\varphi_1}{\sqrt{N}\xi_N}\right).$$

The estimation error is therefore bounded by

$$\|\hat{\boldsymbol{\beta}}_\tau - \boldsymbol{\beta}^*\|_2 \leq \|\hat{\boldsymbol{\Sigma}}\hat{\boldsymbol{\delta}}_\tau\|_2 + \left\|\left(\boldsymbol{\Omega}_{\cdot\mathcal{Q}^*} + \hat{\boldsymbol{\Sigma}}_{\cdot\mathcal{Q}^*}^e\right)\hat{\boldsymbol{\gamma}}_{0,\mathcal{Q}^*}^*\right\|_2 = \mathrm{O}_\mathrm{p}\left(\sqrt{\frac{\tau}{N\xi_N}}\right) + \mathrm{O}_\mathrm{p}\left(\frac{\varphi_1}{\sqrt{N}\xi_N}\right),$$

where the first order asymptotically dominates the second one.

**Part (b).** The result immediately follows from the Cauchy-Schwarz inequality.

### B.6.2 Discussion about Theorem 1

We say the L2-relaxation dual problem (16) satisfies the *compatibility condition* with respect to some $\mathcal{Q}^*$ defined in Lemma 3 (a), if there exists a constant $\phi_{\mathcal{Q}^*}(L) > 0$ with respect to some arbitrary constant $L > 1$, such that for all $\boldsymbol{\delta} \in \mathbb{R}^N$ with $\left\|\boldsymbol{\delta}_{(\mathcal{Q}^*)^c}\right\|_1 \leq L\left\|\boldsymbol{\delta}_{\mathcal{Q}^*}\right\|_1$, we have

$$\|\boldsymbol{\delta}_{\mathcal{Q}^*}\|_1^2 \leq q\boldsymbol{\delta}'\frac{\hat{\boldsymbol{\Sigma}}'\hat{\boldsymbol{\Sigma}}}{N}\boldsymbol{\delta}\bigg/\phi_{\mathcal{Q}^*}(L).$$

We call $\phi_{\mathcal{Q}^*}(L)$ the compatibility constant (or restricted eigenvalue).

The gram decomposition (13) implies

$$\boldsymbol{\delta}'\frac{\hat{\boldsymbol{\Sigma}}'\hat{\boldsymbol{\Sigma}}}{N}\boldsymbol{\delta} \geq \frac{\|\hat{\boldsymbol{\Sigma}}^*\boldsymbol{\delta}\|_2^2}{N} + 2\frac{\boldsymbol{\delta}'(\boldsymbol{\Omega} + \hat{\boldsymbol{\Sigma}}^e)\hat{\boldsymbol{\Sigma}}^*\boldsymbol{\delta}}{N} - \left\|\frac{\boldsymbol{\Omega} + \hat{\boldsymbol{\Sigma}}^e}{\sqrt{N}}\right\|_2^2\|\boldsymbol{\delta}\|_2^2. \tag{B.11}$$

To bound the compatibility constant $\phi_{\mathcal{Q}^*}(L)$ from zero, a feasible and sufficient condition is to find a positive lower bound for the right-hand side of the above inequality for all $\boldsymbol{\delta} \in \mathbb{R}^N$ satisfying $\left\|\boldsymbol{\delta}_{(\mathcal{Q}^*)^c}\right\|_1 \leq L\left\|\boldsymbol{\delta}_{\mathcal{Q}^*}\right\|_1$. Unfortunately, as the positive lower bound cannot be found, we cannot guarantee that *compatibility condition* holds. See the following example.

**Example 2.** Suppose for any $\mathcal{Q}^*$ in Lemma 3 (a), there exist a $i \in (\mathcal{Q}^*)^c$ such that

$$\boldsymbol{\Lambda}'_{\mathcal{Q}^*\cdot}\boldsymbol{\delta}_{\mathcal{Q}^*} + \boldsymbol{\lambda}_i\delta_i = \mathbf{0}_q \tag{B.12}$$

has a non-zero solution satisfying $|\delta_i| \leq L \|\boldsymbol{\delta}_{\mathcal{Q}^*}\|_1$ for all $L > 1$. This equation must have non-zero solutions as $\boldsymbol{\Lambda}'_{\mathcal{Q}^* \cup \{i\},\cdot}$ is column-rank deficient. Embedded with such $\boldsymbol{\delta}_{\mathcal{Q}^*}$ and $\delta_i$, note that $\boldsymbol{\delta} = (\delta_1, \cdots, \delta_N)'$ with $\boldsymbol{\delta}_{(\mathcal{Q}^*)^c \setminus \{i\}} = \mathbf{0}_{N-q-1}$ must also be a solution to $\hat{\boldsymbol{\Sigma}}^* \boldsymbol{\delta} = \mathbf{0}_N$ by the definition of $\hat{\boldsymbol{\Sigma}}^*$, which implies that the right-hand-side of inequality (B.11) cannot be bounded away from 0.

Consider a toy example when $q = 2$ and $N = 3$, where $\hat{\boldsymbol{\Sigma}}_f$ is taken as its population $\boldsymbol{I}_q$, with

$$\boldsymbol{\Lambda} = \begin{pmatrix} 1 & 0 \\ 0 & 1 \\ 1 & 1 \end{pmatrix} \text{ and } \boldsymbol{\lambda}_0 = \begin{pmatrix} 4 \\ 5 \end{pmatrix}, \text{ thus } \hat{\boldsymbol{\Sigma}}^* = \begin{pmatrix} 1 & 0 & 1 \\ 0 & 1 & 1 \\ 1 & 1 & 2 \end{pmatrix} \text{ and } \hat{\boldsymbol{\eta}}^* = \begin{pmatrix} 4 \\ 5 \\ 9 \end{pmatrix}.$$

We first find $\mathcal{Q}^* = \{2, 3\}$ by applying Lemma 3 (a), and then solve for equation (B.12). The solutions $\boldsymbol{\delta}_{\mathcal{Q}^*} = (1, -1)'$ and $\delta_1 = 1$ imply that $|\delta_1| \leq L \|\boldsymbol{\delta}_{\mathcal{Q}^*}\|_1$ holds for all $L > 1$.

The lack of the *compatibility condition* implies that we cannot naively follow the proof for LASSO. It also explains why the convergence rate of the L2-relaxation estimator is slower than those of LASSO under the standard assumptions as in the literature.

### B.6.3 Proof of Corollary 1

Combining Lemma 1 (c) and Theorem 1, by the triangle inequality, we have

$$\|\hat{\boldsymbol{\beta}}_\tau - \boldsymbol{\beta}^0\|_2 \leq \|\hat{\boldsymbol{\beta}}_\tau - \boldsymbol{\beta}^*\|_2 + \|\boldsymbol{\beta}^* - \boldsymbol{\beta}^0\|_2 = o_p(N^{-\frac{1}{2}}) + O(N^{-\frac{3}{2}} \xi_N^{-\frac{3}{2}}) + \psi_{\max} O(N^{-\frac{1}{2}} \xi_N^{-1}),$$

as $N, T_1 \to \infty$. The second term is an $o(N^{-\frac{1}{2}})$ when $N\xi^{\frac{3}{2}} \to \infty$.

## B.7 Proof of Theorem 2

**Part (a).** From the decomposition (19), we have the in-sample MPSE:

$$\mathcal{E}_{\mathcal{T}_1}(e_{t,\tau}^2) = \mathcal{E}_{\mathcal{T}_1}\left(\epsilon_t^{*2}\right) - [\mathcal{E}_{\mathcal{T}_1}(\epsilon_t^*)]^2 + (\hat{\boldsymbol{\beta}}_\tau - \boldsymbol{\beta}^*)' \hat{\boldsymbol{\Sigma}} (\hat{\boldsymbol{\beta}}_\tau - \boldsymbol{\beta}^*)$$
$$- 2 \left\{ \mathcal{E}_{\mathcal{T}_1} \left( [\boldsymbol{x}_t - \mathcal{E}_{\mathcal{T}_1}(\boldsymbol{x}_s)] [\epsilon_t^* - \mathcal{E}_{\mathcal{T}_1}(\epsilon_s^*)] \right) \right\}' (\hat{\boldsymbol{\beta}}_\tau - \boldsymbol{\beta}^*).$$

For the third term, which must be non-negative, we have

$$(\hat{\boldsymbol{\beta}}_\tau - \boldsymbol{\beta}^*)' \hat{\boldsymbol{\Sigma}} (\hat{\boldsymbol{\beta}}_\tau - \boldsymbol{\beta}^*) \leq \sqrt{N} \left( \|\hat{\boldsymbol{\Sigma}}^*\|_{c2} + \|\boldsymbol{\Omega} + \hat{\boldsymbol{\Sigma}}^e\|_{c2} \right) \|\hat{\boldsymbol{\beta}}_\tau - \boldsymbol{\beta}^*\|_2^2 = O_p(\tau/\xi_N)$$

given Lemma B.1 (b), Lemma B.2 (a) and Theorem 1 (a). The second term is bounded as

$$|\mathcal{E}_{\mathcal{T}_1}(\epsilon_t^*)| \leq |\mathcal{E}_{\mathcal{T}_1}(u_{0t})| + \sqrt{N} \|\mathcal{E}_{\mathcal{T}_1}(\boldsymbol{u}_{\mathcal{N}t})\|_\infty \|\boldsymbol{\beta}^*\|_2 = O_p(\varphi_1/\sqrt{\xi_N}). \tag{B.13}$$

From (6) and (20), in the last term,

$$\|\mathcal{E}_{\mathcal{T}_1} \left( [\boldsymbol{x}_t - \mathcal{E}_{\mathcal{T}_1}(\boldsymbol{x}_t)] [\epsilon_t^* - \mathcal{E}_{\mathcal{T}_1}(\epsilon_s^*)] \right)\|_\infty$$

$$\leq \|\boldsymbol{\Lambda}\Gamma_{\mathcal{T}_1}(\boldsymbol{f}_t, u_{0t})\|_\infty + \|\Gamma_{\mathcal{T}_1}(\boldsymbol{u}_{\mathcal{N}t}, u_{0t})\|_\infty$$
$$+ \left\{\sqrt{N}\left[\|\boldsymbol{\Lambda}\Gamma_{\mathcal{T}_1}(\boldsymbol{f}_t, \boldsymbol{u}'_{\mathcal{N}t})\|_\infty + \|\hat{\boldsymbol{\Omega}} - \boldsymbol{\Omega}\|_\infty\right] + \|\boldsymbol{\Omega}\|_{c2}\right\}\|\boldsymbol{\beta}^*\|_2$$
$$= O_p(\varphi_1/\sqrt{\xi_N});$$

hence together with Theorem 1 (b), we have

$$\{\mathcal{E}_{\mathcal{T}_1}\left([\boldsymbol{x}_t - \mathcal{E}_{\mathcal{T}_1}(\boldsymbol{x}_t)]\left[\epsilon_t^* - \mathcal{E}_{\mathcal{T}_1}(\epsilon_s^*)\right]\right)\}'(\hat{\boldsymbol{\beta}}_\tau - \boldsymbol{\beta}^*) = O_p(\varphi_1\sqrt{\tau}/\xi_N).$$

Therefore, the non-negative third term asymptotically dominates the second and last terms as $N, T_1 \to \infty$, since $\varphi_1/\sqrt{\tau} \to 0$. By the triangle inequality, we have the in-sample oracle inequality.

Now consider $\mathcal{E}_{\mathcal{T}_1}(\epsilon_t^{*2})$. By (20), we have

$$\mathcal{E}_{\mathcal{T}_1}\left(\epsilon_t^{*2}\right) - \sigma_0^2 = (\hat{\sigma}_0^2 - \sigma_0^2) + [\mathcal{E}_{\mathcal{T}_1}(u_{0t})]^2 - 2\left[\Gamma_{\mathcal{T}_1}(\boldsymbol{u}_{\mathcal{N}t}, u_{0t}) + \mathcal{E}_{\mathcal{T}_1}(\boldsymbol{u}_{\mathcal{N}t})\mathcal{E}_{\mathcal{T}_1}(u_{0t})\right]'\boldsymbol{\beta}^*$$
$$+ (\boldsymbol{\beta}^*)'\boldsymbol{\Omega}\boldsymbol{\beta}^* + (\boldsymbol{\beta}^*)'\left\{(\hat{\boldsymbol{\Omega}} - \boldsymbol{\Omega}) + \mathcal{E}_{\mathcal{T}_1}(\boldsymbol{u}_{\mathcal{N}t})\left[\mathcal{E}_{\mathcal{T}_1}(\boldsymbol{u}_{\mathcal{N}t})\right]'\right\}\boldsymbol{\beta}^*$$
$$\leq \left|\hat{\sigma}_0^2 - \sigma_0^2\right| + |\mathcal{E}_{\mathcal{T}_1}(u_{0t})|^2$$
$$+ 2\sqrt{N}\left[\|\Gamma_{\mathcal{T}_1}(\boldsymbol{u}_{\mathcal{N}t}, u_{0t})\|_\infty + \|\mathcal{E}_{\mathcal{T}_1}(\boldsymbol{u}_{\mathcal{N}t})\|_\infty |\mathcal{E}_{\mathcal{T}_1}(u_{0t})|\right]\|\boldsymbol{\beta}^*\|_2$$
$$+ \sigma_{\max}^2\|\boldsymbol{\beta}^*\|_2^2 + N\left[\|\hat{\boldsymbol{\Omega}} - \boldsymbol{\Omega}\|_\infty + \|\mathcal{E}_{\mathcal{T}_1}(\boldsymbol{u}_{\mathcal{N}t})\|_\infty^2\right]\|\boldsymbol{\beta}^*\|_2^2$$
$$= O_p(\varphi_1/\xi_N). \tag{B.14}$$

**Part (b).** By rewriting (19) into

$$e_{t,\tau} = \epsilon_t^* - \mathcal{E}_{\mathcal{T}_1}(\epsilon_s^*) - (\boldsymbol{x}_t - \boldsymbol{\mu}_\mathcal{N})'(\hat{\boldsymbol{\beta}}_\tau - \boldsymbol{\beta}^*) + [\mathcal{E}_{\mathcal{T}_1}(\boldsymbol{x}_t) - \boldsymbol{\mu}_\mathcal{N}]'(\hat{\boldsymbol{\beta}}_\tau - \boldsymbol{\beta}^*),$$

we have the OOS MPSE:

$$\mathcal{E}_{\mathcal{T}_2}(e_{t,\tau}^2) = \mathcal{E}_{\mathcal{T}_2}\left(\epsilon_t^{*2}\right) - 2\mathcal{E}_{\mathcal{T}_2}(\epsilon_t^*)\mathcal{E}_{\mathcal{T}_1}(\epsilon_t^*) + [\mathcal{E}_{\mathcal{T}_1}(\epsilon_t^*)]^2$$
$$+ (\hat{\boldsymbol{\beta}}_\tau - \boldsymbol{\beta}^*)'\mathcal{E}_{\mathcal{T}_2}\left((\boldsymbol{x}_t - \boldsymbol{\mu}_\mathcal{N})(\boldsymbol{x}_t - \boldsymbol{\mu}_\mathcal{N})'\right)(\hat{\boldsymbol{\beta}}_\tau - \boldsymbol{\beta}^*)$$
$$+ \left\{[\mathcal{E}_{\mathcal{T}_1}(\boldsymbol{x}_t - \boldsymbol{\mu}_\mathcal{N})]'(\hat{\boldsymbol{\beta}}_\tau - \boldsymbol{\beta}^*)\right\}^2$$
$$- 2\left[\mathcal{E}_{\mathcal{T}_2}\left((\boldsymbol{x}_t - \boldsymbol{\mu}_\mathcal{N})\epsilon_t^*\right)\right]'(\hat{\boldsymbol{\beta}}_\tau - \boldsymbol{\beta}^*) + 2\left[\mathcal{E}_{\mathcal{T}_2}(\boldsymbol{x}_t - \boldsymbol{\mu}_\mathcal{N})\right]'\mathcal{E}_{\mathcal{T}_1}(\epsilon_t^*)(\hat{\boldsymbol{\beta}}_\tau - \boldsymbol{\beta}^*)$$
$$+ 2\left[\mathcal{E}_{\mathcal{T}_1}(\boldsymbol{x}_t - \boldsymbol{\mu}_\mathcal{N})\right]'\left[\mathcal{E}_{\mathcal{T}_2}(\epsilon_t^*) - \mathcal{E}_{\mathcal{T}_1}(\epsilon_t^*)\right](\hat{\boldsymbol{\beta}}_\tau - \boldsymbol{\beta}^*)$$
$$- 2(\hat{\boldsymbol{\beta}}_\tau - \boldsymbol{\beta}^*)'\mathcal{E}_{\mathcal{T}_1}(\boldsymbol{x}_t - \boldsymbol{\mu}_\mathcal{N})\left[\mathcal{E}_{\mathcal{T}_2}(\boldsymbol{x}_t - \boldsymbol{\mu}_\mathcal{N})\right]'(\hat{\boldsymbol{\beta}}_\tau - \boldsymbol{\beta}^*).$$

Consider all terms on the right-hand-side one by one. The summation of the third, fourth and fifth terms must be non-negative with the order of $O_p\left((1+\varphi_2)\tau/\xi_N\right)$, since (B.13),

$$(\hat{\boldsymbol{\beta}}_\tau - \boldsymbol{\beta}^*)'\mathcal{E}_{\mathcal{T}_2}\left((\boldsymbol{x}_t - \boldsymbol{\mu}_\mathcal{N})(\boldsymbol{x}_t - \boldsymbol{\mu}_\mathcal{N})'\right)(\hat{\boldsymbol{\beta}}_\tau - \boldsymbol{\beta}^*)$$
$$\leq \sqrt{N}\left\|\mathcal{E}_{\mathcal{T}_2}\left((\boldsymbol{x}_t - \boldsymbol{\mu}_\mathcal{N})(\boldsymbol{x}_t - \boldsymbol{\mu}_\mathcal{N})'\right)\right\|_{c2}\|\hat{\boldsymbol{\beta}}_\tau - \boldsymbol{\beta}^*\|_2^2$$

$$\leq \sqrt{N}\left[\left\|\mathbf{\Lambda}\mathcal{E}_{\mathcal{T}_2}(\boldsymbol{f}_t\boldsymbol{f}'_t)\mathbf{\Lambda}'\right\|_{c2} + 2\sqrt{N}\left\|\mathbf{\Lambda}\mathcal{E}_{\mathcal{T}_2}(\boldsymbol{f}_t\boldsymbol{u}'_{\mathcal{N}t})\right\|_\infty\right]\|\hat{\boldsymbol{\beta}}_\tau - \boldsymbol{\beta}^*\|_2^2$$
$$+ \sqrt{N}\left[\sqrt{N}\left\|\mathcal{E}_{\mathcal{T}_2}(\boldsymbol{u}_{\mathcal{N}t}\boldsymbol{u}'_{\mathcal{N}t}) - \mathbf{\Omega}\right\|_\infty + \|\mathbf{\Omega}\|_{c2}\right]\|\hat{\boldsymbol{\beta}}_\tau - \boldsymbol{\beta}^*\|_2^2$$
$$= O_p\left((1+\varphi_2)\tau/\xi_N\right),$$

and

$$\left|[\mathcal{E}_{\mathcal{T}_1}(\boldsymbol{x}_t - \boldsymbol{\mu}_{\mathcal{N}})]'(\hat{\boldsymbol{\beta}}_\tau - \boldsymbol{\beta}^*)\right| \leq \|\mathcal{E}_{\mathcal{T}_1}(\boldsymbol{x}_t - \boldsymbol{\mu}_{\mathcal{N}})\|_\infty \|\hat{\boldsymbol{\beta}}_\tau - \boldsymbol{\beta}^*\|_1 = O_p(\varphi_1\sqrt{\tau/\xi_N}),$$

by the fact that $\|\mathcal{E}_{\mathcal{T}_1}(\boldsymbol{x}_t - \boldsymbol{\mu}_{\mathcal{N}})\|_\infty = O_p(\varphi_1)$. While for the other terms, similarly, we know that $|\mathcal{E}_{\mathcal{T}_2}(\epsilon^*_t)| = O_p(\varphi_2/\sqrt{\xi_N})$, and

$$\left|[\mathcal{E}_{\mathcal{T}_2}((\boldsymbol{x}_t - \boldsymbol{\mu}_{\mathcal{N}})\epsilon^*_t)]'(\hat{\boldsymbol{\beta}}_\tau - \boldsymbol{\beta}^*)\right| \leq \|\mathcal{E}_{\mathcal{T}_2}((\boldsymbol{x}_t - \boldsymbol{\mu}_{\mathcal{N}})\epsilon^*_t)\|_\infty \|\hat{\boldsymbol{\beta}}_\tau - \boldsymbol{\beta}^*\|_1$$
$$= O_p\left((N^{-\frac{1}{2}} + \varphi_2)\sqrt{\tau}/\xi_N\right),$$
$$\left|[\mathcal{E}_{\mathcal{T}_2}(\boldsymbol{x}_t - \boldsymbol{\mu}_{\mathcal{N}})]'(\hat{\boldsymbol{\beta}}_\tau - \boldsymbol{\beta}^*)\right| \leq \|\mathcal{E}_{\mathcal{T}_2}(\boldsymbol{x}_t - \boldsymbol{\mu}_{\mathcal{N}})\|_\infty \|\hat{\boldsymbol{\beta}}_\tau - \boldsymbol{\beta}^*\|_1 = O_p(\varphi_2\sqrt{\tau/\xi_N}),$$

since $\|\mathcal{E}_{\mathcal{T}_2}(\boldsymbol{x}_t - \boldsymbol{\mu}_{\mathcal{N}})\|_\infty = O_p(\varphi_2)$, and

$$\|\mathcal{E}_{\mathcal{T}_2}((\boldsymbol{x}_t - \boldsymbol{\mu}_{\mathcal{N}})\epsilon^*_t)\|_\infty = \|\mathbf{\Lambda}\mathcal{E}_{\mathcal{T}_2}(\boldsymbol{f}_t u_{0t})\|_\infty + \|\mathcal{E}_{\mathcal{T}_2}(\boldsymbol{u}_{\mathcal{N}t}u_{0t})\|_\infty + \|\mathbf{\Omega}\|_{c2}\|\boldsymbol{\beta}^*\|_2$$
$$+ \sqrt{N}\left[\|\mathbf{\Lambda}\mathcal{E}_{\mathcal{T}_2}(\boldsymbol{f}_t\boldsymbol{u}'_{\mathcal{N}t})\|_\infty + \|\mathcal{E}_{\mathcal{T}_2}(\boldsymbol{u}_{\mathcal{N}t}\boldsymbol{u}'_{\mathcal{N}t}) - \mathbf{\Omega}\|_\infty\right]\|\boldsymbol{\beta}^*\|_2$$
$$= O_p(\varphi_2/\sqrt{\xi_N}) + O(N^{-\frac{1}{2}}\xi_N^{-\frac{1}{2}}).$$

Therefore, we have

$$\mathcal{E}_{\mathcal{T}_2}(e^2_{t,\tau}) \leq \mathcal{E}_{\mathcal{T}_2}\left((\epsilon^*_t)^2\right) + O_p\left((1+\varphi_2)\tau/\xi_N\right) + O_p\left((N^{-\frac{1}{2}} + \varphi_1^2 + \varphi_2)\sqrt{\tau}/\xi_N\right),$$

where $O_p(\varphi_2\tau/\xi_N)$ is asymptotically dominated by $O_p(\varphi_2\sqrt{\tau}/\xi_N)$ as $\tau \to 0$, and

$$\frac{(N^{-\frac{1}{2}} + \varphi_1^2)\sqrt{\tau}/\xi_N}{\tau/\xi_N} \leq \left[(N\log N)^{-\frac{1}{4}} + \varphi_1^{\frac{3}{2}}\right]\sqrt{\xi_N \cdot \frac{\varphi_1}{\xi_N\tau}} \to 0.$$

Additionally, by (20), we have

$$\mathcal{E}_{\mathcal{T}_2}\left(\epsilon^{*2}_t\right) - \sigma_0^2 = \left[\mathcal{E}_{\mathcal{T}_2}(u^2_{0t}) - \sigma_0^2\right] - 2\left[\mathcal{E}_{\mathcal{T}_2}(\boldsymbol{u}_{\mathcal{N}t}u_{0t})\right]'\boldsymbol{\beta}^*$$
$$+ (\boldsymbol{\beta}^*)'\mathbf{\Omega}\boldsymbol{\beta}^* + (\boldsymbol{\beta}^*)'\left[\mathcal{E}_{\mathcal{T}_2}(\boldsymbol{u}_{\mathcal{N}t}\boldsymbol{u}'_{\mathcal{N}t}) - \mathbf{\Omega}\right]\boldsymbol{\beta}^*$$
$$\leq \left|\mathcal{E}_{\mathcal{T}_2}(u^2_{0t}) - \sigma_0^2\right| + 2\sqrt{N}\|\mathcal{E}_{\mathcal{T}_2}(\boldsymbol{u}_{\mathcal{N}t}u_{0t})\|_\infty\|\boldsymbol{\beta}^*\|_2$$
$$+ \sigma^2_{\max}\|\boldsymbol{\beta}^*\|_2^2 + N\|\mathcal{E}_{\mathcal{T}_2}(\boldsymbol{u}_{\mathcal{N}t}\boldsymbol{u}'_{\mathcal{N}t}) - \mathbf{\Omega}\|_\infty\|\boldsymbol{\beta}^*\|_2^2$$
$$= O(N^{-1}/\xi_N) + O_p(\varphi_2/\xi_N),$$

where the first order is asymptotically dominated by the second one as $N\varphi_2 \to \infty$.

# C Proofs for Section 4

## C.1 Proof of Theorem 3

We first emphasize a fact about $\tau$. Note that as $N, T_1 \to 0$,

$$\frac{\varphi_1/\xi_N}{\xi_N/\log N} + \frac{\varphi_1/\xi_N}{\xi_N/T_1^{1/p}} = (\log N + T_1^{1/p})\varphi_1/\xi_N^2 \to 0;$$

hence we can always find a $\tau$ such that $(\log N + T_1^{1/p})\tau + \varphi_1/\tau = \mathrm{o}(\xi_N)$.

Reorganize test statistic (24) into

$$\hat{Z} = \frac{\sqrt{T_2}(\bar{\Delta} - \Delta_{T_2})}{\sqrt{(T_2/T_1)\hat{\rho}_{(1)}^2 + \hat{\rho}_{(2)}^2}}. \tag{C.1}$$

From (19) and the definition of $d_t^*$, we have the decomposition for the estimated treatment effect as

$$\hat{\Delta}_t = \mathbb{E}(\Delta_t) + d_t^* - \mathcal{E}_{\mathcal{T}_1}(\epsilon_s^*) - [\boldsymbol{x}_t - \mathcal{E}_{\mathcal{T}_1}(\boldsymbol{x}_s)]'(\hat{\boldsymbol{\beta}}_\tau - \boldsymbol{\beta}^*), \quad t \in \mathcal{T}_2; \tag{C.2}$$

and henceforth, the ATE estimator is decomposed by

$$\bar{\Delta} = \Delta_{T_2} + \mathcal{E}_{\mathcal{T}_2}(d_t^*) - \mathcal{E}_{\mathcal{T}_1}(\epsilon_t^*) - [\mathcal{E}_{\mathcal{T}_2}(\boldsymbol{x}_t) - \mathcal{E}_{\mathcal{T}_1}(\boldsymbol{x}_t)]'(\hat{\boldsymbol{\beta}}_\tau - \boldsymbol{\beta}^*). \tag{C.3}$$

By (C.3), the numerator of (C.1) equals

$$\sqrt{T_2}(\bar{\Delta} - \Delta_{T_2}) = \sqrt{T_2}\mathcal{E}_{\mathcal{T}_2}(d_t^*) - \sqrt{T_2/T_1} \cdot \sqrt{T_1}\mathcal{E}_{\mathcal{T}_1}(\epsilon_t^*) - \sqrt{T_2}\left[\mathcal{E}_{\mathcal{T}_2}(\boldsymbol{x}_t) - \mathcal{E}_{\mathcal{T}_1}(\boldsymbol{x}_t)\right]'(\hat{\boldsymbol{\beta}}_\tau - \boldsymbol{\beta}^*).$$

According to Theorem 1, the last term is bounded by

$$\sqrt{T_2}\left|[\mathcal{E}_{\mathcal{T}_2}(\boldsymbol{x}_t) - \mathcal{E}_{\mathcal{T}_1}(\boldsymbol{x}_t)]'(\hat{\boldsymbol{\beta}}_\tau - \boldsymbol{\beta}^*)\right| \leq \sqrt{T_2}\left[\|\mathcal{E}_{\mathcal{T}_2}(\boldsymbol{x}_t)\|_\infty + \|\mathcal{E}_{\mathcal{T}_1}(\boldsymbol{x}_t)\|_\infty\right]\|\hat{\boldsymbol{\beta}}_\tau - \boldsymbol{\beta}^*\|_1$$

$$= \mathrm{O}_\mathrm{p}\left(\left[1 + \sqrt{T_2/(N \wedge T_1)}\right]\sqrt{(\log N)\tau/\xi_N}\right).$$

As $N, T_1, T_2 \to \infty$ with $T_2 = \mathrm{O}(N \wedge T_1)$, for $(\log N)\tau/\xi_N \to 0$, we have

$$\sqrt{T_2}(\bar{\Delta} - \Delta_{T_2}) = \sqrt{T_2}\mathcal{E}_{\mathcal{T}_2}(d_t^*) - \sqrt{T_2/T_1} \cdot \sqrt{T_1}\mathcal{E}_{\mathcal{T}_1}(\epsilon_t^*) + \mathrm{o}_\mathrm{p}(1).$$

In the denominator of (C.1), for $\hat{\rho}_{(1)}^2$, we decompose by the triangle inequality:

$$\left|\hat{\rho}_{(1)}^2 - \rho_{(1)}^2\right| \leq \left|\hat{\rho}_{(1)}^2 - \hat{\rho}_{\epsilon^*}^2\right| + \left|\hat{\rho}_{\epsilon^*}^2 - \rho_{\epsilon^*}^2\right| + \left|\rho_{\epsilon^*}^2 - \rho_{(1)}^2\right|.$$

Observing the decomposition, (i) the first term is the estimation error; (ii) the second term is the sampling error with the order of $\mathrm{o}_\mathrm{p}(1)$, since $\hat{\rho}_{\epsilon^*}^2$ is a consistent estimator of $\rho_{\epsilon^*}^2$ for $h_1 \to \infty$ and $h_1/T_1^{1/4} \to 0$ as $T_1 \to \infty$; and (iii) the third term measures the sequential convergence of the LRV sequence $\{\rho^2_{\epsilon^*} : N \geq N^*\}$ to its limit $\rho^2_{(1)}$ for some sufficiently large $N^*$ as imposed in Assumption 5 (c). Hence we focus on the estimation error in the first term. From (19), we know that

$$\left|\hat{\rho}^2_{(1)} - \hat{\rho}^2_{\epsilon^*}\right| = \left|\sum_{l=-h_1}^{h_1} \mathcal{E}_{\mathcal{T}_1}\left(e_{t,\tau}e_{t+l,\tau} - [\epsilon^*_t - \mathcal{E}_{\mathcal{T}_1}(\epsilon^*_s)]\left[\epsilon^*_{t+l} - \mathcal{E}_{\mathcal{T}_1}(\epsilon^*_s)\right]\right)\right|$$

$$\leq \left|\sum_{l=-h_1}^{h_1} (\hat{\boldsymbol{\beta}}_\tau - \boldsymbol{\beta}^*)'\Gamma_{\mathcal{T}_1}(\boldsymbol{x}_t, \boldsymbol{x}'_{t+l})(\hat{\boldsymbol{\beta}}_\tau - \boldsymbol{\beta}^*)\right| + 2\left|\sum_{l=-h_1}^{h_1} \left[\Gamma_{\mathcal{T}_1}(\boldsymbol{x}_t, \epsilon^*_{t+l})\right]'(\hat{\boldsymbol{\beta}}_\tau - \boldsymbol{\beta}^*)\right|$$

$$\leq N(2h_1 + 1)\left\{\max_{i \in \mathcal{N}} \mathcal{E}_{\mathcal{T}_1}\left([x_{it} - \mathcal{E}_{\mathcal{T}_1}(x_{is})]^2\right)\right\}\|\hat{\boldsymbol{\beta}}_\tau - \boldsymbol{\beta}^*\|^2_2$$

$$+ 2(2h_1 + 1)\sqrt{\left\{\max_{i \in \mathcal{N}} \mathcal{E}_{\mathcal{T}_1}\left([x_{it} - \mathcal{E}_{\mathcal{T}_1}(x_{is})]^2\right)\right\} \mathcal{E}_{\mathcal{T}_1}\left([\epsilon^*_t - \mathcal{E}_{\mathcal{T}_1}(\epsilon^*_s)]^2\right)} \|\hat{\boldsymbol{\beta}}_\tau - \boldsymbol{\beta}^*\|_1$$

$$= O_p(h_1\sqrt{\tau/\xi_N}),$$

where the last inequality holds by the Cauchy-Schwarz inequality and the Gershgorin circle theorem. Together with Theorem 1, we derive the order by the facts that

$$\max_{i \in \mathcal{N}} \mathcal{E}_{\mathcal{T}_1}\left([x_{it} - \mathcal{E}_{\mathcal{T}_1}(x_{is})]^2\right)$$
$$\leq \left\|\boldsymbol{\Lambda}\Gamma_{\mathcal{T}_1}(\boldsymbol{f}_t, \boldsymbol{f}'_t)\boldsymbol{\Lambda}'\right\|_\infty + 2\left\|\boldsymbol{\Lambda}\Gamma_{\mathcal{T}_1}(\boldsymbol{f}_t, \boldsymbol{u}'_{\mathcal{N}t})\right\|_\infty + \|\hat{\boldsymbol{\Omega}} - \boldsymbol{\Omega}\|_\infty + \|\boldsymbol{\Omega}\|_{c2} = O_p(1),$$

and $\mathcal{E}_{\mathcal{T}_1}\left([\epsilon^*_t - \mathcal{E}_{\mathcal{T}_1}(\epsilon^*_t)]^2\right) = O_p(1)$ from (B.13) and (B.14). Finally, for $h_1 = O(T_1^{1/(2p)})$ with $p > 2$ and $T_1^{1/p}\tau/\xi_N \to 0$ as $N, T_1 \to \infty$, we have $\left|\hat{\rho}^2_{(1)} - \hat{\rho}^2_{\epsilon^*}\right| = o_p(1)$.

Similarly, we decompose $\hat{\rho}^2_{(2)}$ into

$$\left|\hat{\rho}^2_{(2)} - \rho^2_{(2)}\right| \leq \left|\hat{\rho}^2_{(2)} - \hat{\rho}^2_{d^*}\right| + \left|\hat{\rho}^2_{d^*} - \rho^2_{d^*}\right| + \left|\rho^2_{d^*} - \rho^2_{(2)}\right|.$$

The second term is of $o_p(1)$ as $\hat{\rho}^2_{d^*}$ is a consistent estimator of $\rho^2_{d^*}$ for $h_2 \to \infty$ and $h_2/T_2^{1/4} \to 0$ as $T_2 \to \infty$. The third term converges to zero as $N \to \infty$ under Assumption 5 (d). For the first term, by taking difference of (C.2) and (C.3), we have

$$\hat{\Delta}_t - \bar{\Delta} = \mathbb{E}(\Delta_t) - \Delta_{T_2} + d^*_t - \mathcal{E}_{\mathcal{T}_2}(d^*_s) - [\boldsymbol{x}_t - \mathcal{E}_{\mathcal{T}_2}(\boldsymbol{x}_s)]'(\hat{\boldsymbol{\beta}}_\tau - \boldsymbol{\beta}^*),$$

which implies that $\left|\hat{\rho}^2_{(2)} - \hat{\rho}^2_{d^*}\right|$ equals to

$$\left|\sum_{l=-h_2}^{h_2} \mathcal{E}_{\mathcal{T}_2}\left((\hat{\Delta}_t - \bar{\Delta})(\hat{\Delta}_{t+l} - \bar{\Delta}) - [d^*_t - \mathcal{E}_{\mathcal{T}_2}(d^*_s)]\left[d^*_{t+l} - \mathcal{E}_{\mathcal{T}_2}(d^*_s)\right]\right)\right|$$

$$\leq \left|\frac{1}{T_2}\sum_{t \in \mathcal{T}_2}\sum_{l=-h_2}^{h_2} \left([\mathbb{E}(\Delta_t) - \Delta_{T_2}]\left[\mathbb{E}(\Delta_{t+l}) - \Delta_{T_2}\right]\right)\right|$$

$$+ \left|\sum_{l=-h_2}^{h_2} (\hat{\boldsymbol{\beta}}_\tau - \boldsymbol{\beta}^*)'\Gamma_{\mathcal{T}_2}(\boldsymbol{x}_t, \boldsymbol{x}'_{t+l})(\hat{\boldsymbol{\beta}}_\tau - \boldsymbol{\beta}^*)\right|$$

$$+ 2 \left| \sum_{l=-h_2}^{h_2} \mathcal{E}_{\mathcal{T}_2} \left( [d_t^* - \mathcal{E}_{\mathcal{T}_2}(d_s^*)] [\mathbb{E}(\Delta_{t+l}) - \Delta_{T_2}] \right) \right|$$

$$+ 2 \left| \sum_{l=-h_2}^{h_2} \{ \mathcal{E}_{\mathcal{T}_2} ([\boldsymbol{x}_t - \mathcal{E}_{\mathcal{T}_2}(\boldsymbol{x}_s)] [\mathbb{E}(\Delta_{t+l}) - \Delta_{T_2}]) \}' (\hat{\boldsymbol{\beta}}_\tau - \boldsymbol{\beta}^*) \right|$$

$$+ 2 \left| \sum_{l=-h_2}^{h_2} \left[ \Gamma_{\mathcal{T}_2}(\boldsymbol{x}_t, d_{t+l}^*) \right]' (\hat{\boldsymbol{\beta}}_\tau - \boldsymbol{\beta}^*) \right|$$

$$\leq \frac{2h_2 + 1}{T_2} \sum_{t \in \mathcal{T}_2} [\mathbb{E}(\Delta_t) - \Delta_{T_2}]^2$$

$$+ N(2h_2 + 1) \left\{ \max_{i \in \mathcal{N}} \mathcal{E}_{\mathcal{T}_2} \left( [x_{it} - \mathcal{E}_{\mathcal{T}_2}(x_{is})]^2 \right) \right\} \|\hat{\boldsymbol{\beta}}_\tau - \boldsymbol{\beta}^*\|_2^2$$

$$+ 2(2h_2 + 1) \sqrt{ \mathcal{E}_{\mathcal{T}_2} \left( [d_t^* - \mathcal{E}_{\mathcal{T}_2}(d_s^*)]^2 \right) \frac{1}{T_2} \sum_{t \in \mathcal{T}_2} [\mathbb{E}(\Delta_t) - \Delta_{T_2}]^2 }$$

$$+ 2(2h_2 + 1) \sqrt{ \left\{ \max_{i \in \mathcal{N}} \mathcal{E}_{\mathcal{T}_2} \left( [x_{it} - \mathcal{E}_{\mathcal{T}_2}(x_{is})]^2 \right) \right\} \frac{1}{T_2} \sum_{t \in \mathcal{T}_2} [\mathbb{E}(\Delta_t) - \Delta_{T_2}]^2 } \|\hat{\boldsymbol{\beta}}_\tau - \boldsymbol{\beta}^*\|_1$$

$$+ 2(2h_2 + 1) \sqrt{ \left\{ \max_{i \in \mathcal{N}} \mathcal{E}_{\mathcal{T}_2} \left( [x_{it} - \mathcal{E}_{\mathcal{T}_2}(x_{is})]^2 \right) \right\} \mathcal{E}_{\mathcal{T}_2} \left( [d_t^* - \mathcal{E}_{\mathcal{T}_2}(d_s^*)]^2 \right) } \|\hat{\boldsymbol{\beta}}_\tau - \boldsymbol{\beta}^*\|_1$$

$$= O_p \left( h_2 / T_2^{1/4} \right) + O_p \left( h_2 \sqrt{\tau / \xi_N} \right),$$

since similarly $\max_{i \in \mathcal{N}} \mathcal{E}_{\mathcal{T}_2} \left( [x_{it} - \mathcal{E}_{\mathcal{T}_2}(x_{it})]^2 \right) = O_p(1)$, and under Assumption 5 (a) we have $\sum_{t \in \mathcal{T}_2} [\mathbb{E}(\Delta_t) - \Delta_{T_2}]^2 = O(\sqrt{T_2})$ and $\mathcal{E}_{\mathcal{T}_2} \left( [d_t^* - \mathcal{E}_{\mathcal{T}_2}(d_s^*)]^2 \right) = O_p(1)$. Therefore, as $N, T_1, T_2 \to \infty$ with $T_2 = O(T_1)$, for $h_2 = O(T_2^{1/(2p)})$ with $p > 2$ and $T_1^{1/p} \tau / \xi_N \to 0$, we have $\left| \hat{\rho}_{(2)}^2 - \hat{\rho}_{d^*}^2 \right| = o_p(1)$.

Given the above analyses, since $\rho_{(1)}^2$ and $\rho_{(2)}^2$ are positive and finite, we conclude that as $N, T_1, T_2 \to \infty$ and $T_2 = O(T_1)$, for $T_2/T_1 \to r < \infty$,

$$\hat{Z} = \frac{\sqrt{T_2} \mathcal{E}_{\mathcal{T}_2}(d_t^*) - \sqrt{T_2/T_1} \cdot \sqrt{T_1} \mathcal{E}_{\mathcal{T}_1}(\epsilon_t^*) + o_p(1)}{\sqrt{\rho_{(2)}^2 + (T_2/T_1) \rho_{(1)}^2 + o_p(1 + T_2/T_1)}}$$

$$= \frac{\sqrt{T_2} \mathcal{E}_{\mathcal{T}_2}(d_t^*) - \sqrt{T_2/T_1} \cdot \sqrt{T_1} \mathcal{E}_{\mathcal{T}_1}(\epsilon_t^*)}{\sqrt{\rho_{(2)}^2 + r \rho_{(1)}^2}} + o_p(1).$$

Assumption 5 (c)–(d) ensure that $\sqrt{T_2} \mathcal{E}_{\mathcal{T}_2}(d_t^*) - \sqrt{T_2/T_1} \cdot \sqrt{T_1} \mathcal{E}_{\mathcal{T}_1}(\epsilon_t^*)$ asymptotically follows a normal distribution with zero mean and variance being

$$\text{Asym.Var} \left( \sqrt{T_2} \mathcal{E}_{\mathcal{T}_2}(d_t^*) - \sqrt{T_2/T_1} \cdot \sqrt{T_1} \mathcal{E}_{\mathcal{T}_1}(\epsilon_t^*) \right)$$

$$= \rho_{(2)}^2 + r \rho_{(1)}^2 - 2 \text{Asym.Cov} \left( \sqrt{T_2} \mathcal{E}_{\mathcal{T}_2}(d_t^*), \sqrt{T_2/T_1} \cdot \sqrt{T_1} \mathcal{E}_{\mathcal{T}_1}(\epsilon_t^*) \right)$$

$$= \rho_{(2)}^2 + r \rho_{(1)}^2,$$

where the convariance vanishes as under Assumption 5 (e), we have

$$\text{Cov}\left(\sqrt{T_2}\mathcal{E}_{\mathcal{T}_2}(d_t^*), \sqrt{T_2/T_1} \cdot \sqrt{T_1}\mathcal{E}_{\mathcal{T}_1}(\epsilon_t^*)\right) = \frac{1}{T_1}\sum_{s\in\mathcal{T}_1}\sum_{t\in\mathcal{T}_2}\mathbb{E}(d_t^*\epsilon_s^*) = \text{o}(1).$$

Therefore, we confirms $\hat{Z} \xrightarrow{\text{d}} N(0,1)$.

## C.2  Proof of Theorem 4

Similarly to the previous proof, we first verify for the order of $\tau$. When

$$\frac{\varphi_1/\xi_N}{\xi_N/(M\log N)} = (M\log N)\varphi_1/\xi_N^2 \to 0$$

as $M, N, T_1 \to 0$, there exists a $\tau$ such that $(M\log N)\tau + \varphi_1/\tau = \text{o}(\xi_N)$.

For the multiple-treated-unit setting, we must also have some argument. Here we show the preliminaries without repeating the proof; instead, we compare between the assumptions. Similar to Lemma 1 (b), under Assumption 1 and 6 (a), for all $i \in \mathcal{M}$, we have $\|\beta_i^*\|_2 = \text{O}(N^{-\frac{1}{2}}\xi_N^{-\frac{1}{2}})$. Similar to Theorem 1, under Assumption 1–3 and 6 (a)–(c), as $N, T_1 \to \infty$, if $\tau + \varphi_1/\tau = \text{o}(\xi_N)$, then for all $i \in \mathcal{M}$, we have $\|\hat{\beta}_{i,\tau} - \beta_i^*\|_2 = \text{O}_\text{p}\left(\sqrt{\tau/(N\xi_N)}\right)$ and $\|\hat{\beta}_{i,\tau} - \beta_i^*\|_1 = \text{O}_\text{p}\left(\sqrt{\tau/\xi_N}\right)$.

By the decomposition of $e_{it,\tau}$ in (27), the time-varying individual treatment effect is

$$\hat{\Delta}_{it} = \Delta_{it} + \epsilon_{it}^* - \mathcal{E}_{\mathcal{T}_1}(\epsilon_{is}^*) - [\boldsymbol{x}_t - \mathcal{E}_{\mathcal{T}_1}(\boldsymbol{x}_s)]'(\hat{\boldsymbol{\beta}}_{i,\tau} - \boldsymbol{\beta}_i^*), \quad i \in \mathcal{M}, t \in \mathcal{T}_2,$$

and henceforth the ATE at each $t \in \mathcal{T}_2$ is

$$\bar{\Delta}_t = \Delta_{M,t} + \frac{1}{M}\sum_{i\in\mathcal{M}}[\epsilon_{it}^* - \mathcal{E}_{\mathcal{T}_1}(\epsilon_{is}^*)] - [\boldsymbol{x}_t - \mathcal{E}_{\mathcal{T}_1}(\boldsymbol{x}_s)]'\left[\frac{1}{M}\sum_{i\in\mathcal{M}}(\hat{\boldsymbol{\beta}}_{i,\tau} - \boldsymbol{\beta}_i^*)\right]. \quad \text{(C.4)}$$

The numerator of test statistic (29) equals to

$$\sqrt{M}(\bar{\Delta}_t - \Delta_{M,t}) = \frac{1}{\sqrt{M}}\sum_{i\in\mathcal{M}}\epsilon_{it}^* - \frac{1}{\sqrt{M}}\sum_{i\in\mathcal{M}}\mathcal{E}_{\mathcal{T}_1}(\epsilon_{is}^*) - [\boldsymbol{x}_t - \mathcal{E}_{\mathcal{T}_1}(\boldsymbol{x}_s)]'\left[\frac{1}{\sqrt{M}}\sum_{i\in\mathcal{M}}(\hat{\boldsymbol{\beta}}_{i,\tau} - \boldsymbol{\beta}_i^*)\right],$$

where

$$\left|\frac{1}{\sqrt{M}}\sum_{i\in\mathcal{M}}\mathcal{E}_{\mathcal{T}_1}(\epsilon_{it}^*)\right| \leq \left|\frac{1}{\sqrt{M}}\sum_{i\in\mathcal{M}}\mathcal{E}_{\mathcal{T}_1}(u_{it})\right| + \left|[\mathcal{E}_{\mathcal{T}_1}(\boldsymbol{u}_{\mathcal{N}t})]'\left(\frac{1}{\sqrt{M}}\sum_{i\in\mathcal{M}}\boldsymbol{\beta}_i^*\right)\right|$$

$$\leq \sqrt{M}\|\mathcal{E}_{\mathcal{T}_1}(\boldsymbol{u}_{\mathcal{M}t})\|_\infty + \sqrt{MN}\|\mathcal{E}_{\mathcal{T}_1}(\boldsymbol{u}_{\mathcal{N}t})\|_\infty \max_{i\in\mathcal{M}}\|\boldsymbol{\beta}_i^*\|_2$$

$$= \text{O}_\text{p}(\varphi_1\sqrt{M/\xi_N}),$$

and

$$\left|[\boldsymbol{x}_t - \mathcal{E}_{\mathcal{T}_1}(\boldsymbol{x}_s)]'\left[\frac{1}{\sqrt{M}}\sum_{i\in\mathcal{M}}(\hat{\boldsymbol{\beta}}_{i,\tau} - \boldsymbol{\beta}_i^*)\right]\right| \leq \sqrt{M}\|\boldsymbol{x}_t - \mathcal{E}_{\mathcal{T}_1}(\boldsymbol{x}_s)\|_\infty \max_{i\in\mathcal{M}}\|\hat{\boldsymbol{\beta}}_{i,\tau} - \boldsymbol{\beta}_i^*\|_1$$

$$= O_p(\sqrt{(M \log N)\tau/\xi_N}).$$

Hence, by the triangle inequality, we have

$$\sqrt{M}(\bar{\Delta}_t - \Delta_{M,t}) = \frac{1}{\sqrt{M}} \sum_{i \in \mathcal{M}} \epsilon_{it}^* + o_p(1),$$

as $M, N, T_1 \to \infty$, since $(M \log N)\tau/\xi_N \to 0$ and

$$\frac{\varphi_1 \sqrt{M/\xi_N}}{\sqrt{(M \log N)\tau/\xi_N}} = \sqrt{[(N \wedge T_1) \log N]^{-\frac{1}{2}} \xi_N \cdot \frac{\varphi_1}{\xi_N \tau}} \to 0.$$

While for the denominator of the statistic (29), we first define

$$\hat{V}_{\epsilon^*}^2 := \frac{1}{M} \sum_{i \in \mathcal{M}} \sum_{j \in \mathcal{M}} \Gamma_{\mathcal{T}_1}(\epsilon_{it}^*, \epsilon_{jt}^*)$$

as a bridge between $\hat{V}^2$ and $V_{\epsilon^*}^2$. Decomposition (27) gives

$$\hat{V}^2 - \hat{V}_{\epsilon^*}^2 = \frac{1}{M} \sum_{i \in \mathcal{M}} \sum_{j \in \mathcal{M}} \left[ \mathcal{E}_{\mathcal{T}_1}(e_{it,\tau} e_{jt,\tau}) - \Gamma_{\mathcal{T}_1}(\epsilon_{it}^*, \epsilon_{jt}^*) \right]$$

$$= \frac{1}{M} \sum_{i \in \mathcal{M}} \sum_{j \in \mathcal{M}} (\hat{\boldsymbol{\beta}}_{i,\tau} - \boldsymbol{\beta}_i^*)' \hat{\boldsymbol{\Sigma}} (\hat{\boldsymbol{\beta}}_{j,\tau} - \boldsymbol{\beta}_j^*) - \frac{2}{M} \sum_{i \in \mathcal{M}} \sum_{j \in \mathcal{M}} [\Gamma_{\mathcal{T}_1}(\boldsymbol{x}_t, \epsilon_{it}^*)]' (\hat{\boldsymbol{\beta}}_{j,\tau} - \boldsymbol{\beta}_j^*),$$

where

$$\frac{1}{M} \left| \sum_{i \in \mathcal{M}} \sum_{j \in \mathcal{M}} (\hat{\boldsymbol{\beta}}_{i,\tau} - \boldsymbol{\beta}_i^*)' \hat{\boldsymbol{\Sigma}} (\hat{\boldsymbol{\beta}}_{j,\tau} - \boldsymbol{\beta}_j^*) \right| \leq M \sqrt{N} \|\hat{\boldsymbol{\Sigma}}\|_{c2} \max_{i \in \mathcal{M}} \|\hat{\boldsymbol{\beta}}_{i,\tau} - \boldsymbol{\beta}_i^*\|_2^2 = O_p(M\tau/\xi_N),$$

$$\frac{1}{M} \left| \sum_{i \in \mathcal{M}} \sum_{j \in \mathcal{M}} [\Gamma_{\mathcal{T}_1}(\boldsymbol{x}_t, \epsilon_{it}^*)]' (\hat{\boldsymbol{\beta}}_{j,\tau} - \boldsymbol{\beta}_j^*) \right| \leq M \left[ \max_{i \in \mathcal{M}} \|\Gamma_{\mathcal{T}_1}(\boldsymbol{x}_t, \epsilon_{it}^*)\|_\infty \right] \left[ \max_{j \in \mathcal{M}} \|\hat{\boldsymbol{\beta}}_{j,\tau} - \boldsymbol{\beta}_j^*\|_1 \right]$$

$$= O_p(M\varphi_1 \sqrt{\tau}/\xi_N),$$

since $\|\hat{\boldsymbol{\Sigma}}\|_{c2} \leq \|\hat{\boldsymbol{\Sigma}}^*\|_{c2} + \|\boldsymbol{\Omega} + \hat{\boldsymbol{\Sigma}}^e\|_{c2} = O_p(\sqrt{N})$ and

$$\max_{i \in \mathcal{M}} \|\Gamma_{\mathcal{T}_1}(\boldsymbol{x}_t, \epsilon_{it}^*)\|_\infty \leq \|\boldsymbol{\Lambda} \Gamma_{\mathcal{T}_1}(\boldsymbol{f}_t, \boldsymbol{u}_{\mathcal{M}t}')\|_\infty + \|\Gamma_{\mathcal{T}_1}(\boldsymbol{u}_{\mathcal{N}t}, \boldsymbol{u}_{\mathcal{M}t}')\|_\infty$$

$$+ \left[ \sqrt{N} \|\boldsymbol{\Lambda} \Gamma_{\mathcal{T}_1}(\boldsymbol{f}_t, \boldsymbol{u}_{\mathcal{N}t}')\|_\infty + \sqrt{N} \|\hat{\boldsymbol{\Omega}} - \boldsymbol{\Omega}\|_\infty + \|\boldsymbol{\Omega}\|_{c2} \right] \max_{i \in \mathcal{M}} \|\boldsymbol{\beta}_i^*\|_2$$

$$= O_p(\varphi_1/\sqrt{\xi_N}).$$

Hence $|\hat{V}^2 - \hat{V}_{\epsilon^*}^2| = O_p(M\tau/\xi_N)$, since $\varphi_1/\sqrt{\tau} \to 0$ as $N, T_1 \to \infty$. On the other hand, by the triangle inequality, note that

$$|\hat{V}_{\epsilon^*}^2 - V_{\epsilon^*}^2| \leq M \max_{i,j \in \mathcal{M}} \left| \Gamma_{\mathcal{T}_1}(\epsilon_{it}^*, \epsilon_{jt}^*) - \mathbb{E}(\epsilon_{it}^* \epsilon_{jt}^*) \right|$$

$$\leq M \|\Gamma_{\mathcal{T}_1}(\boldsymbol{u}_{\mathcal{M}t}, \boldsymbol{u}_{\mathcal{M}t}') - \boldsymbol{\Omega}_{\mathcal{M}}\|_\infty$$

$$+ M\left[2\sqrt{N}\|\Gamma_{\mathcal{T}_1}(\boldsymbol{u}_{\mathcal{N}t},\boldsymbol{u}'_{\mathcal{M}t})\|_\infty \max_{i\in\mathcal{M}}\|\boldsymbol{\beta}^*_i\|_2 + N\|\hat{\boldsymbol{\Omega}} - \boldsymbol{\Omega}\|_\infty \max_{i\in\mathcal{M}}\|\boldsymbol{\beta}^*_j\|_2^2\right]$$
$$= O_p(M\varphi_1/\xi_N).$$

Therefore, additionally with $V^2 = \lim_{M,N\to\infty} V^2_{\epsilon^*}$ under Assumption 6 (e), we have
$$\left|\hat{V}^2 - V^2\right| \le \left|\hat{V}^2 - \hat{V}^2_{\epsilon^*}\right| + \left|\hat{V}^2_{\epsilon^*} - V^2_{\epsilon^*}\right| + \left|V^2_{\epsilon^*} - V^2\right| = o_p(1),$$
as $M, N, T_1 \to \infty$ and $M\tau/\xi_N \to 0$.

To sum up, since $V$ is bounded from zero, we finally conclude that as $M, N, T_1 \to \infty$, for all $t \in \mathcal{T}_2$,
$$\hat{Z}_t = \frac{\sqrt{M}(\bar{\Delta}_t - \Delta_{M,t})}{\hat{V}} = \frac{M^{-\frac{1}{2}}\sum_{i\in\mathcal{M}}\epsilon^*_{it} + o_p(1)}{V + o_p(1)} = \frac{M^{-\frac{1}{2}}\sum_{i\in\mathcal{M}}\epsilon^*_{it}}{V} + o_p(1) \xrightarrow{d} N(0,1),$$
where the convergence in distribution holds under Assumption 6 (e).

# D  Additional Simulation Results

## D.1  Estimation and Prediction

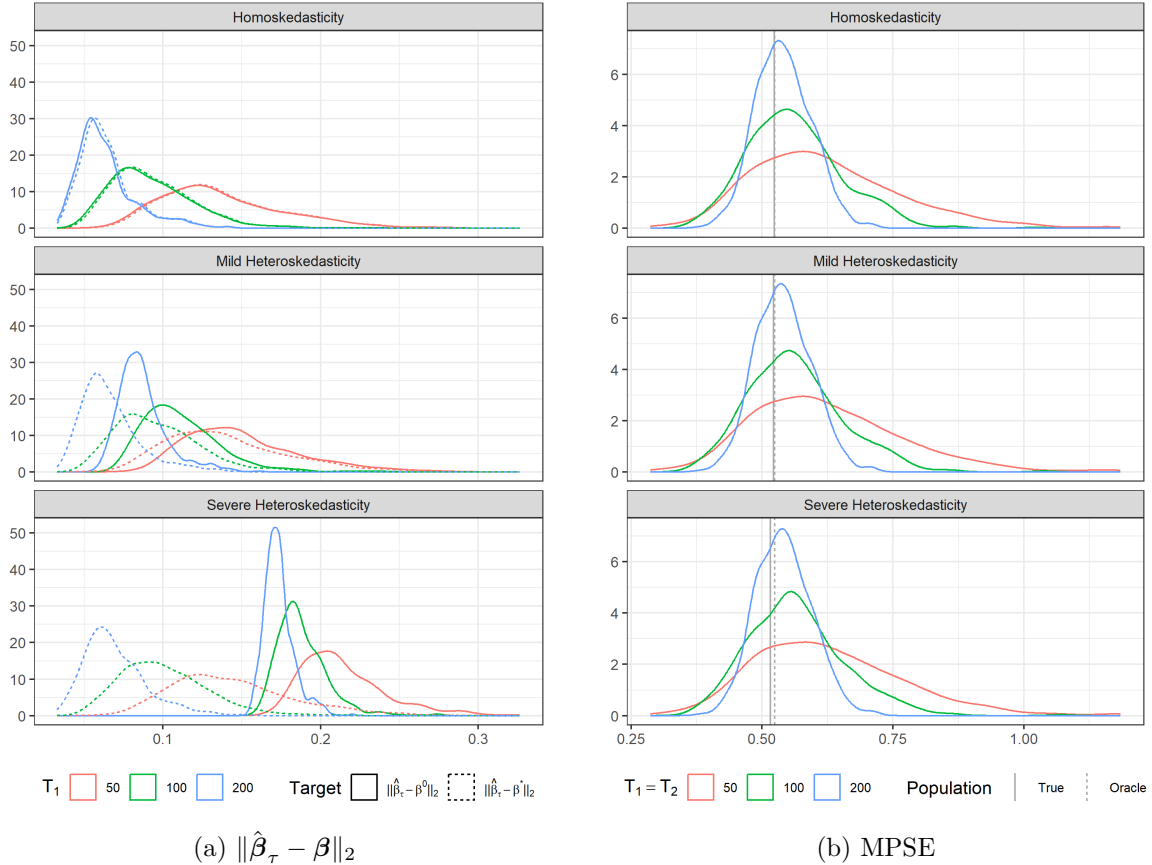

(a) $\|\hat{\boldsymbol{\beta}}_\tau - \boldsymbol{\beta}\|_2$  (b) MPSE

Figure D.1: Simulated probability density (DGP: *strong factors*, $N = 100$)

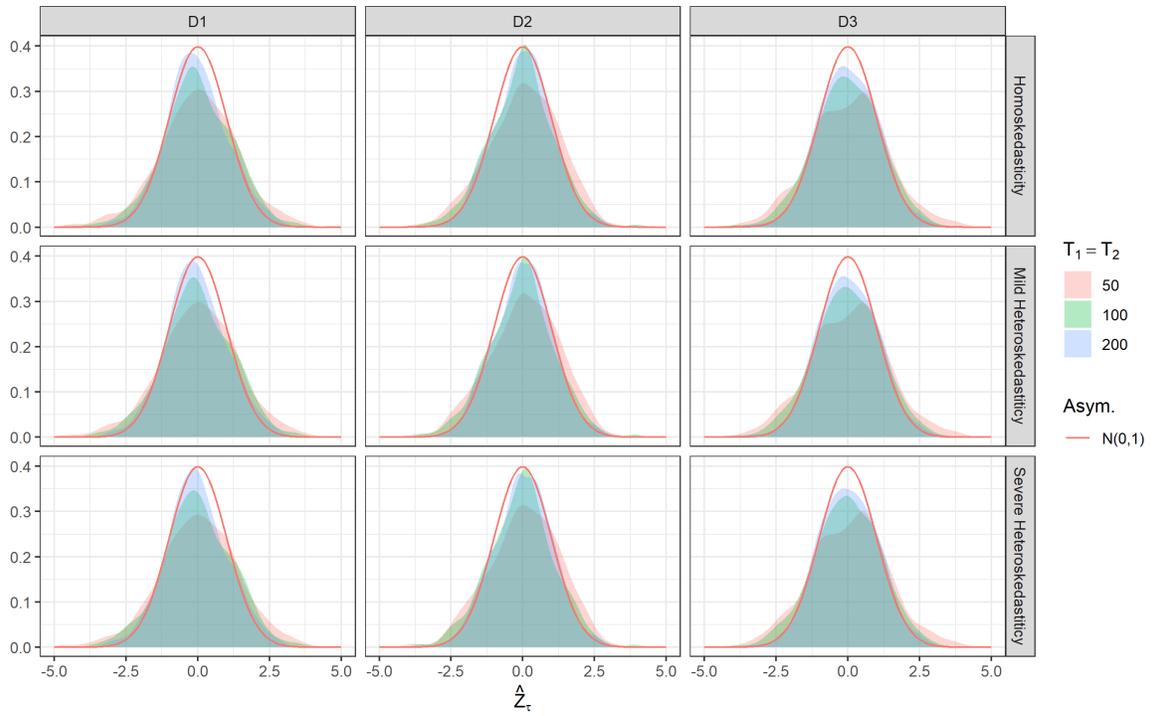

(a) Single treated unit

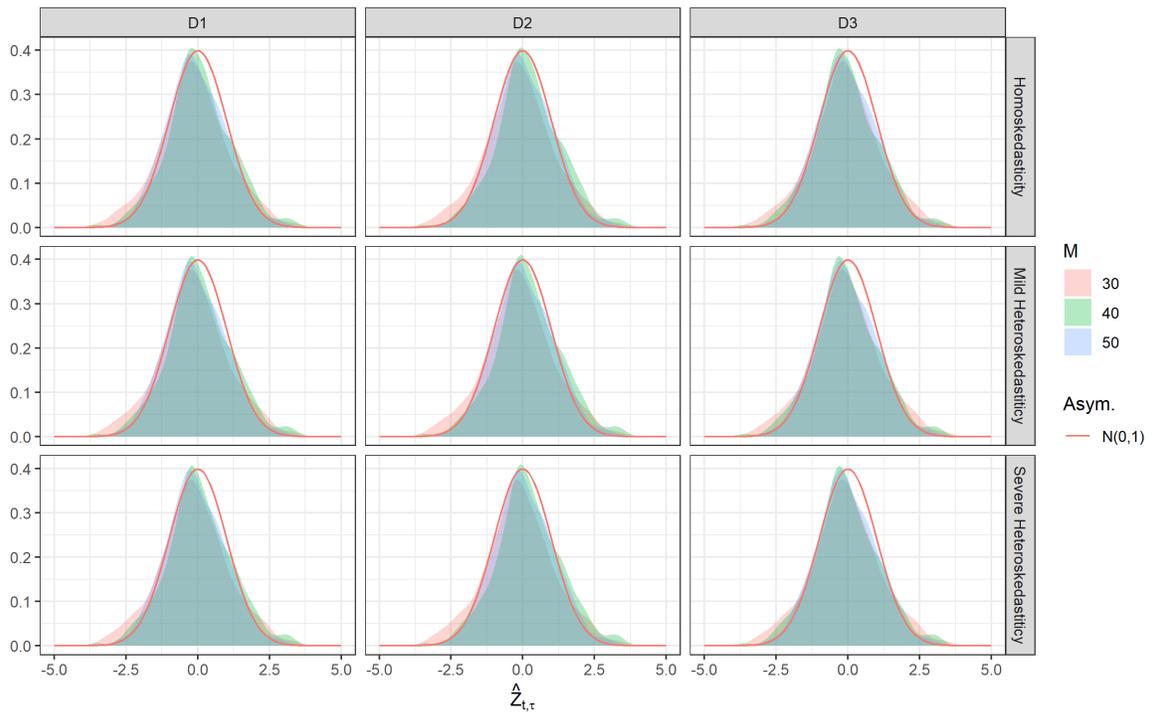

(b) Many treated units

Figure D.2: Simulated probability density of test statistic under $H_0$

Figure D.1 (a) compares the distributions of simulated $\|\hat{\boldsymbol{\beta}}_\tau - \boldsymbol{\beta}^*\|_2$ and $\|\hat{\boldsymbol{\beta}}_\tau - \boldsymbol{\beta}^0\|_2$. In all panels, it can be found that $\|\hat{\boldsymbol{\beta}}_\tau - \boldsymbol{\beta}^*\|_2$ converges to zero as the sample size goes large, which exhibits L2-consistency as suggested by Theorem 1. Furthermore, for any fixed sample size, clearly the severity of heteroskedasticity separates $\hat{\boldsymbol{\beta}}_\tau$ and $\boldsymbol{\beta}^0$, as implied by Corollary 1.

Figure D.1 (b) illustrates that the OOS MPSE is not affected by heteroskedasticity. The dotted vertical lines represent the oracle MSE $\mathbb{E}(\epsilon_t^{*2})$, while the solid vertical lines stand for the true variance $\mathbb{E}(\epsilon_t^2)$, and their gap is controlled by $\psi_{\max}$ as in (21). Corroborating Theorem 2, the simulated MPSE of L2-relaxation prediction is mainly determined by the oracle MSE, which is robust to heteroskedasticity. The simulation result provides finite sample evidence that although the consistency of the L2-relaxation estimator to $\boldsymbol{\beta}^0$ can be contaminated by heteroskedasticity, its prediction is robust as it approaches the oracle target, which rules out the variance-covariance structure of idiosyncratic errors.

## D.2 Panel Data Approach

The panels in Figure D.2 (a) show that the standard normal distribution approximates the simulated densities of the single-treated-unit test statistic (24) under the null hypothesis across different designs very well. Figure D.2 (b) demonstrates similar implications for the test statistic (29) with many treated units and only one post-treatment period. It shows that the CLT works along the cross section nicely under moderate $M$.

# E   Revisit PDA Empirical Examples

## E.1   Hong Kong Handover

The original empirical application of Hsiao et al. (2012) evaluates the impact of the political and economic integration of Hong Kong with mainland China. Hong Kong signed the Closer Economic Partnership Arrangement (CEPA) with mainland China, which started implementation in January 2004. HCW's dataset contains quarterly YoY real GDP growth rates ($T_1 = 44$) from a single treated unit (Hong Kong) and 24 control units (other countries). Obviously, this is a $N < T_1$ example.

We first use the data before treatment to compare the OOS prediction performance of different methods and to conduct the pre-treatment placebo test. For all data before 2004, we treat the subsample before 2002 as the training sample (36 months), while testing on the subsample after 2002 (8 months). As summarized in Table E.1, by comparing the RMPSE and OOS $R^2$, L2-relaxation performs the best among all listed methods. We also conduct the placebo test based on the pre-treatment OOS ATE estimated by each method, where L2-relaxation, LASSO, ridge, and PCA do not reject the null hypothesis that there is no treatment effect with non-zero mean.

Table E.1: OOS prediction performance and placebo test

| Rank | Method | RMPSE | OOS $R^2$ | ATE | $t$-stat. | $p$-value |
|---|---|---|---|---|---|---|
| 1 | L2-relaxation | 0.9978 | 0.8108 | -0.1078 | -0.8516 | 0.3944 |
| 2 | LASSO | 1.0325 | 0.7974 | 0.4412 | 1.4967 | 0.1345 |
| 3 | Ridge | 1.0949 | 0.7722 | 0.0491 | 0.1577 | 0.8747 |
| 4 | PCA ($PC_{p1}$) | 1.1261 | 0.7590 | 0.3184 | 1.0711 | 0.2841 |
| 5 | SCM | 1.2837 | 0.6869 | 0.7758 | 1.6613 | 0.0967 * |
| 6 | OLS | 1.4784 | 0.5847 | -0.7929 | -1.9216 | 0.0547 * |
| 7 | FS | 1.6426 | 0.4873 | 1.2587 | 7.7482 | 0.0000 *** |

Note: * $p$-value < 0.1, ** $p$-value < 0.05, *** $p$-value < 0.01.

We then replicate HCW's example by applying L2-relaxation. Figure E.1 plots the realized Hong Kong's real GDP growth rate and the counterfactual prediction. The gap between the observed post-treatment series and the counterfactual prediction is quite large, with positive treatment effects benefiting from the CEPA implementation. The ATE over the post-treatment period is about 2.65(%), with the corresponding $t$-statistic being 8.3524, which clearly rejects the zero mean ATE null hypothesis.

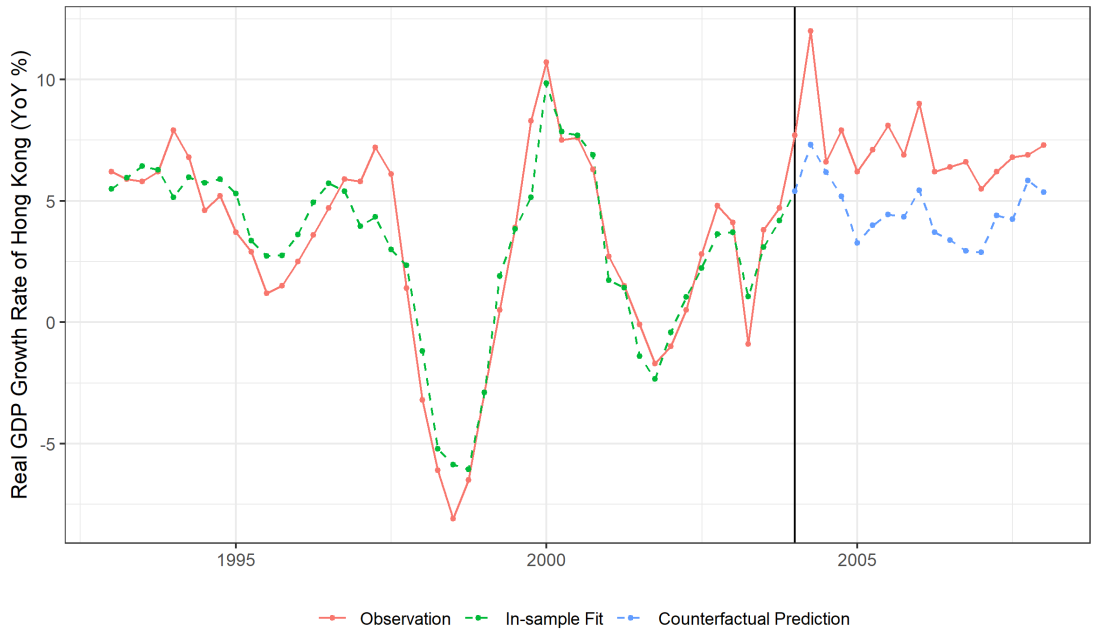

Figure E.1: Counterfactual prediction for Hong Kong real GDP growth rate

## E.2 Luxury Watch Import

Shi & Huang (2023) uses China's import data from the United Nations' *Comtrade* database

to evaluate the impact of the anti-corruption campaign that started in January 2013 after the "Eight-point" policy announcement. Instead of applying PDA methods to the monthly growth rates in Shi & Huang (2023), considering the seasonality in trade data, we use the YoY growth rates, at the expense of reducing the length of pre-treatment time series by one year. As the monthly import data for China are only available since 2010, we only have 24 months of YoY growth rates for China's import before treatment. The treated unit is China's import growth rate of luxury watches denominated in U.S. dollars (USD), and the control units are China's import growth rates of 87 other non-luxury commodities. Unlike the previous example, in this case, $N$ is much larger than $T_1$.

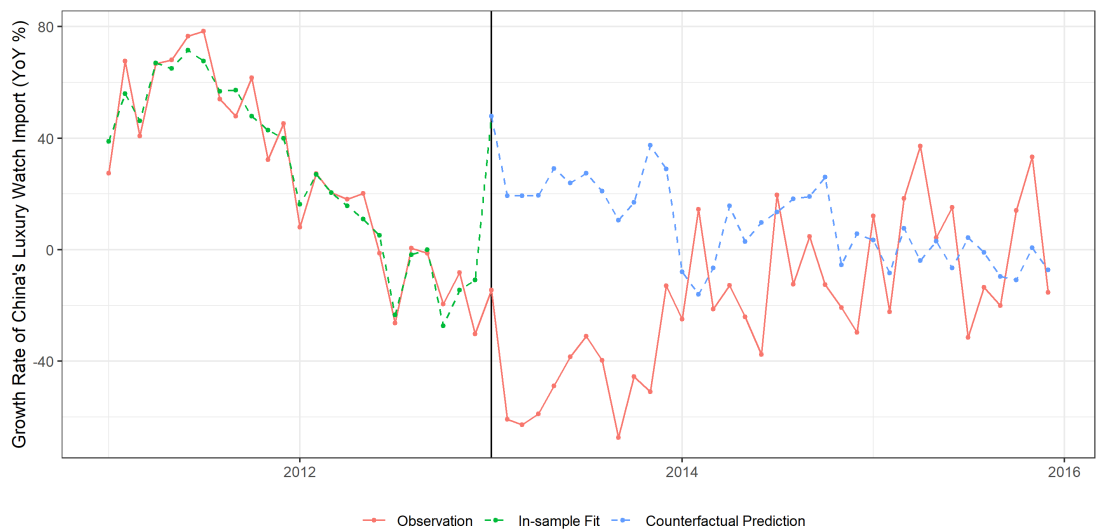

(a) Growth rate of China's luxury watch import

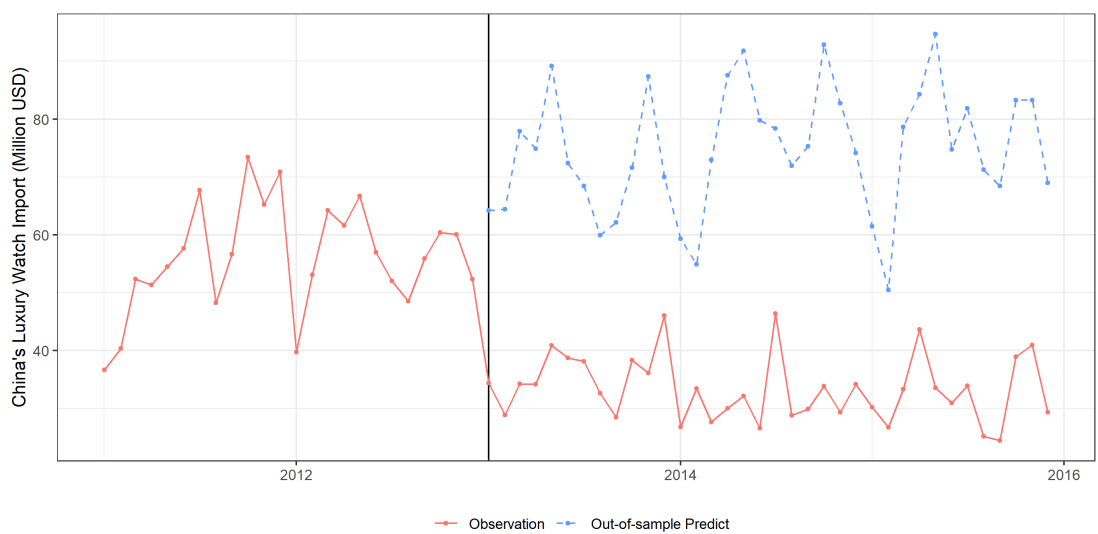

(b) China's luxury watch import

Figure E.2: Counterfactual prediction for China's luxury watch import

As $T_1 = 24$ being short, the pre-treatment OOS test is hardly applicable. We employ L2-relaxation for the PDA evaluation. The upper panel in Figure E.2 plots the realized growth rate of China's luxury watch import (YoY) and the counterfactual prediction, while the lower panel plots the import amount in USD. From all panels, the differences between the actual post-treatment series and the counterfactual predicts are significant with negative treatment effects in the first year after the policy implementation. The ATE over the post-treatment period is -27.92(%), with the corresponding $t$-statistic being -2.5493, which rejects the zero mean ATE null hypothesis with $p$-value = 0.0108.